# Participatory Mapping of Local Green Hydrogen Cost-Potentials in Sub-Saharan Africa

C. Winkler[1,2], H. Heinrichs,[1,*], S. Ishmam[1,2], B. Bayat [3,4], A. Lahnaoui [1], S. Agbo[5], E. U. Peña Sanchez[1,2], D. Franzmann[1,2], N. Oijeabou[1], C. Koerner[1], Y. Michael[1], B. Oloruntoba[3], C. Montzka[3], H. Vereecken[3], H. Hendricks Franssen[3], J. Brendt[5], S. Brauner[1] , W. Kuckshinrichs [6], S. Venghaus [1,7], D. Kone[7] , B. Korgo[8,11] , K. Ogunjobi[8] , J. Olwoch[9], V. Chiteculo [9] Z. Getenga[10], J. Linßen,[1] and D. Stolten[1,2]

1 Forschungszentrum Jülich GmbH, Institute of Energy and Climate Research – Jülich Systems Analysis (IEK-3), 52425 Jülich, Germany
2 RWTH Aachen University, Chair for Fuel Cells, Faculty of Mechanical Engineering, 52062 Aachen, Germany
3 Institute of Bio- and Geosciences: Agrosphere (IBG-3), Forschungszentrum Jülich GmbH, 52425 Jülich, Germany
4 Center for Remote Sensing and GIS Research, Faculty of Earth Sciences, Shahid Beheshti University, Tehran, Iran
5 Forschungszentrum Jülich GmbH, Corporate Development, (UE), 52425 Jülich, Germany.
6 Forschungszentrum Jülich GmbH, Institute of Energy and Climate Research – System Analysis and Technology Evaluation (IEK-STE), 52425 Jülich, Germany
7 RWTH Aachen University, School of Business and Economics, 52062 Aachen, Germany
8 West African Science Service Centre on Climate Change and Adapted Land Use (WASCAL), Agostino Road, PMB CT 504, Accra, Ghana
9 Southern African Science Service Centre for Climate Change and Adaptive Land Management (SASSCAL), 28 Robert Mugabe Avenue, Windhoek, Namibia.
10 Machakos University, Machakos County, Kenya
11 University Joseph KI-ZERBO, Ouagadougou, Burkina-Faso

* Corresponding author: h.heinrichs@fz-juelich.de

## Abstract

Green hydrogen is a promising solution within carbon free energy systems with Sub-Saharan Africa being a possibly well-suited candidate for its production. However, green hydrogen in Sub-Saharan Africa is not yet investigated in detail. This work determines the green hydrogen cost-potential for green hydrogen within this region. Therefore, a potential analysis for PV, wind and hydropower, groundwater analysis, and energy systems optimization are conducted. The results are evaluated under local socio-economic factors. Results show that hydrogen costs start at 1.6 EUR/kg in Mauritania with a total potential of ~259 TWh/a under 2 EUR/kg in 2050. Two third of the regions experience groundwater limitations and need desalination at surplus costs of ~1% of hydrogen costs. Socio-economic analysis show, that green hydrogen deployment can be hindered along the Upper Guinea Coast and the African Great Lakes, driven by limited energy access, low labor costs in West Africa, and high labor potential in other regions.

**Keywords (max 6):** land eligibility, renewable energy, sustainable groundwater, desalination, energy system optimization, socio-economic indicators

# 1 Introduction

Even though in comparison to the rest of the world, Africa contributes only 4% of carbon emissions and, hence, global warming, yet it bears the biggest brunt of its impact (Serdeczny et al. 2017). Incidences of severe drought across Sub-Saharan Africa in 2023 resulted in millions of people going hungry (Pavlidis et al. 2024). Countries like Malawi and Zambia declared national emergencies in 2023 with 9 and 6 million people impacted by the drought respectively (Mutsaka and Imray 2024). The changing rainfall pattern, extreme heat waves, and rising temperature are clear indicators for a changing climate in the region (Amjath-Babu et al. 2016). Projections show that East Africa is at a higher risking of flooding while Southern Africa experiences the most severe decrease in precipitation with attendant drought and food shortages (Serdeczny et al. 2017).

Despite this, Sub-Saharan Africa holds huge renewable energy potential ranging from abundant solar and wind energy to hydropower (Hafner, Tagliapietra, and de Strasser 2018), while the region still has more than half its population without access to sustainable and clean energy (IRENA 2024). The per capita energy consumption is one of the lowest in the world, varying even more significantly among different socio-economic groups (Khraief, Omoju, and Shahbaz 2016). In parallel, conventional sources such as coal and fossil fuel still dominate the regions energy markets, contributing also dominantly to the pollution and the carbon emission in the subregion (Mukhtar, Adun, and Cai 2023). Coal for example still accounts for 60% of the installed electricity capacity in Southern Africa (Chowdhury, Deshmukh, and Armstrong 2022). With growing population and GDP, electricity consumption in the sub-region is projected to reach a peak demand of 115GW by 2040 (SAPP 2017). Hence, cost-effective and climate-friendly energy sources must be sought to match the growing demand. In addition, this is one of the important steps necessary to accelerate economic development in the region and improve the quality of life of the people. Against this background, the downward trend in PV and wind energy capital costs present them as options towards attaining the carbon target of the region (Spalding-Fecher, Senatla, and Yamba 2017).

As the variable nature of the electricity feed-in from wind turbines and solar photovoltaics requires energy storages across several time horizons, the subject of green hydrogen to balance seasonal variations of variable renewable energies and to decarbonize challenging sectors (Egeland-Eriksen, Hajizadeh, and Sartori 2021) has continued to be on the headlines in the last couple of years. This has also been the case in Sub-Saharan Africa. Hydrogen when produced green is believed to be the oil of the future, in one way in addressing the issue of decarbonization and climate change and in another way ensuring sustainable energy supply (Ballo, Koffi, and Korgo 2022). With huge renewable energy potential, Sub-Saharan Africa has the possibility of producing green hydrogen to contribute to its decarbonization agenda, export green hydrogen across the world and transition to green energy.

Groundwater assessments results have been published at global, continental, and regional scales. On a global scale, early groundwater recharge studies, i.e., the amount of precipitation that reaches the aquifers, by L'vovič (L'vovič 1979) laid the groundwork using baseflow components to map global groundwater recharge. Further advancements in providing global groundwater recharge information were made by hydrologists. For instance, Döll et al. (Döll, Lehner, and Kaspar 2002) generated a global groundwater recharge map employing the hydrological modeling (Alcamo et al. 2003; Döll, Kaspar, and Lehner 2003). Groundwater studies at the regional level have yielded significant findings across various parts of Africa, including southern Africa (Abiye 2016; Xu and Beekman 2003), northern Africa (Edmunds and

Wright 1979; Guendouz et al. 2003; Sturchio et al. 2004), and western Africa (Edmunds and Gaye 1994; Favreau et al. 2009; Leblanc et al. 2008; Leduc, Favreau, and Schroeter 2001). Additionally, a detailed and long-term groundwater recharge map for the entire African continent, spanning from 1970 to 2019, was developed using ground-based measurements (MacDonald et al. 2021). They produced long-term groundwater recharge map for Africa, showing a statistical relationship between long-term average rainfall and groundwater recharge across the continent and underscored the influence of climate and terrestrial factors on local-scale recharge dynamics. Their findings provide a robust dataset of ground-based estimates, offering valuable insights into groundwater renewability and serving as a baseline for assessing water security in Africa. Recently, a relatively high-resolution (i.e., 10 km) groundwater recharge and sustainable yield maps have been generated spanning five decades (1965-2014) in Africa to assess groundwater sustainability and promote its sustainable use (Bayat et al. 2023). However, still remains a strong interest in not only quantifying groundwater recharge and sustainable yield by means of physically based modeling under current climatic conditions but also in estimating projected recharge under various climate change scenarios, particularly for the African continent. This is especially important for scientists, water managers, and local communities, as it marks a significant step towards promoting sustainable groundwater use in Africa. The groundwater part of current study focuses specifically on these two aspects: assessing the sustainable yield of groundwater under both present and future climate projections. To the best of our knowledge, high resolution projected groundwater sustainable yield maps to quantify the climate change impacts specifically for Africa have not been published to date.

In the realm of hydrogen generation, limited studies have been conducted for African countries. Franzmann et al. (2023) investigated the cost-potentials for liquid hydrogen based on a holistic energy system optimization approach including the whole infrastructure until an export harbor. They illustrated for select countries (Namibia and Libya) that liquid hydrogen exports from Africa exceeding 1 PWh annually could be feasible, with costs starting at 2 EUR/kgH$_2$ in 2050 and local gaseous hydrogen costs starting at 1.50 EUR/kgH$_2$. Meanwhile, IRENA (2022) established local gaseous hydrogen costs for the African continent, starting at 1.1 USD/kgH2 in 2050. Additionally, Mukelabai et al. (2022) highlighted the pivotal role of water limitations in Sub-Saharan Africa for electrolysis-based green hydrogen production from renewables. There also exist studies, investigating hydrogen export potential of single plant projects on a sub-country scale. In the example case of Nigeria, Kamil et al. (Kamil, Samuel, and Khan 2024) determine the green hydrogen potential of PV plants in Nigeria in combination with ammonia production simultaneously. Ayodele and Munda (Ayodele and Munda 2019) investigates the cost of green hydrogen production for 15 different wind parks for South Africa being at 1.4 USD/kgH2 to ~40 USD/kgH2 for the cost year of 2019. Even as hydrogen generation costs for Sub-Saharan Africa are still uncertain in literature, its economic relevance is perceived as high (Cardinale 2023). Therefore, a detailed and consistent analysis of green hydrogen production costs for Sub-Saharan African is needed to allow for direct comparison of sub-regions and decision support to develop a green hydrogen economy in Sub-Saharan Africa.

Regarding the socio-economic impact of green hydrogen project using composite indicator, the focus was on electrification, electricity access via renewable energy, and their effects on employment and society. Most studies concentrate on national-level analysis, with only a few focusing on higher resolution local impact in the Sub-Saharan African case. Overall, recent research proved that energy projects have positively impacted local populations in Sub-Saharan Africa. These benefits include increased income levels (Lawal et al. 2020), improved education (Litzow, Pattanayak, and Thinley 2019; Sovacool and Ryan 2016), and job creation

(Ravillard et al. 2021; Shirley et al. 2019). Furthermore, electrification has been linked to poverty reduction by enabling income-generating activities and thus enhancing the quality of life (Khandker, Barnes, and Samad 2012; Vernet et al. 2019). Additionally, energy projects improve health outcomes by reducing indoor air pollution from traditional biomass cooking (Barron and Torero 2017).

Most existing literature focused primarily at the national level and only on the socio-economic impacts of power generation. For instance, (Ondraczek, Komendantova, and Patt 2015) studied the socio-economic impacts of solar power projects in East Africa. The impact of wind energy projects on local development was assessed by (Rao 2019). (Kirchherr and Charles 2016) investigated the economic growth, population displacement effects and environmental impacts of large-scale hydropower projects. While (Mariita 2002) discussed the socio-economic impact of geothermal power plants on poor rural communities in Kenya. When it comes to green hydrogen projects, there is agreement that these projects can stimulate local economies, however, most studies have focused on the region's readiness to adopt green hydrogen technologies (Nnachi, Richards, and Hamam 2024; Brauner et al. 2023). Against this background, the detailed socio-economic impact analysis aims, on the one hand, to assess at a higher spatial resolution the locations of renewable energy projects with the highest potential for promoting energy access. On the other hand, it evaluates the impact of green hydrogen project locations on job creation through direct local employment analysis.

Developing a green hydrogen economy in Sub-Saharan Africa is a multifaceted challenge calling for joint efforts of different disciplines within one joint approach. Hence, this study applies a multidisciplinary approach (Ishmam et al. 2024) to 31 countries in Sub-Saharan Africa. By this we aim to provide a comprehensive basis for decision support for green hydrogen projects in the analyzed region. This is further supported by making the obtained results available via a web-based GUI and within the Appendix and within the Supplementary.

## 2 Methodology

The underlying approach for deriving cost-potentials of green hydrogen for selected regions in Sub-Saharan Africa combines several assessments like land eligibility and placement for open-field photovoltaic and onshore wind turbines. Also assessed are local preferences, renewable energy potentials, sustainable water supply, and local green hydrogen potentials as well as socio-economic impacts (see Figure 1). While each step is further described in the following subsections the full level of detail can be found in Ishmam et al. (Ishmam et al. 2024).

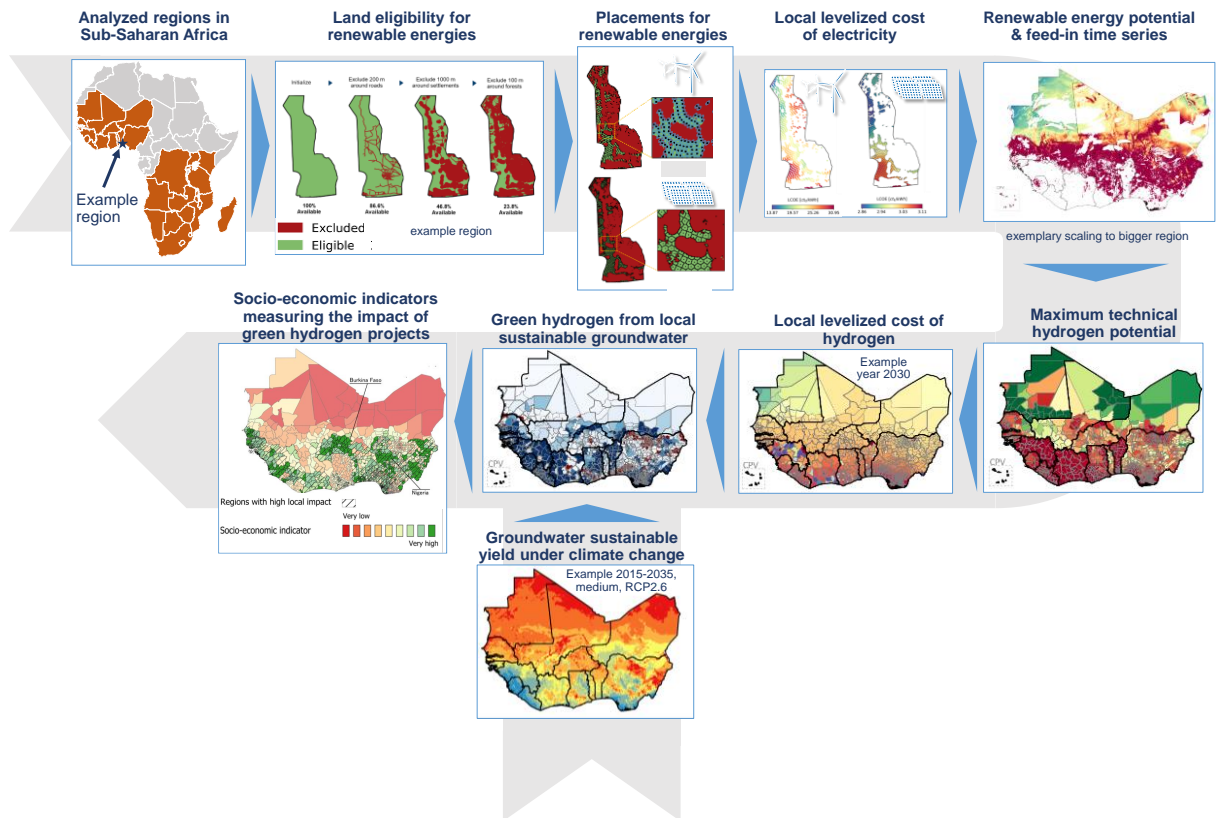

**Figure 1: Stepwise overview of applied methodology based on Ishmam et al. (Ishmam, et al. 2024)**

## 2.1 Land eligibility assessment for open-field photovoltaic and onshore wind turbines

An analysis of land eligibility is carried out to identify suitable areas for renewable energy facilities based on 33 criteria and buffer distances for onshore wind turbines and open-field PV. Generally, the analysis considers several criteria, including land use, topography, environmental constraints, and social and economic criteria, which are used to exclude areas where wind turbines and PV modules cannot or should not be placed. The land eligibility criteria for this study were carefully selected from the body of current literature with additional input from local partners and regional stakeholders.

The land eligibility model GLAES (GLAES 2024) and the open-source general-purpose geospatial toolkit GeoKit (GeoKit 2024) were used. The resulting suitable areas serve as the foundation for evaluating the potential for renewable energy. For a more detailed explanation of the criteria used and the methodology, please refer to Ishmam et al. (Ishmam et al. 2024).

## 2.2 Renewable energy potential assessment

PV and onshore wind energy assessments consist of hourly-resolved energy production simulations over 20 years (2000-2019) using reanalysis weather data from ERA5 (Hersbach et al. 2018), the Global Solar Atlas[i] (World Bank Group and Solargis s.r.o. 2023) or the Global

---

[i] "Data obtained from the Global Solar Atlas 2.0, a free, web-based application is developed and operated by the company Solargis s.r.o. on behalf of the World Bank Group, utilizing Solargis data, with funding provided by the Energy Sector Management Assistance Program (ESMAP)."

Wind Atlas (Davis et al. 2023) for PV and onshore wind accordingly. The selected 20-year time span allows to derive robust levelized cost of electricity at each individual location, which was derived from the preceding land eligibility assessment. The simulations were performed using the RESKit python model (Ryberg et al. 2019). Hydropower represents an important power source for many African nations as of today and was therefore considered after consultation with the project partners. For the hydropower assessment, the "normal" generation scenario and hydropower fleets larger than 10MW of generation capacity for conventional (with dam) and run-of-river hydropower plants derived by Sterl et al. (Sterl et al. 2021) were used. The source time series were linearly resampled to hourly to match PV and onshore wind. The techno-economic parameters used to obtain the levelized cost of electricity and further details are provided in the methodology publication (Ishmam et al. 2024).

## 2.3 Sustainable water supply assessment

For water availability assessment, first, groundwater sustainable yield has been calculated for 2020, 2030, and 2050 under RCP2.6 and RCP8.5 scenarios. As an alternative water supply option for regions without sustainable groundwater yield costs for seawater desalination and water transport via pipeline was considered.

### 2.3.1 Sustainable groundwater supply

To ensure a sustainable water supply for green hydrogen production, we quantified the long-term (2015 - 2100) groundwater sustainable yield. The averages from 2015 – 2035, 2015 – 2045, and 2036 – 2065 are considered as representative of 2020, 2030, and 2050, respectively under RCP2.6 and RCP8.5 scenarios. The determination of groundwater sustainable yield has been carried out, taking into consideration three key aspects (Bayat et al., 2023): simulated groundwater recharge, estimated environmental flow (i.e., minimum ecological water requirement), and all sectoral water consumptions. For groundwater sustainable yield, we considered two climate scenarios: RCP2.6, which is known as optimistic indicating a limited increase in greenhouse gas concentrations, and RCP8.5, which is known as pessimistic indicating a longer increase of greenhouse gas concentrations to higher values. Climate scenarios are hypothetical representations of future climatic conditions based on greenhouse gas (GHG) emissions, used to assess vulnerability to climate change (Tramberend et al. 2021). These hypothetical representations are captured in different setups designed to represent a range of possible emissions trajectories and corresponding radiative forcing levels called representative concentration pathways (RCP). Four main RCPs exist specifying particular radiative forcing levels by the year 2100 (Akinsanola et al. 2015). Noteworthy are the implications of the RCPs on temperature. The projected global mean surface temperature change ranges from 0.3°C to 0.7°C for the period 2016-2035, and by the end of the 21st century, it is likely to hit 1.7°C for RCP2.6, and 4.8°C for RCP 8.5 compared to the historical industrial period. This study utilizes RCP 2.6 and RCP 8.5, representing optimistic and pessimistic emission scenarios, respectively. Additionally, we have calculated three different scenarios for groundwater sustainable yield: (i) a conservative case assuming only 10% of the annual recharge is allowed for green hydrogen production, (ii) a medium case where up to 40% of the annual recharge can be used and, (iii) an extreme case where about 70% of annual recharge is allowed for the green hydrogen production. Lastly, the levelized cost of groundwater extraction were modeled geospatially and assigned to each region individually.

### 2.3.2 Desalinating seawater and water transport

For regions with insufficient amounts of sustainable groundwater yields seawater desalination and water transport via pipeline is considered as an alternative. For this the cost of water transport to the centroid of each region based on electricity cost for pumping as well as distance to shore and elevation and the cost for seawater desalination are implemented in the models to derive the green hydrogen potentials. As long as the sustainable groundwater potential is not exceeded, the model can select the cheapest water provision option.

## 2.4 Local green hydrogen potential assessment

Based on the renewable cost-potentials and the water supply options, the cost-potential curves for local green hydrogen are calculated. Each region is based on the "GID-2" definition from GADM (Database of Global Administrative Areas) (GADM 2023) and is modeled as an independent energy system within the ETHOS.FINE framework (Welder et al. 2018; Groß et al. 2024). Each regional energy system can utilize onshore wind turbines, open-field PV, existing hydropower plants up to their maximum regional potential as a renewable electricity input. The green hydrogen production is then modeled by using PEM electrolysis and LI-ion battery storages for a cost-optimal balance of potential curtailment of renewable energies. The cost-potential curve is achieved by conducting cost minimizations for different expansion steps of the energy systems based on approaches like in Franzmann et al. (Franzmann et al. 2023). Further details regarding the cost assumptions can be found in Ishmam et al. (Ishmam, et al. 2024).

National electricity and hydrogen demand are prioritized over potential export schemes to ensure a just energy transition (Brauner et al. 2023) and increase acceptance (Loehr et al. 2022). Figure 2 illustrates schematically how a country-specific share of the energy potentials needs to be set aside to first cover local energy demands including their future projected growth. The national demands for electricity and hydrogen are projected to the respective year based on the "Net Zero 2050" scenario from the NGFS Climate Scenarios Database (Richters, Bertram, Kriegler, Al Khourdajie, et al. 2022). In the exemplary case demonstrated below, 52% of the energetic potential would need to be set aside for local demands, thereof 33% for electricity demand and 19% for hydrogen demand. To quantify both electricity demand and hydrogen together, relative to the available hydrogen potential, the electricity demand value is converted to the equivalent value in hydrogen quantity by applying the conversion efficiency. The comparison between potential and demand is performed at national level under the assumption that national energy policies will seek to cover demand first. An application at sub-national level is not meaningful as the distortion by inter-regional energy exchange could not be quantified whereas international energy trade balances are usually accessible.

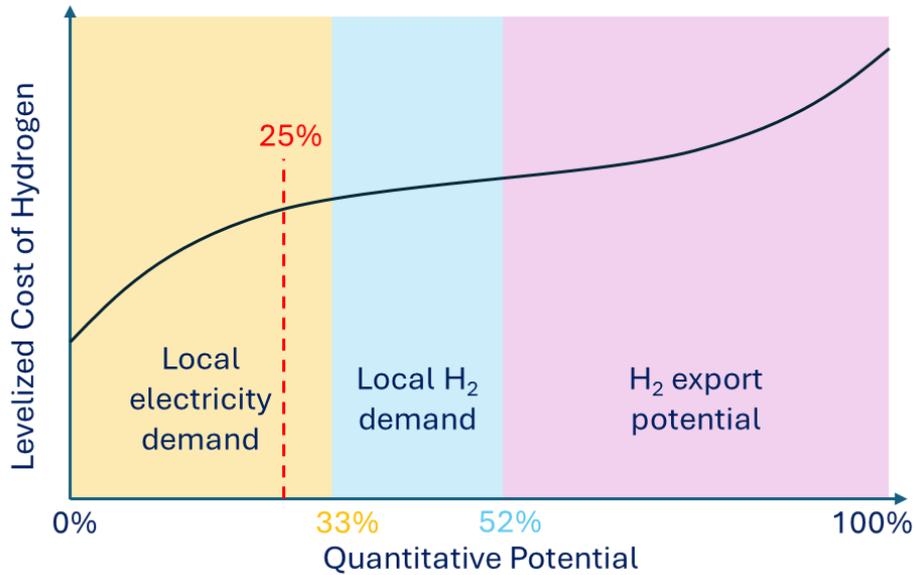

**Figure 2: Exemplary Cost-Potential Curve with 25% potential expansion shown in red and exemplary energy shares set aside to cover local electricity and hydrogen demand first.**

## 2.5 Socio-economic impact assessment

Socio-economic impacts were assessed by constructing composite indicators to provide a comprehensive and quantifiable measure of green hydrogen impact on sustainable development goals. The focus of the study was on datasets that provide an indication of sustainable development performance, hence, the sustainable development goals (SDGs) most impacted by green hydrogen projects were chosen as a reference, whether directly or indirectly. In particular, data was collected and cross-checked from a variety of sources to address concerns about incomplete or inadequate data, both during data selection and subsequent analysis. To evaluate the feasibility of using socio-economic indicators to measure the specified SDGs, additional research was conducted, along with spatial data analysis. Prioritization of social development goals based on local visions, as revealed by stakeholder interviews conducted during the project, was also reflected in the study (Brauner et al., 2023). The socio-economic indicator is aggregated based on three sub-indicators as developed in Ishmam et al. (Ishmam et al. 2024) including access to energy, macroeconomic effects and other indirect effects (biomass dependence and poverty).

## 3 Results

This chapter shows key results and findings along the steps of the applied approach. Starting with the land eligibility, the renewable energy potentials for all considered technologies depending on the eligible areas are presented. Afterwards, the sustainable groundwater availability is shown, and the local green hydrogen potentials are described based on all previously presented results. This is complemented by the local socio-economic impact of the green hydrogen production.

## 3.1 Land eligibility assessment

In the following the distribution of the local preferences collected and the resulting land areas eligible for open-field photovoltaics, onshore wind turbines and geothermal power plants

(where applicable) are presented. Within these results eligibility patterns and underlying drivers are analyzed.

### 3.1.1 Land eligibility assessment for open-field photovoltaic and onshore wind turbines

The local preferences for the set of 33 land eligibility criteria, collected and processed accordingly are displayed in Figure 3 with buffer values normalized per criterion. The highest values per criterion are represented by 1 while the lowest ones by 0. The absolute values of the buffer values received and subsequently applied are listed in the supplementary Table S1 and Table S2. Each grey dot within the figure represents the buffer value obtained from countries that responded to the survey. The distribution of local preferences for each criterion highlights the distinct inclinations of regional stakeholders, including community members, governmental bodies, and international institutions in every country.

Upon comparing the buffer values for open-field PV and onshore wind turbines, it is evident that open-field PV has greater variations. Notably, the buffer values for the criteria "Airports," "Historical Sites," "Military Areas," "Natural Habitats," "Biospheres," "Bird Areas," and "Natural Monuments" show the largest distributions. Conversely, the buffer values selected for onshore wind for the criteria "Military Areas," "Coastline," and "Protected Landscape" show the largest deviations.

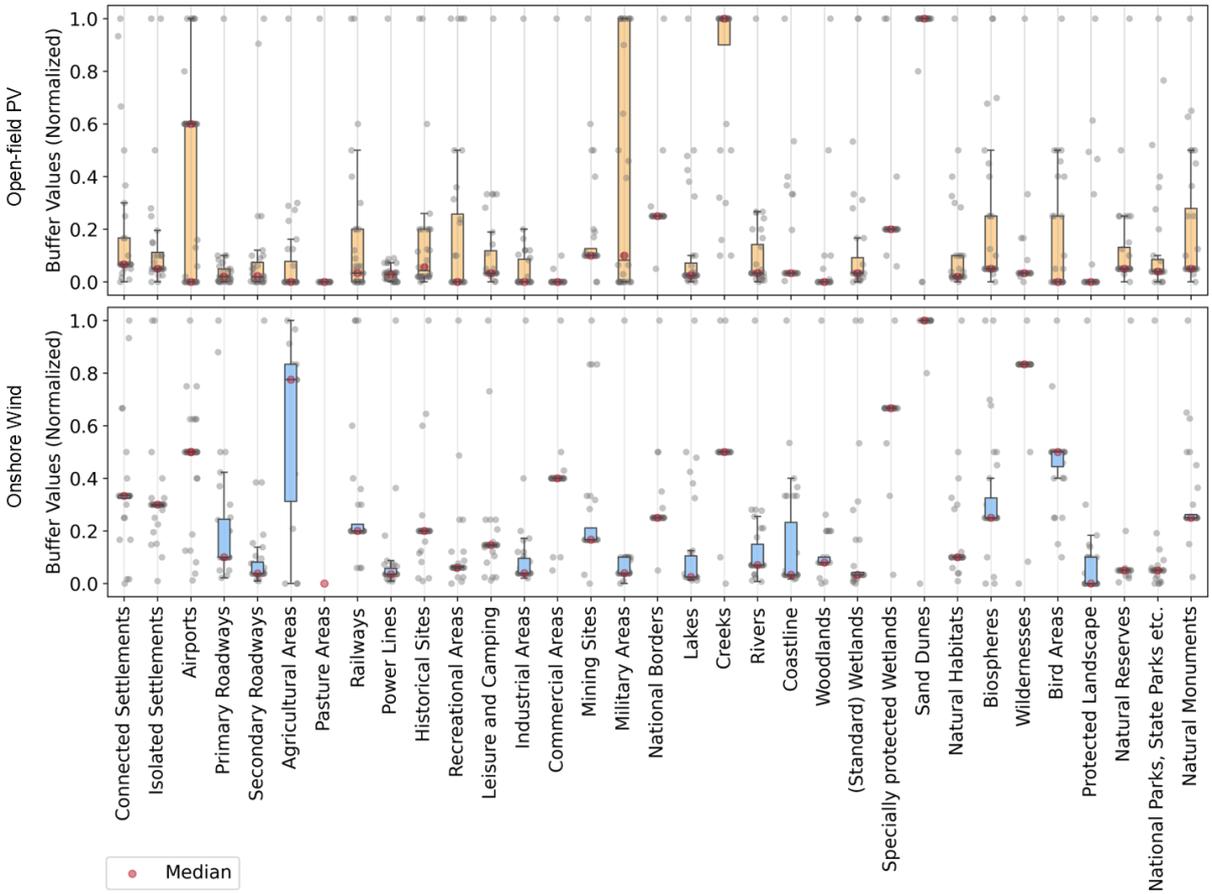

**Figure 3: Distribution of buffer values received for each criterion from partner countries and their resulting medians**

In order to gain a comprehensive understanding of the process behind determining a country's total land eligibility, it is necessary to examine each individual criterion that excludes land separately. Through this method, it becomes clear that certain criteria, such as "Woodlands", "Isolated Settlements", "Connected Settlements", "Agricultural Areas", and "Pasture Areas", lead to the exclusion of large areas of land (over 40 - 50%) when considering open-field PV (see Figure 4). On the other hand, for onshore wind, the majority of excluded land area (over 40 - 50%) in most countries can be attributed to "Woodlands" and "Isolated Settlements", as illustrated in Figure 5. Very low overall eligibilities in countries such as Guinea resulted from the combined high exclusions of land area by criteria such as "Agricultural Areas", "Pasture Areas", "Secondary Roadways" and "Woodlands".

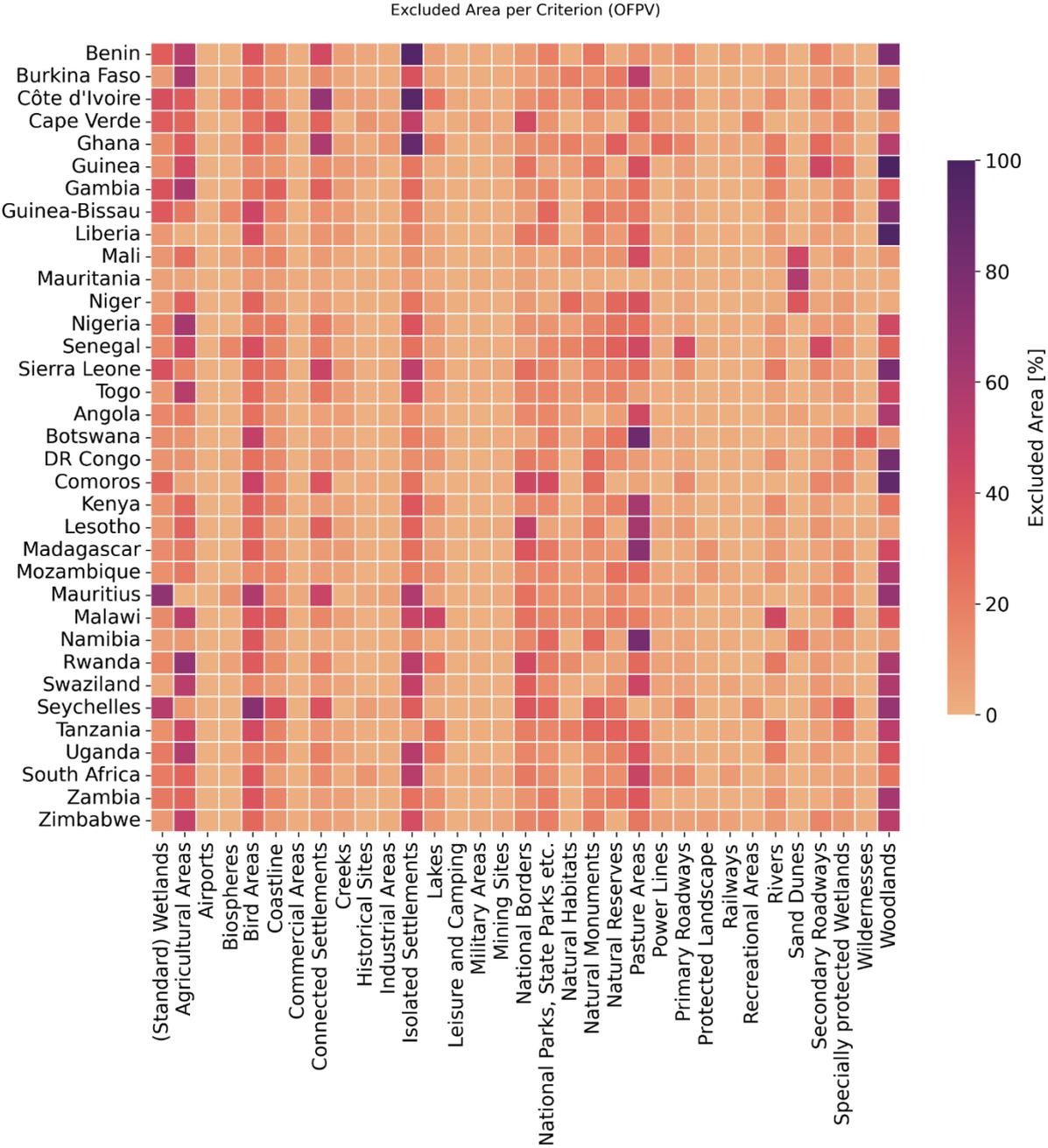

**Figure 4: Excluded area per criterion and country in western, southern, and eastern Africa for open-field PV**

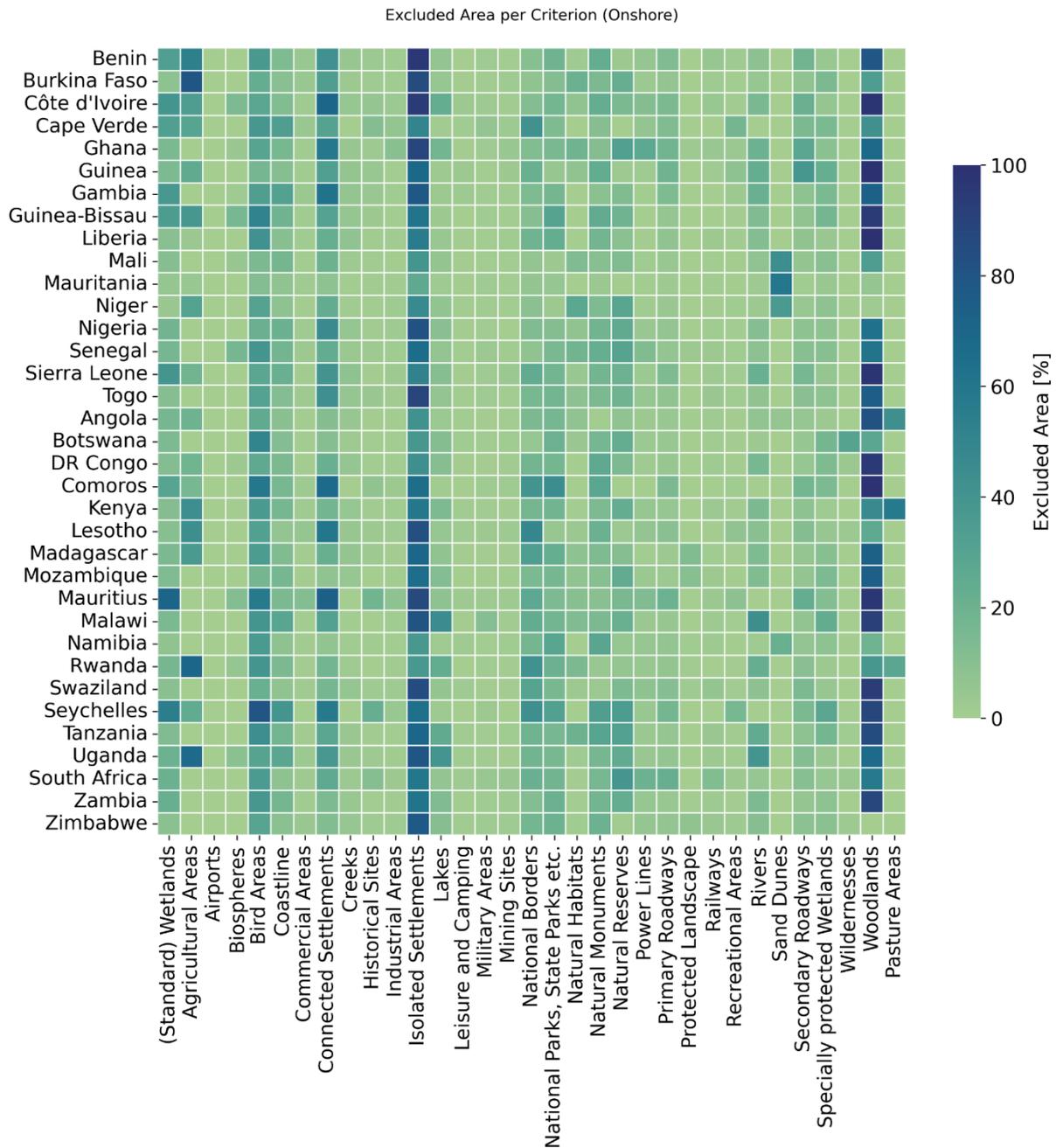

**Figure 5: Excluded area per criterion and country in western, southern, and eastern Africa for onshore wind**

Figure 6 (a) provides an overview of the potential land areas that are identified as potentially eligible for open-field PV parks in West, Southern and East Africa. In the case of West Africa, approximately 36% of the available land is eligible for open-field PV, with larger areas of eligibility in the northern part of the region and smaller patches throughout the southern part. Eligibility rates per country vary from ~0.1% in Guinea to around 50% in Niger, Mali, and Mauritania. In Southern and East Africa, larger areas of eligibility are found in the central and southern parts of the region. Approximately 24% of the available land in Southern and East Africa is eligible for open-field PV, with eligibility rates per country ranging from ~0.1% in Seychelles to around 36% in Mozambique. Protected forests and pastures are key constraints to the eligible land areas in both regions, while sand dunes in western Africa also exclude large areas from eligibility.

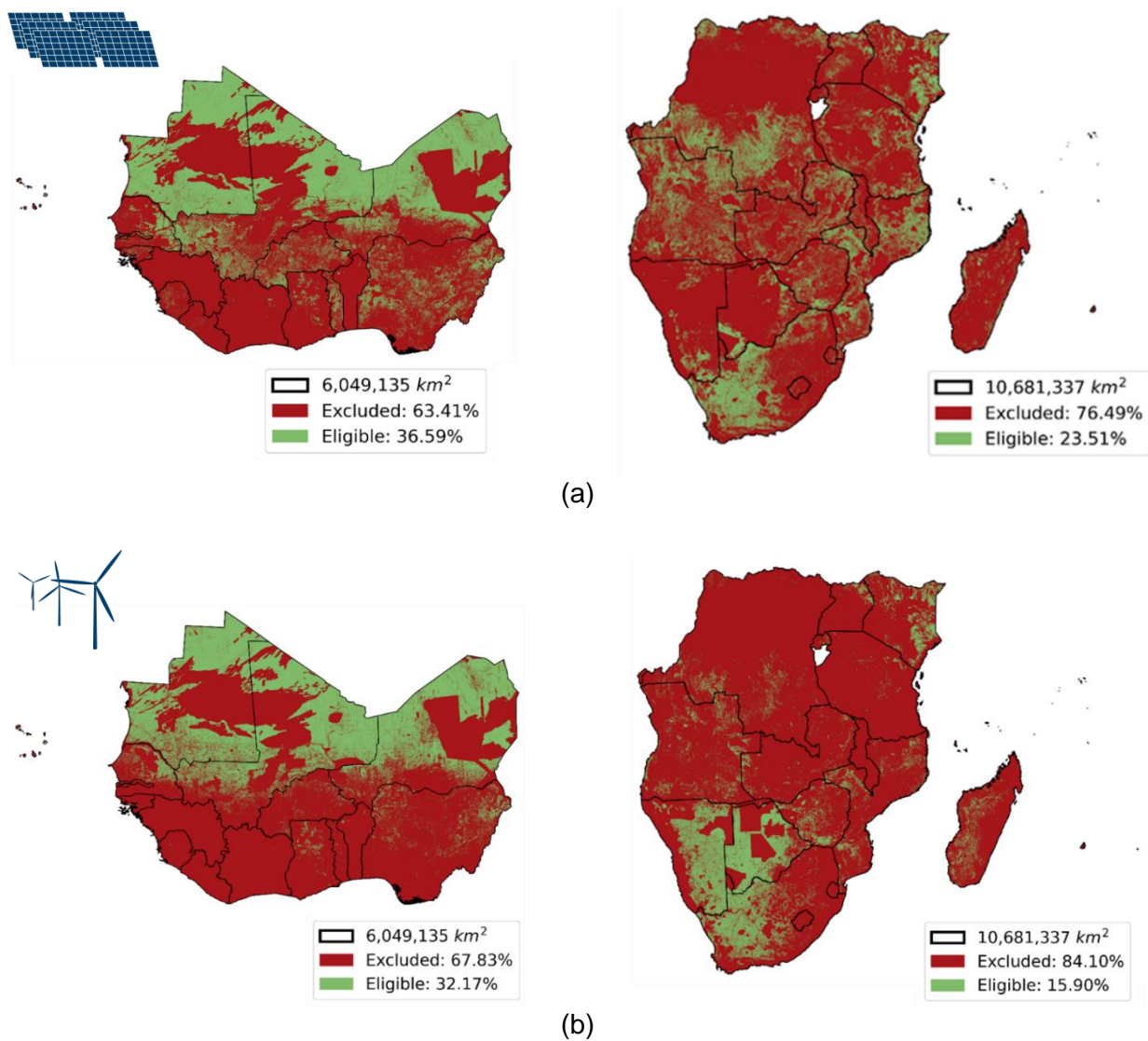

**Figure 6: Land eligibility within west (left) and south-eastern Africa (right) according to the local preferences for (a) open-field PV and (b) onshore wind parks**

Figure 6 (b) displays the areas of land eligible for onshore wind turbines, taking into account local preferences. West Africa boasts a 32% eligibility rate, whereas Southern Africa roughly has a 16% eligibility rate. Within West Africa, eligibility ranges from 0.06% (in Cote d'Ivoire and Guinea) to approximately 55% (in Mauritania). In Southern and East Africa, eligibility rates range from 0% (in Seychelles) to about 50% (in Botswana), with a particularly high eligibility rate in the border region between Namibia, Botswana, and South Africa. However, it's worth noting that the largest shares of land areas are excluded due to isolated and connected settlements, forests, and pastures, in both cases. Detailed eligibility results per country and per technology are summarized in Table 1. These land eligibilities represent potentials limited by technical, sociological and ecological criteria, referred to as "geographical potential" in the following. Generally, it is not possible to exploit them to their maximum. It should be noted that economic exclusions were not considered here on purpose, as the economic viability of renewable energy plants depends significantly on the systemic context. A location with a given elevated energy cost may be uneconomical in a country or region with abundant low-cost resources and low demand yet may be of critical importance in a country with high-demand and limited potentials and constrained cross-border transmission capacities.

Table 1: Overall land eligibility for countries in western, southern and eastern Africa

| Country | Onshore Wind Eligiblity | | Open-field PV Eligiblity | |
|---|---|---|---|---|
| | % | km$^2$ | % | km$^2$ |
| Benin | 0.7 | 793 | 0.7 | 786 |
| Burkina Faso | 6.1 | 16652 | 25.5 | 69790 |
| Cote d'Ivoire | 0.1 | 203 | 0.8 | 2515 |
| Cape Verde | 22.1 | 905 | 21.2 | 866 |
| Ghana | 5.8 | 13749 | 6.5 | 15484 |
| Guinea | 0.1 | 143 | 0.1 | 299 |
| Gambia | 4.6 | 493 | 21.4 | 2278 |
| Guinea-Bissau | 0.2 | 68 | 13.2 | 4471 |
| Liberia | 0.002 | 2 | 2.2 | 2084 |
| Mali | 50.5 | 632070 | 48.7 | 609590 |
| Niger | 48.4 | 573744 | 49.9 | 590072 |
| Nigeria | 8.9 | 81385 | 22.3 | 202245 |
| Senegal | 18.4 | 36188 | 21.7 | 42787 |
| Sierra Leone | 0.5 | 366 | 8.1 | 5849 |
| Togo | 2.5 | 1437 | 33.5 | 19099 |
| Mauritania | 55.6 | 579261 | 63.2 | 585476 |
| Angola | 10.5 | 131131 | 33.9 | 1270166 |
| Botswana | 50.4 | 291008 | 12.7 | 220637 |
| DR Congo | 3.9 | 92003 | 21.0 | 1465824 |
| Comoros | 0.2 | 3 | 3.3 | 164 |
| Lesotho | 1.5 | 457 | 7.9 | 7223 |
| Madagascar | 13.3 | 78669 | 9.2 | 162343 |
| Mozambique | 11.7 | 92247 | 35.5 | 839530 |
| Mauritius | 0.2 | 5 | 5.9 | 362 |
| Malawi | 0.03 | 30 | 14.4 | 50983 |
| Namibia | 44.8 | 369242 | 9.1 | 224865 |
| Swaziland | 1.2 | 209 | 13.4 | 7009 |
| Seychelles | 0 | 0 | 0.1 | 2 |
| Tanzania | 1.7 | 15544 | 17.3 | 488083 |
| South Africa | 29.8 | 363378 | 34.9 | 1277897 |
| Zambia | 5.2 | 38755 | 24.6 | 553623 |
| Zimbabwe | 19.4 | 75937 | 23.1 | 271203 |
| Kenya | 20.3 | 117872 | 31.2 | 151617 |
| Rwanda | 9.0 | 2275 | 6.0 | 153 |
| Uganda | 6.3 | 14950 | 23 | 39149 |

### 3.1.2 Land eligibility assessment for geothermal power plants in Kenya
Due to the central role of geothermal energy in the Kenyan energy strategy (Ogola, Davidsdottir, and Fridleifsson 2012) geothermal power potentials were assessed within the scope of the study as well. The land eligibility assessment yielded an average eligible land area of 42% across the whole country. Most of these areas concentrate in the sparsely populated North and North-East of the country. The exclusions in the South-West and South are mainly driven by human habitations, where a relatively large safety distance of 2 km should be observed to avoid structural damages by small earthquakes or geological settlements. The central to western part of the country is ineligible also due to mountain slopes, and all across the country, large contiguous exclusion zones for nature protected areas can be seen in Figure 7. Areas with high and extremely high water stress according to the Aqueduct water risk atlas (World Resources Institute (WRI) 2023) were excluded as well due to the water usage of petrothermal plants during stimulation and water losses during operation (Tester 2006).

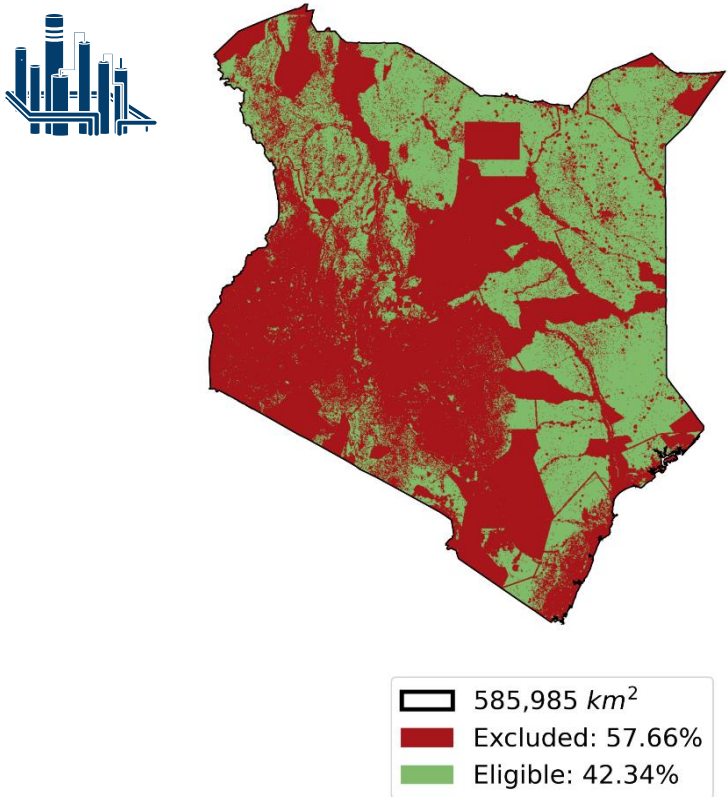

**Figure 7: Land eligibility for geothermal power plants in Kenya**

## 3.2 Renewable energy potential assessment

The following section discusses the available capacity potentials for the main renewable energy sources, namely open-field PV, onshore wind, hydropower, and geothermal. Additionally, levelized cost of electricity patterns are analyzed, shedding light on the underlying drivers and connections.

### 3.2.1 Open-Field Photovoltaics

The total installable capacity potential of open-field photovoltaics amounts to 107 TW in West Africa and to 123 TW in Southern and East Africa respectively. The pattern in Figure 8 mirrors the land eligibility restrictions that were discussed in section 3.1.1. The vast majority of the potential concentrates in the desert regions of the Sahara and the Nama Karoo region in South Africa, extending into Namibia and Botswana. Other notable capacity concentrations are found in the shrubs and bushland of Angola, extending into the Southern provinces of the Democratic Republic of Congo.

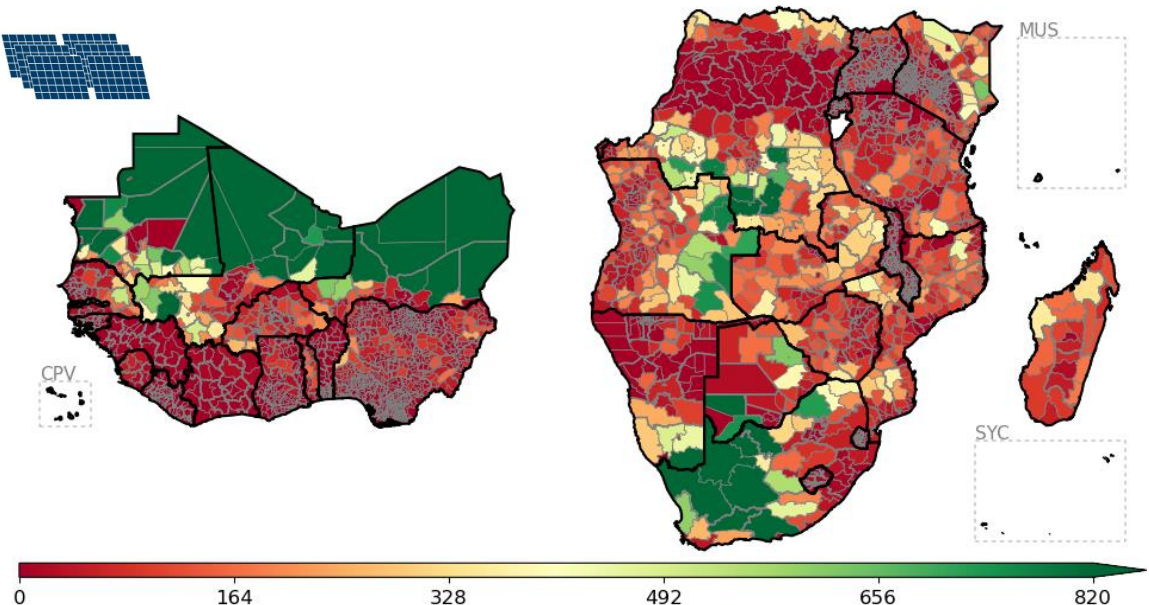

**Figure 8: Open-field photovoltaics capacity potential [GW] in West (left) and Southern/East Africa (right)**

Whilst installable potential can be found in all countries, there are some clear zones with larger installable capacity. For instance, West Africa has the greatest accumulation of potential in its Northern areas whereas Southern and East Africa have them in the south. Compared to these desert provinces, which reach the Terawatt range, capacity potentials in most of the coastal and equatorial countries are much smaller in comparison. It should be noted though that the capacity potential is directly linked with the country's available area, which favors potentials in countries with large areas. In addition, densely populated countries tend to also have reduced areas available for PV installations. These countries tend to be in southern West Africa and many equatorial countries. National aggregate potentials (see Appendix Section 5.2 or on the online GUI (Jülich Systems Analysis and IBG-3, Forschungszentrum Juelich 2024)) should therefore always be consulted as well.

Based on the solar farm locations determined in the placement-distribution step after the land eligibility assessment, hourly time series and average energy yields were simulated for 20 years, using ERA-5 (Hersbach et al. 2018) and Global Solar Atlas[ii] (World Bank Group and Solargis s.r.o. 2023) weather data input. The average yield was then used together with the techno-economic parameter assumptions to calculate levelized cost of electricity for every location, once for every reference decade from 2020 through 2050.

---

[ii] "Data obtained from the Global Solar Atlas 2.0, a free, web-based application is developed and operated by the company Solargis s.r.o. on behalf of the World Bank Group, utilizing Solargis data, with funding provided by the Energy Sector Management Assistance Program (ESMAP)."

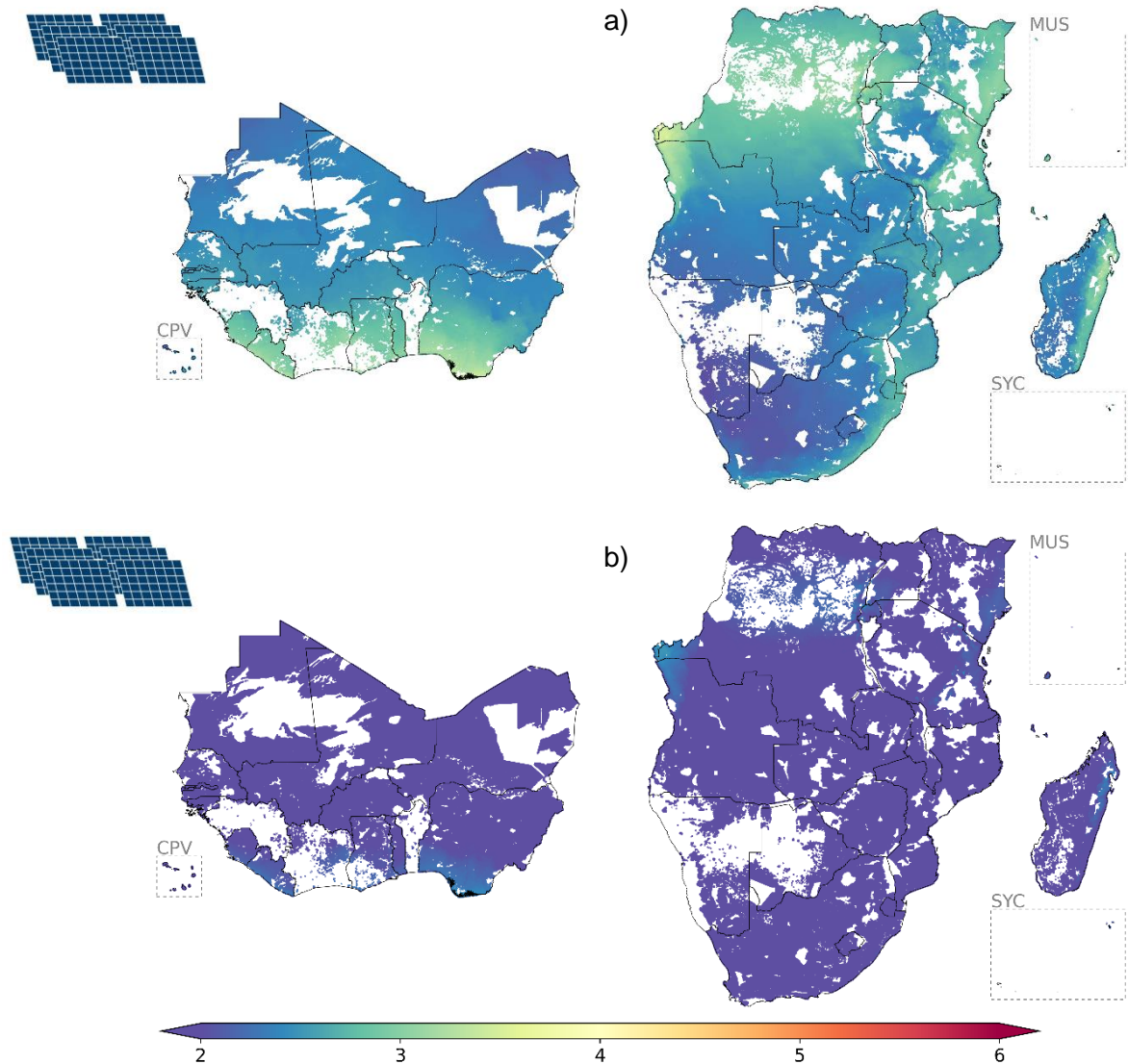

**Figure 9: Levelized cost of electricity [Ct€/kWh] from open-field PV for two years: a) 2030 and b) 2050.**

The plant locations in Figure 9 clearly reflects the land eligibility patterns in Figure 6. The simulation results reveal achievable levelized cost of electricity (LCOE) between 3.5 Ct€/kWh down to 2 Ct€/kWh by 2030 in the best locations in the Nama Karoo ecoregion in South Africa and Namibia, extending into Botswana. In West Africa, the cheapest solar electricity can be produced in the northern desert regions of Niger and Mauritania. Full-load hours in this region may be 50% higher than along the southern coast, where clouds and higher precipitation in the regions near the equator are the principal reason for the reduced energy yield (Schneider, Bischoff, and Haug 2014). The above difference translates into an additional 800 FLH/a or a reduction in levelized cost of up to 1.5 Ct€/kWh. This concentration of vast available lands and high full load hours leads to large scale low-cost solar electricity potentials in the northern parts of West Africa, whereas demand is located mainly along the Southern coast. Also, in Southern and East Africa, reduced full-load hours with resulting higher LCOE can be observed near the equator, such as in the Congo basin particularly towards the Congolese and northern Angolan Atlantic coast, and to a lesser extent also all along the Pacific coast. The Congo basin can be explained by clouds and higher precipitation due to the ITZC (Schneider, Bischoff, and Haug 2014) whilst the second is caused by weather influences due to the Indian ocean (Blau and Ha 2020).

Until 2050, levelized cost of electricity are expected to decrease further by about 25% or 0.5-1.0 Ct€/kWh, then falling below the 2 Ct€/kWh mark in most regions throughout the study extent. Consequently, the LCOE difference between the high- and low-cost regions is narrowed to approximately 1 Ct€/kWh, ultimately causing a progressive yet slow reduction of the competitive advantage of the desert regions.

### 3.2.2 Onshore Wind Turbines

The total installed onshore wind capacity found is 15.4 TW in West Africa and 14.6 TW in Southern and East Africa respectively. Similarly to the open-field photovoltaic pattern, also onshore wind turbine capacities concentrate in the sparsely populated and vegetated desert areas in the North of West Africa and in the arid border area between South Africa but also Namibia and Botswana. Also here, capacity differences are very large between regions, meaning that also regions with seemingly limited potential displayed in shades of red to orange in Figure 10 may reach up to 32 GW, which is a very high absolute value for a district level region. Such regions exist in most countries, most notably across Southern Africa.

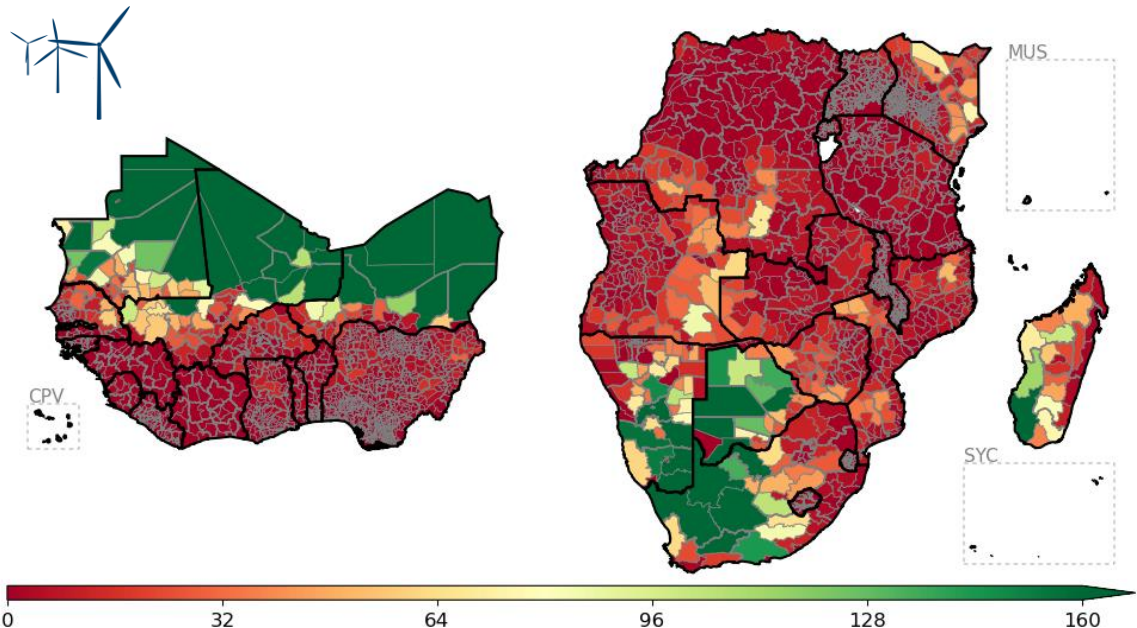

**Figure 10: Onshore wind capacity potential [GW] in West (left) and Southern and East Africa (right)**

An analysis of the simulated levelized cost of electricity for onshore wind yields considerably higher electricity cost on average compared to solar power in the same regions. The amount of onshore wind installations, whose LCOE falls within the same range as PV in 2030 (2 Ct€/kWh up to 3.5 Ct€/kWh) is about 5% and 9% percent of the total (71.9 GW in West Africa and 1.3 TW in Southern and East Africa) respectively. In 2050, due to the proportionally larger capacity cost reductions expected for PV, the amount of onshore installations at the corresponding PV LCOE range (1.5 Ct€/kWh up to 3 Ct€/kWh) is reduced further to 64.3 GW and 970 GW respectively. This means that only about 5% or 9% of the regional onshore wind potential has a comparable LCOE to PV respectively[iii] and its competitiveness is expected to reduce overtime. A more representative onshore wind LCOE might be from under 3 Ct$_€$/kWh up to 6 Ct$_€$/kWh where 10.9 TW in West Africa and 1.7 TW in Southern and East Africa can be installed in 2030. The rest of the installations (30% in West Africa and 89% in Southern and East Africa) are in unfavorable locations, whose LCOE more than doubles that of PV, with the latter being more constant across the study area (see Figure 11). Interestingly, the most favorable wind power cost regions coincide with the best solar locations in several cases, most notably the North of West Africa, where again large wind potentials meet very low LCOE. Particularly north-western Mauritania has excellent wind resources, together with some locations in Kenya and Cabo Verde. To a lesser extent, this applies also to the North of Niger and Mali, several locations in South Africa and Madagascar. Very low wind energy yield is again found in the subtropical belt along the Southern coast of West Africa, similarly to the Congo basin but also in most landlocked regions in Southern and East Africa apart from the ones listed above. Towards 2050, again a reduction in levelized cost of electricity can be observed also for wind power installations, about 12% in average, less significant though compared to the above discussed solar cost reductions.

---

[iii] This percentage refers to the for the total potential of the regions (Western and south and eastern Africa respectively) of Wind compared to PV installation capacity (not generation). Individual countries and sub-regions will have their own proportions.

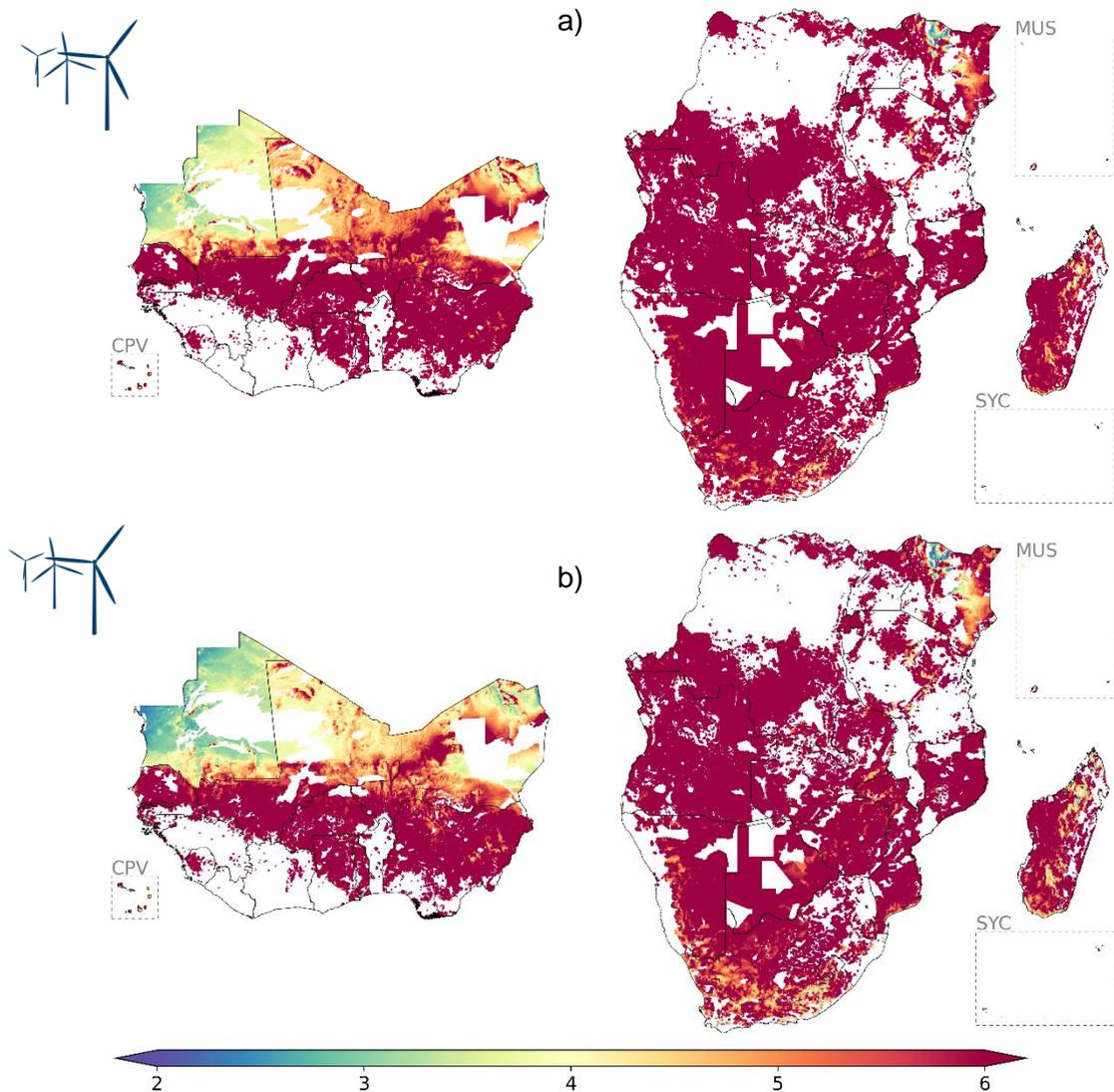

**Figure 11: Levelized cost of electricity [Ct€/kWh] from onshore wind for two years: a) 2030 and b) 2050**

### 3.2.3 Hydropower

Other than in the cases of open-field photovoltaics and onshore wind, hydropower potentials are based mainly on an assessment of existing hydropower plants together with locations under construction or in planning stages (Sterl et al. 2021). In 2050, 23 GW in West Africa and 61 GW in Southern and East Africa were found. These future potentials represent almost 17 GW and 47 GW additional installations as the current hydropower fleet respectively. The chosen approach takes into account the fact that theoretical and also technical hydropower potential assessments tend to be overly large, their realization would lead to significant ecological and social impacts (Schäffer, Korpås, and Bakken 2023). By relying on practically relevant and hence feasible locations, a tradeoff between sensitive capacity additions and the advantages of hydropower for low-carbon energy generation could be achieved.

**Error! Reference source not found.** illustrates in general terms which countries have the most hydropower potential. It becomes apparent that the previously less competitive countries are now the most prominent hydropower locations: DR Congo and Nigeria reach the largest feasible capacities with approximately 22 and 10 GW respectably. This can also be seen when comparing the location of hydropower plans in Figure 13.

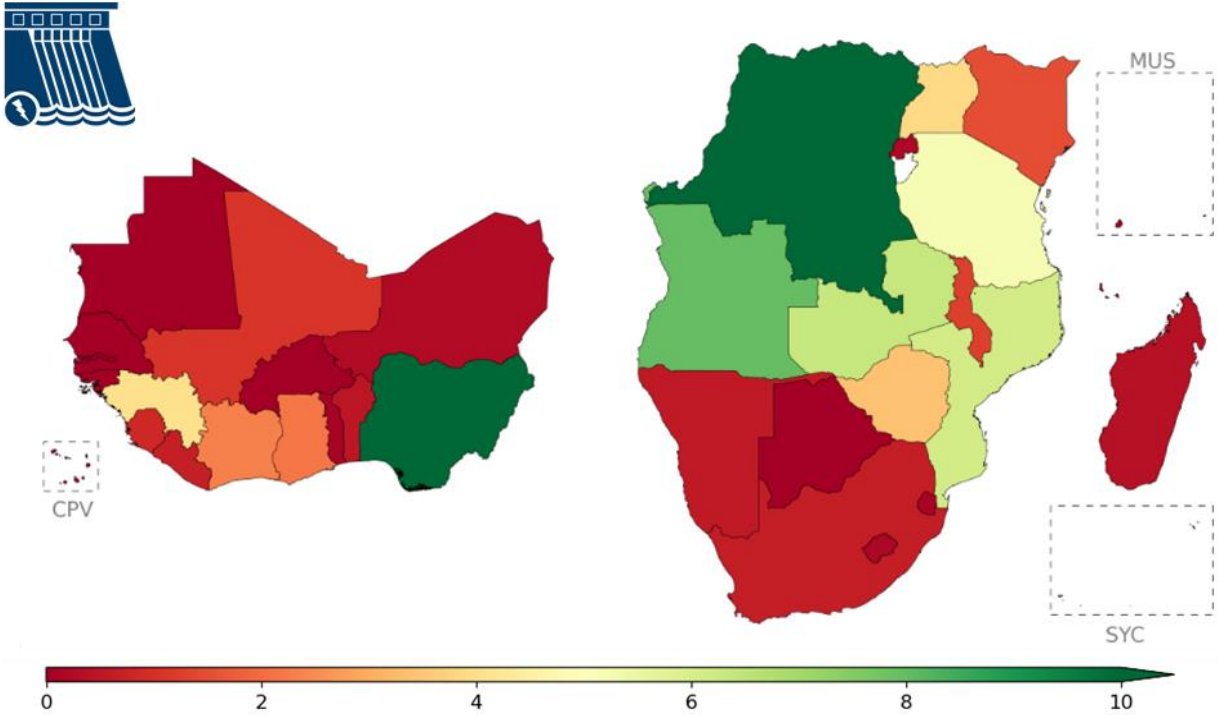

**Figure 12: Hydropower capacity potential [GW] in 2030 in West (left) and Southern and East Africa (right)**

Countries like Angola, Zambia, Mozambique, and to a lesser extent Tanzania and the much smaller Guinea follow suit. Whilst the precipitation surplus along the equatorial belt may be a limiting factor when it comes to solar potentials (Schneider, Bischoff, and Haug 2014), it proves beneficial for hydropower potentials. Countries like Ghana, Côte d'Ivoire or Zimbabwe and Uganda also make use of this opportunity. The hydropower capacity in the previously outstanding electricity cost regions in the Sahara, all over the Southern tip of Africa and Kenya is, however, marginal now due to the predominantly dry climate in these regions.

Similar patterns are revealed when looking at the achievable levelized cost of electricity (see Figure 13). Since the hydropower costs were taken as constant over time, there is no change in LCOE and only 2050, the year with the most locations, is shown in the figure. The lowest cost can be achieved in the Congo basin, where hydropower is very competitive also compared with other renewable energy sources. Main reasons are the large discharge quantities in combination with a comparably low seasonality in the rain forest. Especially in East Congo, mountainous terrain additionally allows favorable pressure heads.

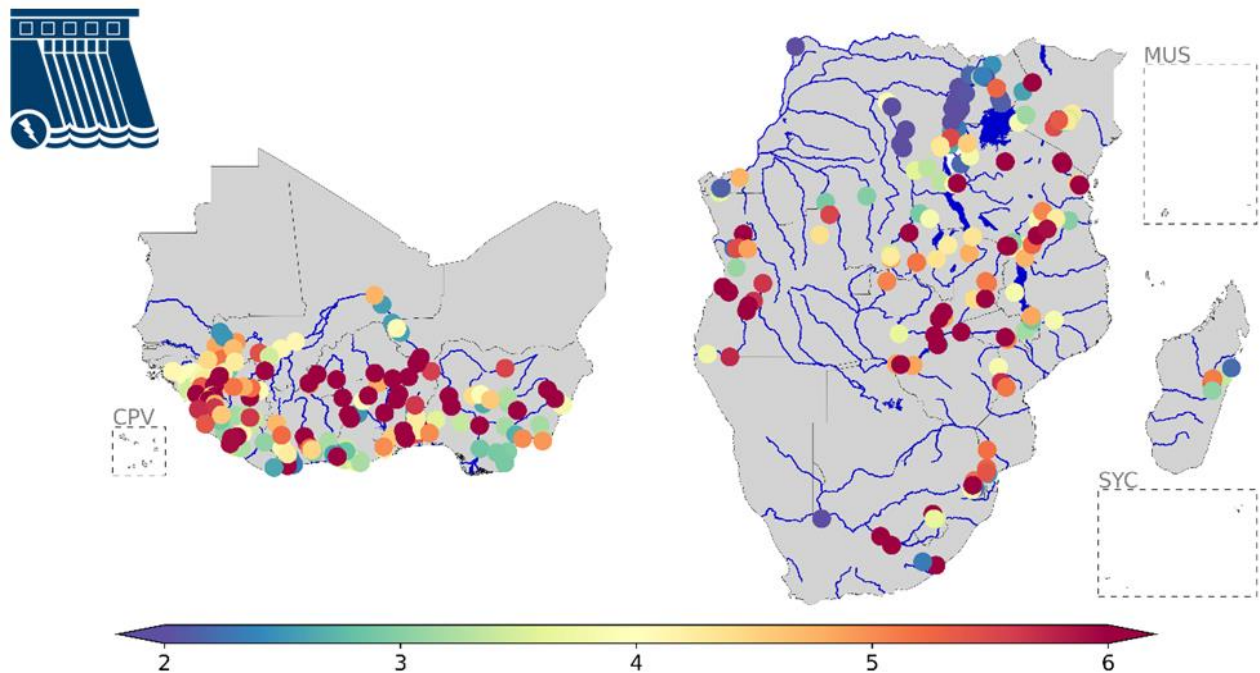

**Figure 13: Levelized cost of electricity [Ct€/kWh] from hydropower for the year 2050.**

In regions with higher seasonality, the hydropower plants cannot operate at full capacity during dry season, especially when no significant reservoir storage is available. Therefore, annual full load hours are reduced and levelized cost rises. This effect can be shown in the savannas of Southern Africa, but also as a South-North gradient in the West African coastal countries. Several hydropower plants along the Niger river in dry areas in Mali and Niger may serve as an example here. The opposite to this is the very large streams where only a minor part of the discharge is required for the hydropower plant for instance those located in the Congo basin and Lake Victoria.

### 3.2.4 Geothermal Power

Due to the great relevance of geothermal energy for the Kenyan energy strategy, geothermal power potentials for Kenya were assessed within the scope of the study. Capacity potentials are greatest in the northern regions (see Figure 14). The very low capacities in the southern and south-western regions of Kenya can be attributed to high population densities, in the south also to the exclusion of protected areas (see section 3.1.1). The central regions are affected mainly by mountain slope exclusion. Overall, the capacity potential in Kenya for geothermal power amounts to slightly over 187 GW. This capacity potential is about 116 times that of hydropower installations in Kenya (1.6 GW), about 20% of onshore wind (895 GW), or nearly 2.5% of PV (7.5 TW).

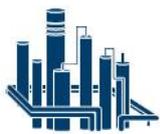
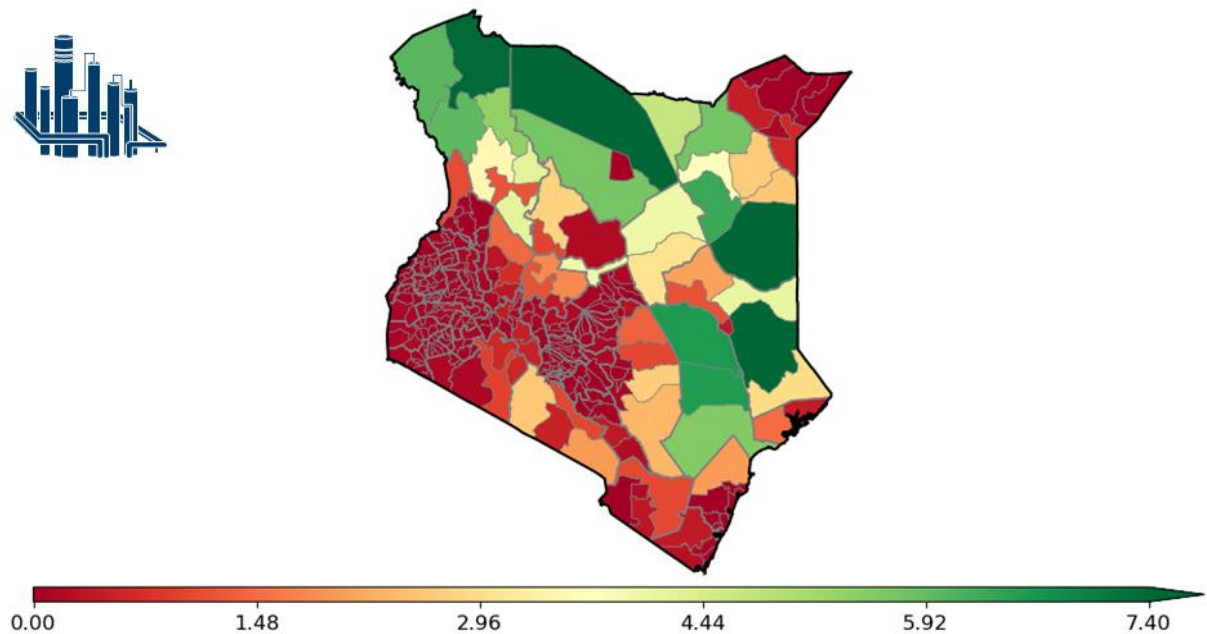

Figure 14: Regional geothermal capacity potentials [GW] in Kenya

Levelized cost of electricity are lowest in the North-East along the Great Rift Valley starting from ~6.8 Ct$_€$/kWh, but to a slightly lesser extent also East of the central mountain ranges towards the Somali border and partly. As these low-cost regions align well with the greatest capacity potentials, it can be stated that the main geothermal potentials in Kenya are located in Turkana and Wajir county, to a slightly lesser extent also in Garissa and Marsabit. Also, in the South of the Rift Valley province, interesting locations can be identified, however, the capacity potential is comparably low in this area at costs between 6.8-15 Ct$_€$/kWh. Beyond these regions, the cost-potential curve rises sharply. In general, geothermal power is more costly than wind or solar resources in Kenya, but it has the advantage of being dispatchable. Subsequent processes are therefore able to reach near-ideal utilization rates. The economic viability of geothermal power for hydrogen production in Kenya hence depends on the achievable full load hours based on solar and wind power. 70% of the total potential are feasible below 10 Ct$_€$/kWh.

## 3.3 Sustainable water supply assessment

For the water availability assessment, we considered the average groundwater sustainable yield for 2020 (2015 - 2035), 2030 (2015 – 2045), and 2050 (2036 - 2065) and seawater desalination including water transport.

**Groundwater sustainable yield in 2020**

The long-term average (2015 – 2035) groundwater sustainable yield maps representative for the year 2020 are presented in Figure 15, considering two climate scenarios: RCP2.6 (Figure 15a & c & e), and RCP8.5 (Figure 15b & d & f). For each climate scenario, three cases are investigated: conservative (Figure 15a & b), medium (Figure 15c & d), and extreme conditions (Figure 15e & f) based on the scenarios described in Ishmam et al. (Ishmam et al. 2024) The results show that there is a notable rise in groundwater sustainable yield from the conservative (Figure 15a & b) to the extreme (Figure 15a & b) scenario. The medium scenario (Figure 15c & d) located between the conservative and extreme cases could be considered as the most favorable scenario for green hydrogen production. In West Africa, the coastal parts of the region (e.g., Gambia, Guinea, Sierra Leone, Liberia, and Cote d'Ivoire) as well as those located in the Southern part (e.g., Nigeria, Benin, Ghana) consistently have higher groundwater sustainable yield for RCP2.6 and RCP8.5 and all three cases. The regional analysis (Table 2) demonstrates that in the selected West Africa region, the average groundwater sustainable yield for the year 2020 RCP2.6 (RCP8.5) would be 7.4 (6.9) mm yr$^{-1}$ (conservative scenario), 34.9 (33) mm yr$^{-1}$ (medium scenario), and 63 (59.7) mm yr$^{-1}$ (extreme scenario) based on the method and scenarios described in Ishmam et al. (Ishmam et al. 2024).

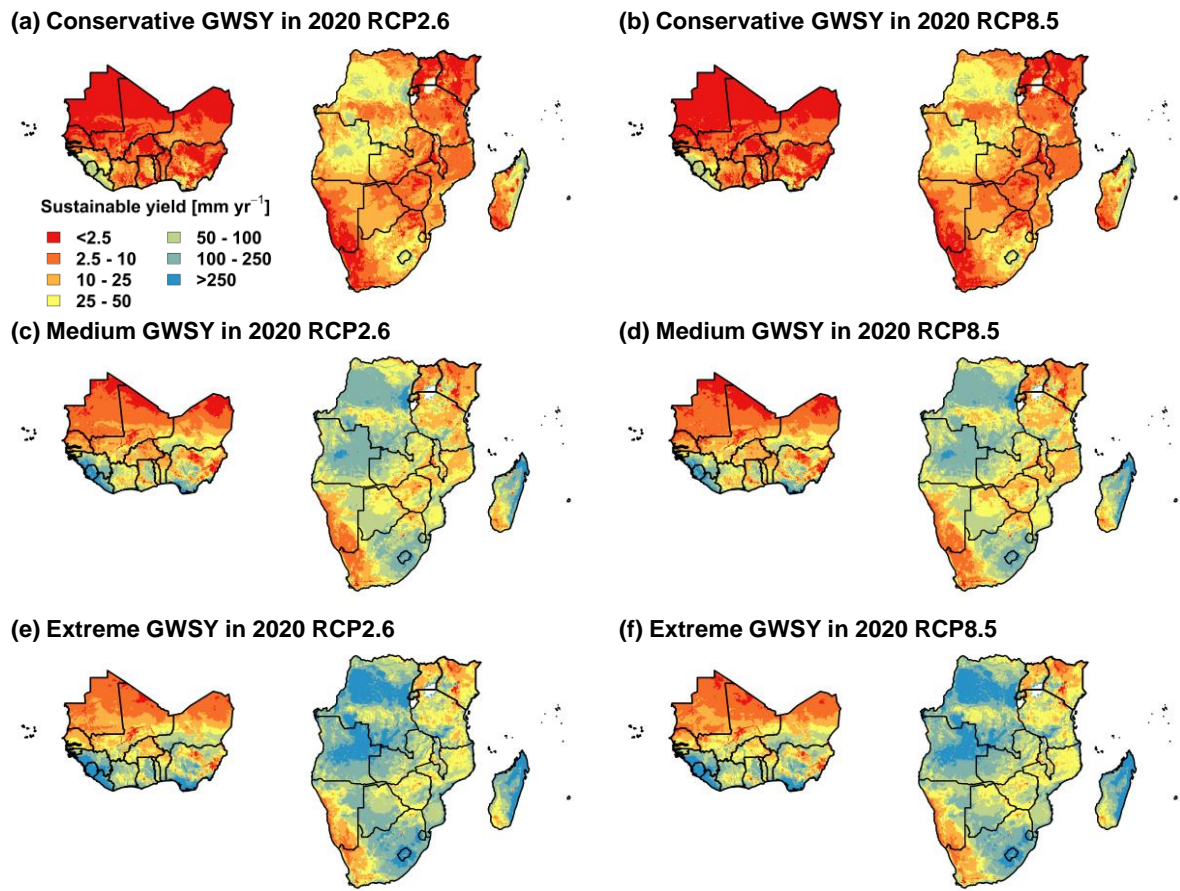

**Figure 15. The groundwater sustainable yield for 2020 (the average of 2015 - 2035) under two climate scenarios: RCP2.6 (left panels) and RCP8.5 (right panels). Each scenario considers three cases: conservative (a & b), medium (c & d), and extreme conditions (e & f) (compare method in Ishmam et al. (Ishmam et al. 2024)).**

In Southern East Africa, the results suggest that countries such as Angola, Democratic Republic of Congo, Zambia, Tanzania, and Madagascar consistently exhibit high sustainable yield across all scenarios (Figure 15). In addition, we found that the southern part of the region (e.g., South Africa, Lesotho, Namibia, Botswana) had more groundwater sustainable yield in 2020 (RCP2.6 and RCP8.5). In 2020 the average groundwater sustainable yield for Southern East Africa (

Table 3) according to the RCP2.6 (RCP8.5) scenarios would be 17.9 (17.1) mm yr$^{-1}$ (conservative scenario), 75.8 (72.6) mm yr$^{-1}$ (moderate scenario), and 134.2 (128.6) mm yr$^{-1}$ (extreme scenario).

**Groundwater sustainable yield in 2030**

Regarding 2030, the sustainable yield of groundwater based on Ishmam et al. (Ishmam et al. 2024) is averaged over the 2015 – 2045 period and presented in Figure 16 under two different climate scenarios: RCP2.6 (Figure 16a & c & e) and RCP8.5 (Figure 16b & d & f). It also illustrates three scenarios: conservative (Figure 16a & c), medium (Figure 16b & d), and extreme (Figure 16e & f). In West Africa, similar to 2020, the higher groundwater sustainable yield values are present along the western coastal regions (such as Gambia, Guinea, Sierra Leone, Liberia, and Cote d'Ivoire), as well as in the Southern part (e.g., Nigeria, Benin, Ghana). We found that the average groundwater sustainable yield in 2030 would be slightly lower than the one in 2020 under both RCP2.6 and RCP8.5 (Table 2). Taking into account the entire region, our findings highlight the potential for groundwater sustainable yield in West Africa under RCP2.6, with average values of 6.8 mm yr$^{-1}$ (conservative scenario), 33.3 mm yr$^{-1}$ (medium scenario), and 60.6 mm yr$^{-1}$ (extreme scenario). For RCP8.5, the groundwater sustainable yield changed to 6.4 mm yr$^{-1}$ (conservative scenario), 31.8 mm yr$^{-1}$ (medium scenario), and 58 mm yr$^{-1}$ (extreme scenario). Regarding Southern East Africa, as can be seen from Figure 16, the higher sustainable yield values are found in Angola, the Democratic Republic of Congo, Zambia, Tanzania, and Madagascar especially in medium and extreme scenarios. Similar to West Africa region, the average groundwater sustainable yield in 2030 RCP2.6 and RCP8.5 has slightly decreased compared to 2020 RCP2.6 and RCP8.5 (

Table 3). Across the entire Southern African region, we found average rates of 17.3 (16.8) mm yr$^{-1}$ (conservative scenario), 73.7 (71.9) mm yr$^{-1}$ (medium scenario), and 130.8 (127.6) mm yr$^{-1}$ (extreme scenario) under RCP2.6 (RCP8.5).

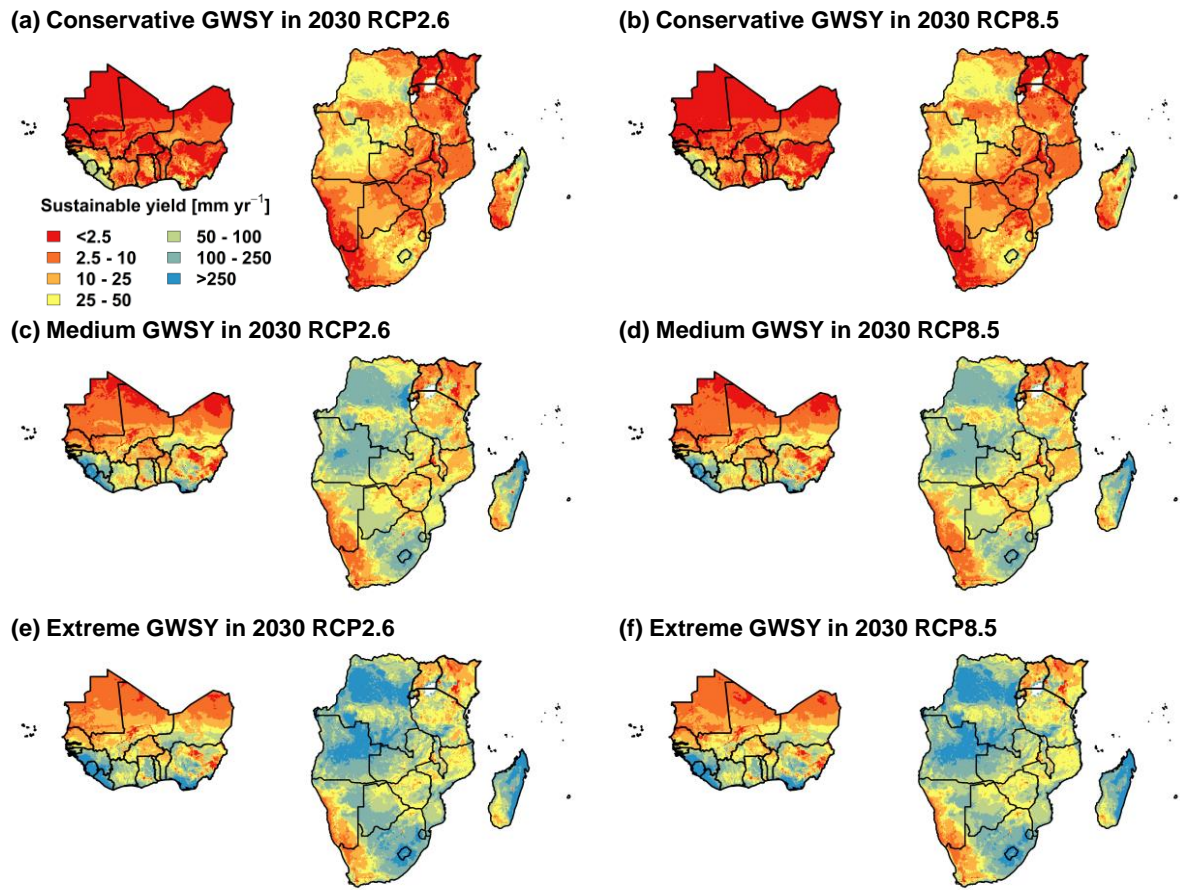

**Figure 16. The groundwater sustainable yield for 2030 (the average of 2015 - 2045) under two climate scenarios: RCP2.6 (left panels) and RCP8.5 (right panels). Each scenario considers three cases: conservative (a & b), medium (c & d), and extreme conditions (e & f) (based on method and scenario definition in Ishmam et al. (Ishmam et al. 2024).**

Table 2. The average estimates of groundwater sustainable yield in the West Africa region for 2020 (2015-2035), 2030 (2015-2045), and 2050 (2036-2065) considering two climate scenarios: RCP2.6 and RCP8.5 under conservative, medium, and extreme conditions (based on method and scenario definition in Ishmam et al. (Ishmam et al. 2024)).

| Scenario | West Africa groundwater sustainable yield [mm yr$^{-1}$] | | | | | |
| --- | --- | --- | --- | --- | --- | --- |
| | 2020 | | 2030 | | 2050 | |
| | RCP2.6 | RCP8.5 | RCP2.6 | RCP8.5 | RCP2.6 | RCP8.5 |
| Conservative | 7.4 | 6.9 | 6.8 | 6.4 | 5.6 | 5.1 |
| Medium | 34.9 | 33 | 33.3 | 31.8 | 29.6 | 27.2 |
| Extreme | 63 | 59.7 | 60.6 | 58 | 55.1 | 50.9 |

Table 3. The average estimates of groundwater sustainable yield in the Southern East Africa region for 2020 (2015-2035), 2030 (2015-2045), and 2050 (2036-2065) considering two climate scenarios: RCP2.6 and RCP8.5 under conservative, medium, and extreme conditions (based on method and scenario definition in Ishmam et al. (Ishmam et al. 2024)).

| Scenario | Southern East Africa groundwater sustainable yield [mm yr$^{-1}$] | | | | | |
| --- | --- | --- | --- | --- | --- | --- |
| | 2020 | | 2030 | | 2050 | |
| | RCP2.6 | RCP8.5 | RCP2.6 | RCP8.5 | RCP2.6 | RCP8.5 |
| Conservative | 17.9 | 17.1 | 17.3 | 16.8 | 15.6 | 15.4 |
| Medium | 75.8 | 72.6 | 73.7 | 71.9 | 68.6 | 67.8 |
| Extreme | 134.2 | 128.6 | 130.8 | 127.6 | 122.3 | 121 |

**Groundwater sustainable yield in 2050**

We calculated the long-term average (2036-2065) groundwater sustainable yield as representative for 2050 (Figure 17) for the climate scenarios RCP 2.6 (Figure 17a & c & e) and RCP8.5 (Figure 17b & d & f), and for three cases: conservative (Figure 17a & b), moderate (Figure 17c & d), and extreme (Figure 17e & f). The results demonstrate that the average groundwater sustainable yield in 2050 is clearly lower than in 2020 under both RCP2.6 and RCP8.5 and for all conservative, medium, and extreme cases (Table 2 and

Table 3). The regional analysis (Table 2) gives an average West Africa groundwater sustainable yield in 2050 RCP2.6 of 5.6 mm yr$^{-1}$ (conservative scenario), 29.6 mm yr$^{-1}$ (medium scenario), and 55.1 mm yr$^{-1}$ (extreme scenario). Moreover, groundwater sustainable yield in 2050 for RCP8.5 is lower than for RCP2.6 for all three scenarios, particularly over the Northern part of the region. The results show that the Southern part of the region (e.g., Nigeria, Benin, Ghana) consistently receives less sustainable yield in all three scenarios (Figure 17b & d & f) than in 2020 for RCP8.5. In 2050 for RCP8.5, the region is expected to exhibit average groundwater sustainable yield values of 5.1 mm yr$^{-1}$ (conservative scenario), 27.2 mm yr$^{-1}$

(medium scenario), and 50.9 mm yr$^{-1}$ (extreme scenario), which is less than in 2020 for RCP8.5 (Table 2).

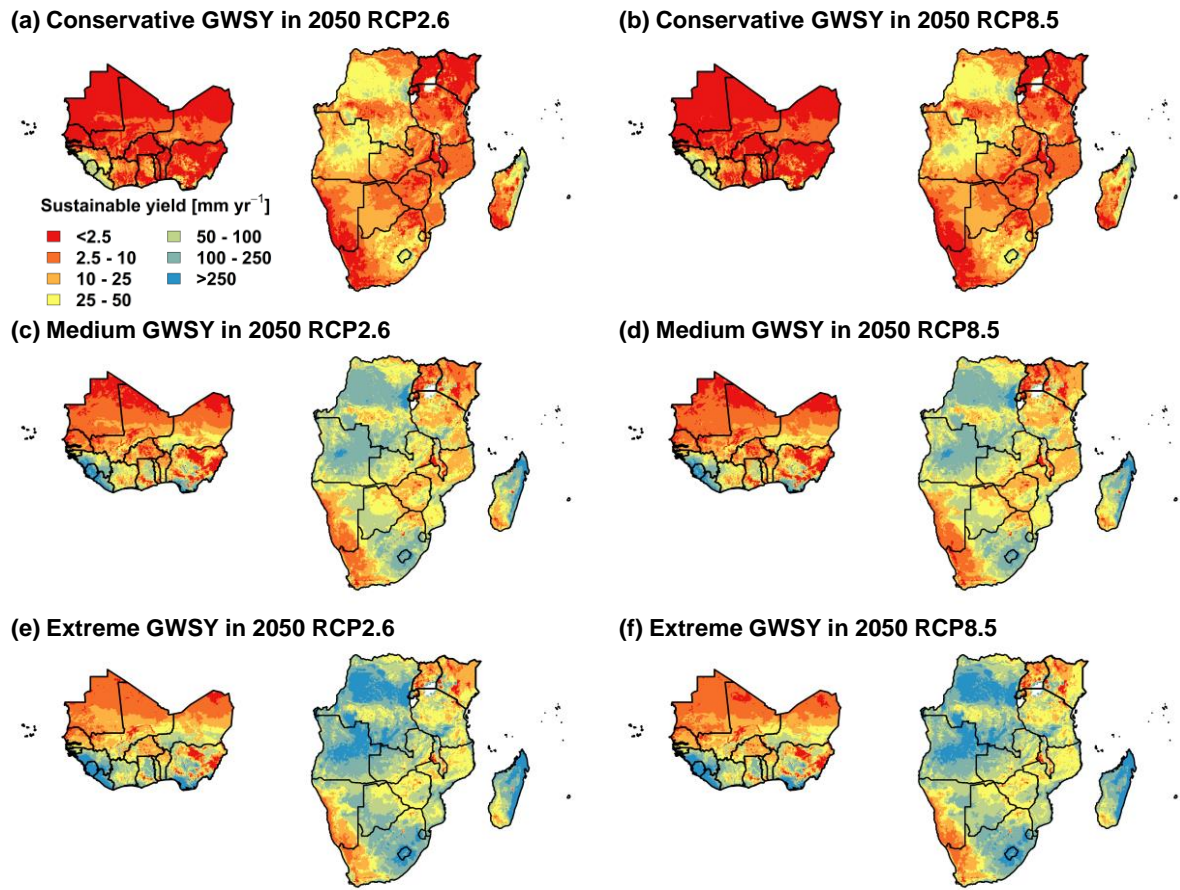

**Figure 17: The groundwater sustainable yield for 2050 (the average of 2036 - 2065) under two climate scenarios: RCP2.6 (left panels) and RCP8.5 (right panels). Each scenario considers three cases: conservative (a & b), medium (c & d), and extreme conditions (e & f) (based on method and scenario definition in Ishmam et al.** (Ishmam et al. 2024)**).**

For Southern East Africa the spatial pattern of groundwater sustainable yield for the three scenarios in 2050 under RCP2.6 (Figure 17a & c & e), and RCP8.5 (Figure 17b & d & f) is similar to 2020, but the absolute amount of sustainable yield is expected to be lower in 2050 than in 2020, both for RCP2.6 and RCP8.5. Considering 2050 RCP2.6, the average groundwater sustainable yield for the Southern East region is projected to be 15.6 mm yr$^{-1}$ (under conservative conditions), 68.6 mm yr$^{-1}$ (under moderate conditions), and 122.3 mm yr$^{-1}$ (under extreme conditions) which is lower than under the RCP2.6 scenario in 2020 (

Table 3). Furthermore, the region is projected to show average groundwater sustainable yield values of 15.4 mm yr$^{-1}$ (conservative scenario), 67.8 mm yr$^{-1}$ (medium scenario), and 121 mm

yr$^{-1}$ (extreme scenario) for RCP8.5, all of which are lower than those observed in 2020 under RCP8.5 (

Table 3).

## 3.4 Local green hydrogen potential assessment

In the following, quantitative hydrogen potentials will be analyzed and put in relation to the hydrogen amounts producible from local sustainable groundwater potentials. Cost of hydrogen and the cost contribution of water will be presented, and reasons for hydrogen cost differences will be discussed. Further information will be provided by means of cost-potential curves, which will then be explained in detail based on a profound analysis of the region- and time-dependent optimal system designs to produce green hydrogen.

### 3.4.1 Degree of potential expansion

For the analysis of the levelized cost of hydrogen (LCOH) as well as for system design considerations, the dependency of the cost upon the degree of potential expansion is discussed first. The lowest-cost potentials are exploited first, with cost increasing along the cost-potential curve when the quantity of produced hydrogen is increased. Typically, e.g. for regions with a large quantity of wind turbines, cost start comparably low and rise sharply at the beginning, with decreasing slope within the first quartile. Over the second and third quartile they increase very moderately and increase sharply towards the end again (compare exemplary distribution in Figure 2). Whilst very high expansion degrees towards the end of the curve describe economically unrealistic solutions, the very low expansion degrees describe entry level cost. These are defined by the most favorable locations and system configurations. Typically, these would also be locations that might be occupied by first movers, which may have implications for the economy of scale and learning rate assumptions. In an approach with independent regions and high geospatial resolution, however, the economy of scale and learning effects in any given region may depend not only on the region itself but also on installations in neighboring regions or the whole country. It is therefore not possible to reliably quantify the cost effects of first-mover projects in an atlas approach, which is why cost at very low potential expansion degrees must be interpreted with care. In addition to the above, simulation tools for renewable energy are always subject to a statistical error. Whilst the median and average values are very reliable, potential outliers might affect entry level cost further. Due to these reasons, the following analysis focusses on a potential expansion rate of 25% when comparing levelized cost of hydrogen and system designs. The cost in this range is still comparably low, yet very reliable and rather stable against increase of the production quantity.

### 3.4.2 Quantitative Hydrogen Potentials

The following quantitative hydrogen potentials were calculated based on the local preference exclusions defined with the national specialists and renewable energy simulations. Hydrogen potentials were then determined based on optimal system designs per each district level node. The total maximum hydrogen potentials in West Africa together with Southern and East Africa amount to just over 400 000 TWh/a, more than twice the equivalent of today's global primary energy consumption (165 319 TWh/a in 2021 (bp 2022)). The cumulative projected local electricity and hydrogen demands in 2050 for the analyzed regions based on the "Net Zero 2050" scenario from the NGFS Climate Scenarios Database (Richters, Bertram, Kriegler, Anz, et al. 2022; Richters, Bertram, Kriegler, Al Khourdajie, et al. 2022) amount to only 0.5% of the total potentials. The focus of this analysis is hence not on the total quantitative potentials of the whole project region, but rather on regional and national potential constraints as well as on available quantities at defined production cost.

The distribution of the regional potentials per area shown in Figure 18: Hydrogen potential per area [kWh/(a*m²)] at district level in 2030, limited only by energy.Figure 18 avoids distortion of the potential values by the region size and reflects the limitations and results described in the land eligibility and renewable energy potentials sections. The greatest potentials are found in the sparsely populated desert regions in the Sahara and Nama Karoo regions, but very significant area potentials are found all over the study area. Exceptions to that are the densely populated and forested areas along the Southern coast of West Africa, the Congo basin and rangeland areas in Southern Africa with low wind potentials. It is worth noting, however, that the required area potentials may be very low even for industrialized countries like Germany where an area potential of under 6 kWh/m² would be sufficient to cover the total projected final energy demand of 2164 TWh/a in 2045 (Stolten et al. 2021). About 54% of the project regions exceed this threshold.

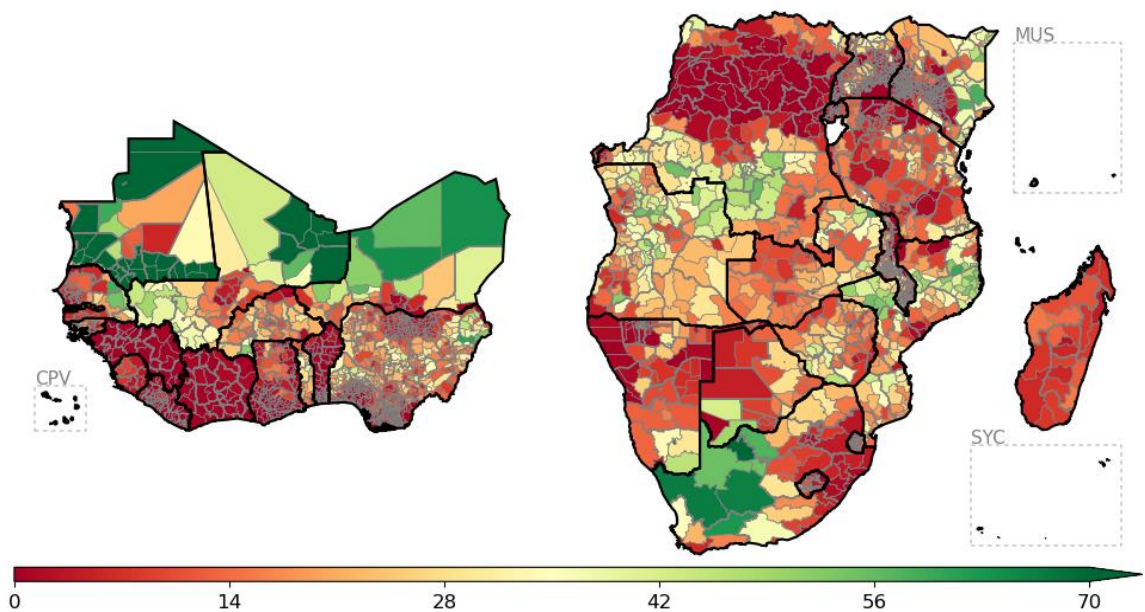

**Figure 18: Hydrogen potential per area [kWh/(a*m²)] at district level in 2030, limited only by energy.**

The absolute hydrogen potentials per area show a similar pattern (see (Jülich Systems Analysis and IBG-3, Forschungszentrum Juelich 2024) for regional values), further amplified by the fact that remote desert regions tend to have much larger areas.

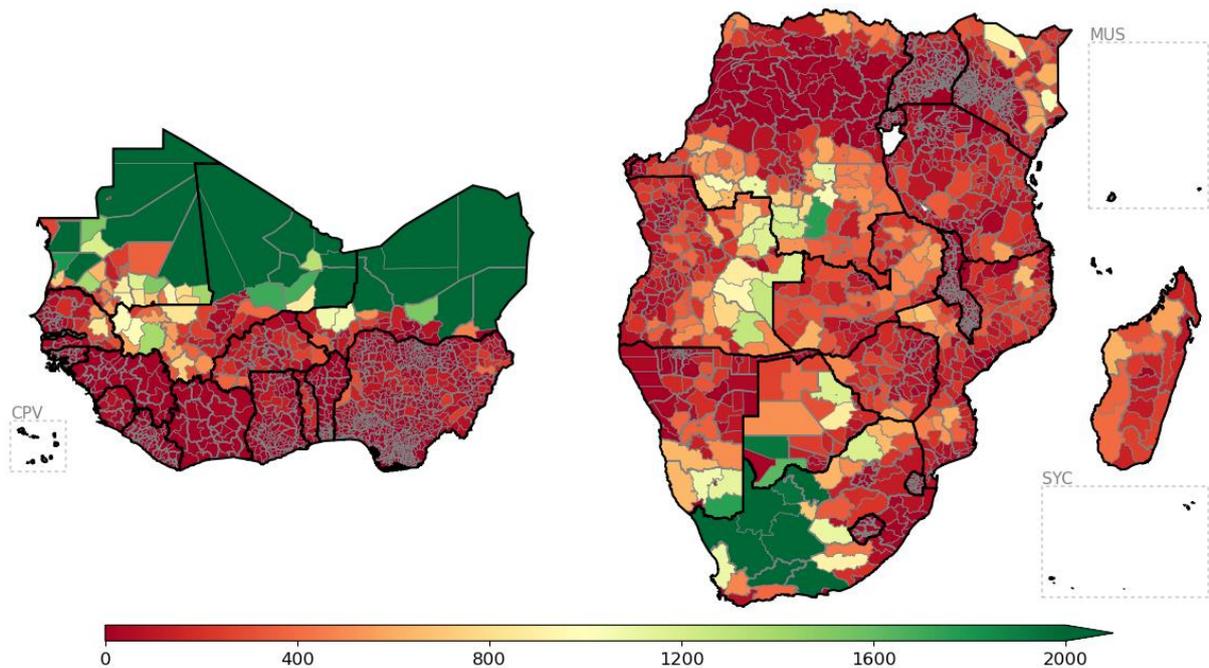

**Figure 19: Absolute technical hydrogen potentials [TWh/a] at district level, limited only by energy.**

The national and total potentials were calculated, and local electricity and hydrogen demands were considered as detailed in section 2.4. National aggregate values are listed in the appendix in section 5.1. Defined by the land eligibility and theoretical energy potential limitations, only 3 countries out of the 35 countries under consideration may not be able to supply their own electricity and hydrogen demand in 2050, again considering a 50% feasibility threshold for the maximum potential. This number is reduced to 2 out of 35 countries in the unlikely case of a 100% expansion of the potential. However, this situation could change with regulatory decisions, which may have great relative impact on potentials especially in regions with very low overall eligibility, such as in island nations like the Seychelles or Mauritius. In the case of the Republic of Guinea in particular, energy import from neighboring countries would be a viable and efficient solution. Corresponding transmission infrastructure projects are already being implemented within the West African Power Pool (WAPP) framework (West African Power Pool 2024).

### 3.4.3 Impact of Sustainable Water Yield on Hydrogen Potentials

When hydrogen production based on local sustainable groundwater is considered, the quantitative potentials are massively reduced (see Figure 20). Only 42% of all regions have sufficient sustainable groundwater for an expansion degree of 25% of the maximum energetic potential. Especially the low-cost and high-potential desert regions are affected due to both high potentials and limited amount of precipitation. The share of project regions with sufficient groundwater is reduced to 28% when only considering regions with an average LCOH below 2.7 EUR/kg in 2030. Overall, the hydrogen potential based on sustainable groundwater amounts to roughly 16% of the total technical hydrogen potential limited only by energy constraints. This share may be reduced further if water-intensive cooling methods are used instead of air cooling. The impact of the cooling method on water consumption cannot be quantified generally though since it depends on the exact cooling method and environmental conditions, but also on the re-use options of the cooling water return flow. When weighted by sustainable water potential only, the average levelized cost of hydrogen is 4-5% higher than the overall LCOH without water limitations.

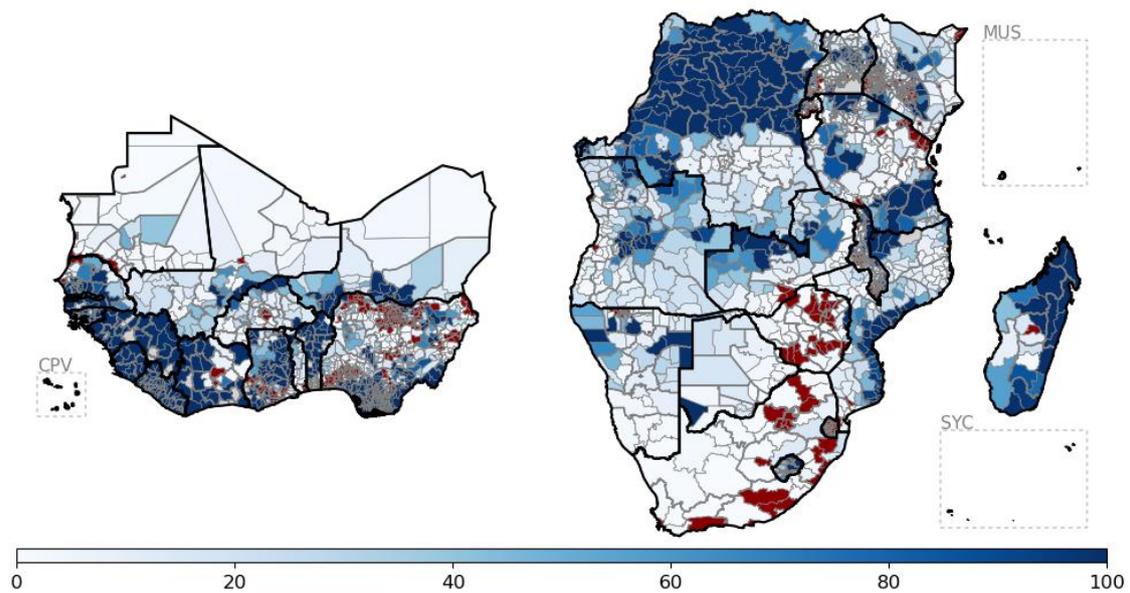

**Figure 20: Percentage share of hydrogen [%] producible from local sustainable groundwater.**

However, this does not explicitly imply that sustainable hydrogen production is not feasible beyond the limit of sustainable groundwater potential. In specific cases, surface water may be used where available throughout the year, but its sustainability must be assessed on a case-by-case basis. Particular attention should also be paid to the seasonality of water availability for hydrogen production in such cases. Water may be required especially in the dry season in systems that are dominated by solar photovoltaics. Water for sustainable hydrogen may also come from alternate water sources such as seawater desalination, as long as renewable energy is consumed for the treatment and transport and that the water intake and brine outlet are optimized for minimal ecological impact. According to the present assessment, 100% exploitation of the overall technical potential would require 84% of the consumed water to come from such alternative sources. In the case of desalinated water, this would correspond to 755 new large-scale desalination plants with a capacity of 367 000 m³/d each suggested as future reference plant in Heinrichs et al. (Heinrichs et al. 2021) or 1846 plants of the 2020 UAE reference size suggested by Eke et al. (Eke et al. 2020). The additional cost for the water treatment is not significant, given the high value of hydrogen and the relatively low amount of water that is needed to produce it via electrolysis: Even in north-eastern Mali, where the levelized cost of water including pipeline transport reaches nearly 2.50 €/m³ in combination with comparably low LCOH, the share of water cost does not exceed ~1% of the total levelized cost of hydrogen. The focus will therefore rather be on the feasibility of pipeline water transport over such long distances and across country borders as soon as large-scale sustainable hydrogen production in potential-rich inland countries such as Niger and Mali should be realized. Otherwise, power export from such countries to coastal locations via power grid combined with electrolysis near the coast would also be an option. However, transnational power lines would also require high planning effort and cooperation and would come at a higher cost compared to pipelines. Countries with large low-cost potentials near the coast include for example Mauritania, South Africa, Namibia, Kenia, and Madagascar, but also small island nations such as Cabo Verde.

### 3.4.4 Levelized Cost of Hydrogen (LCOH)

First, levelized cost of hydrogen mainly based on onshore wind, open-field photovoltaics and geothermal power supply shall be discussed as these represent the large-scale, widely untapped renewable energy resources. Regions dominated by hydropower often represent extremely competitive yet local cost outliers and shall be discussed afterwards.

The lowest levelized cost of hydrogen without hydropower contribution at 25% of the regional maximum potential are found in Mauritania with the cheapest region just below 2 €/kg in 2030, decreasing to 1.6 €/kg in 2050. If the actual entry level cost is considered, LCOH around 1.8 €/kg in 2030 and 1.5 €/kg in 2050 may be possible without hydropower as long as economy of scale still applies. The most economical hydrogen production is located in the Sahara region particularly towards the Mauritanian coast and in the Nama Karoo region in the border area between Southern Africa, Namibia and Botswana (see Figure 21), but also in smaller countries such as Cabo Verde and Lesotho. Cheap hydrogen can also be produced in Kenya and to a lesser degree in Madagascar. When comparing the overall potential weighted averages in 2050 between the project regions, West Africa has a slight cost advantage (~1.9 €/kg) due to the quantitatively large low-cost potentials in the Sahara in comparison to Southern and East Africa (~2 €/kg) with its higher cost-potentials near the equator. The overall potential-weighted average levelized cost of hydrogen across the whole study extent at 25% of the respective regional technical potential is at ~2.7 €/kg in 2030 and ~1.9 €/kg in 2050.

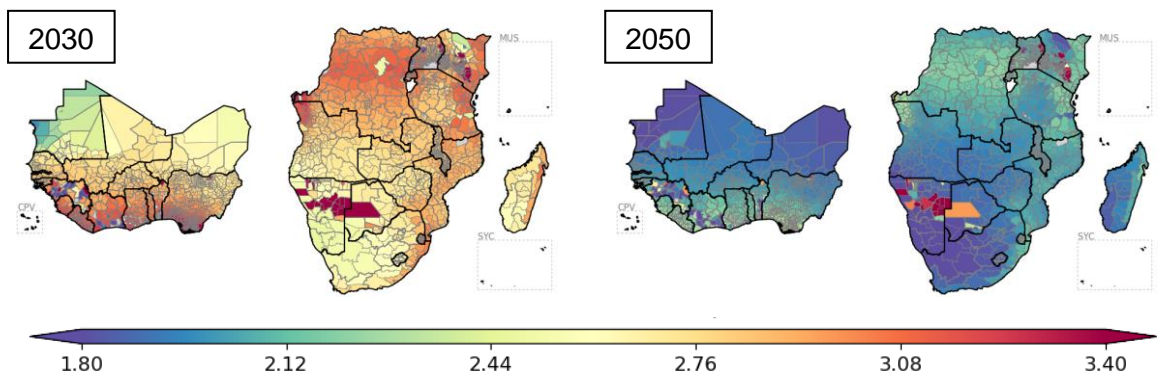

Figure 21: Regional Levelized Cost of Hydrogen [EUR/kg] at 25% potential expansion in the years 2030 (left) and 2050 (right)

A direct comparison of the levelized cost of hydrogen at 25% expansion between 2030 and 2050 (Figure 22 on the left) shows that the LCOH reduction over the decades is greatest in regions with solar-dominated hydrogen production (see section 3.2.1). Most of the cheapest hydrogen regions in 2030 feature mixed solar-wind generation fleets, however, and experience lower cost reductions over time. This means that other, solar-driven regions are able to catch up, the solar- and wind mixed regions in North-Eastern Mauritania remain amongst the cheapest though. 64% of the average cost reductions based on the assumed techno-economic parameters of the technologies between 2030 and 2050 are achieved in the first decade until 2040.

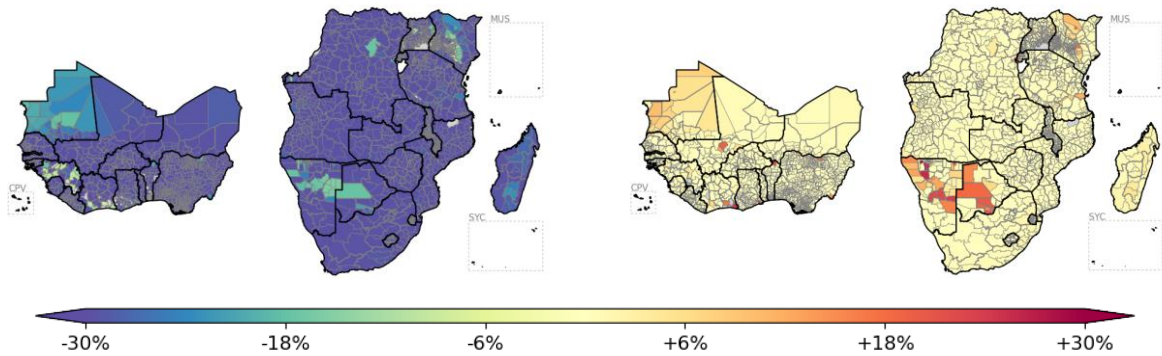

**Figure 22: Percentage change [%] of levelized cost between 2030 and 2050 at 25% of technical potential (left) and between 25% and 50% of technical potential in 2030 (right).**

In order to quantify the dependency of the levelized cost of hydrogen on the production quantity, the relative cost increase of the LCOH at 50% potential expansion compared to the standard potential expansion of 25% is analyzed for the year 2030 (see Figure 22 on the right). The differences are marginally higher yet very similar in 2050. The results indicate that LCOH in the wind-dominated low-cost hydrogen regions are most dependent on the production quantity. Levelized cost of hydrogen in regions with mainly wind power generation rise by a median of ca. 6% when production capacity is doubled from 25% to 50% of the technical potential. The situation is very different in solar-dominated systems, independent of the LCOH level, where cost does practically not change with the production capacity increase (~0.1% median cost increase). Production quantity increases may also have significant impact on average LCOH in regions where the capacity limits of the cheapest generation technology are exceeded, and more costly alternatives must then be used.

When looking at the hydrogen cost map in Figure 21, several cost outliers become apparent that shall be discussed here as well. On the one hand, regions mainly in central Namibia and Botswana, but also isolated cases all over the study extent stand out because of their high levelized cost of hydrogen. These regions are mainly located in low-wind areas and cannot accommodate sufficient photovoltaic capacity due to land eligibility restrictions. In the exemplary case of the Kalahari in Namibia and Botswana, the main reasons are rangeland protection for the fauna and pastures for cattle farming. Especially the latter can be combined with wind turbines, but not well with solar open-field photovoltaics, leading to elevated combined cost driven by high wind shares with high LCOE in such regions. Similar effects may occur in regions relying nearly exclusively on geothermal potentials in Kenya.

On the other hand, isolated very low-cost regions can be made out all over the map, but mainly in southern West Africa. These regions are dominated by hydropower, which is able to generate extremely competitive levelized cost of hydrogen. Especially in southern West Africa, overall hydrogen potentials are low due to limited land eligibility and comparably small region sizes, leading to high shares of low-cost hydropower in the mix and hence very low overall LCOH. The main advantage of hydrogen from hydropower is the high utilization rate of hydropower plants and electrolyzer especially at low expansion grades, when the usable river discharge is comparable low and hence can be very constant throughout the year. This generates very low cost of electricity and hydrogen. In the case of conventional reservoir hydropower, the discharge profile can be further smoothened by storing and releasing excess water, however, investment cost is higher. The lowest entry-level LCOH in 2030 is achieved by run-of-rover hydropower around 1.1 EUR/kg in 2030 and 0.9 EUR/kg in 2050. This is feasible only in the best locations though, and only for very limited potentials. In most regions and at higher potential expansion degrees, hydrogen based on run-of-river hydropower is more expensive than reservoir hydropower starting at ca. 1.5 EUR/kg in 2030 and 1.3 EUR/kg in 2030.

These values should be considered only as theoretical lower bounds though: In reality, reservoir hydropower plants usually play an important flexibility role in the electricity grid, which cannot easily be replaced. In the vast majority of all cases, hydrogen could hence only be produced from excess electricity, which would significantly reduce electrolysis full load hours. Currently, electrolysis cost contributes 25% of LCOH (median value) in low-cost conventional hydropower regions (35% in run-of-river hydropower systems), a reduction of the electrolyzer utilization rate by 50% would hence lead to a hydrogen cost increase of 25-35%. This systemic effect cannot be reflected in an atlas approach though where regions are independent of each other. The economic viability and the quantitative potential of green hydrogen produced from hydropower must therefore always be assessed on a plant-by-plant basis in an (inter-) national energy system model approach, taking into account temporal patterns of demand and other generation capacities in the system.

### 3.4.5 Hydrogen Cost-Potential Curves

Combining the above findings on quantitative hydrogen potentials and levelized cost into cost-potential curves allows to further analyze quantity-dependent aspects. When looking at the West, Southern and East African curves in Figure 23, it becomes apparent that the overall hydrogen potentials in the West Africa respectively the Southern and East Africa project regions are nearly identical at ca. 200 000 TWh/a. It has to be noted that this is more a theoretical number as it is unlikely that all renewable energies will only be used for hydrogen production. West Africa can offer slightly cheaper levelized cost of hydrogen on average in 2030, mainly caused by the vast areas and energetically favorable conditions in Mauritania, to a lesser degree also in Niger and Mali, all with similar technical potentials around 56 000 TWh/a. Next to the Sahara countries and Cabo Verde in West Africa, also South Africa, Lesotho, Namibia, and Botswana have very low-entry level cost-potentials, the largest quantities thereof in South Africa.

While the cost-potential curves in Niger and Mali are very comparable and rather flat in most parts, indicating very stable cost in wide ranges, the pattern in Mauritania is different. Very low entry cost around 2.0 €/kg in 2030 and high-cost slopes indicate the use of good wind locations early on in the potentials' expansion. In South Africa in turn, a significant share of the more expensive wind potentials is added towards higher expansion degrees, causing near constant solar-driven cost in the first half and a steady increase of the cost slope in the second half.

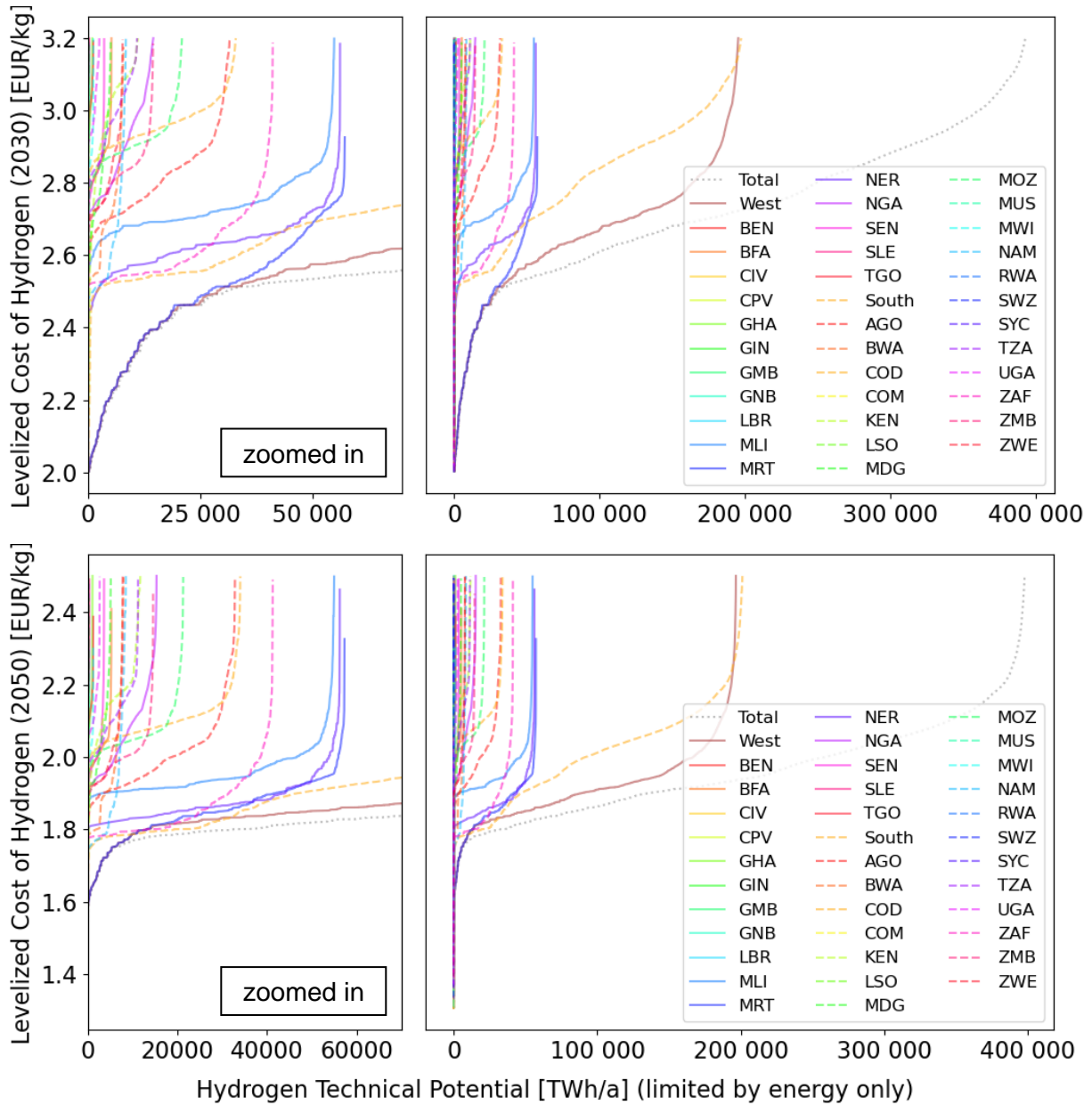

**Figure 23: National cost-potential curves for West Africa (continuous) and Southern and East Africa (dashed lines) in the year 2030 (top) and 2050 (bottom), zoom on country level curves on the left.**

When comparing the 2030 values to the 2050 data in above Figure 23, it becomes apparent that the overall cost level is lower in 2050 and the cost spread between entry level and high potential expansion degrees is reduced significantly. The average cost dependency on production quantity is hence lower in 2050. Together with an overall cost reduction, this is caused also by higher shares of solar power in the systems in low expansion steps due to solar PV cost decreasing over time. When considering only the low-cost technical potentials across the whole project region, the producible green hydrogen under 2.5 €/kg in 2030 amounts to ~31 TWh/a. By 2050, nearly 260 TWh/a can be produced under 2.0 €/kg.

### 3.4.6 Hydrogen Production System Design

An analysis of the region-specific designs of an optimal hydrogen production system reveals significant differences that explain the cost patterns observed before. Figure 24 shows that solar photovoltaics represent the main or sole power source in most regions across the whole study area. Notable exceptions include the Sahara region, Madagascar, Northern Kenya and multiple regions in Southern West Africa as well as some regions in Namibia and Botswana. The latter can be explained by the lack of photovoltaic potential in the Kalahari region where either rangeland for the natural fauna or pastures for cattle herding forbid open-field photovoltaics in many regions. Instead, onshore wind turbines with locally very high LCOE are used, leading to high overall hydrogen cost in such regions. Due to limited solar and wind potentials in south-western West Africa regions, cheap hydropower can produce significant shares of the local hydrogen. In other regions with hydropower, this relative share is reduced by the greater overall potential. Due to the importance of geothermal power in Kenya, it was considered as an additional power source for hydrogen here. Its role is very limited though, it is mainly used in regions where the solar and wind potentials are very low. It may, however, become more economical once constant hydrogen output stream is required where dispatchable generation capacities avoid additional storage cost compared to volatile electricity generation.

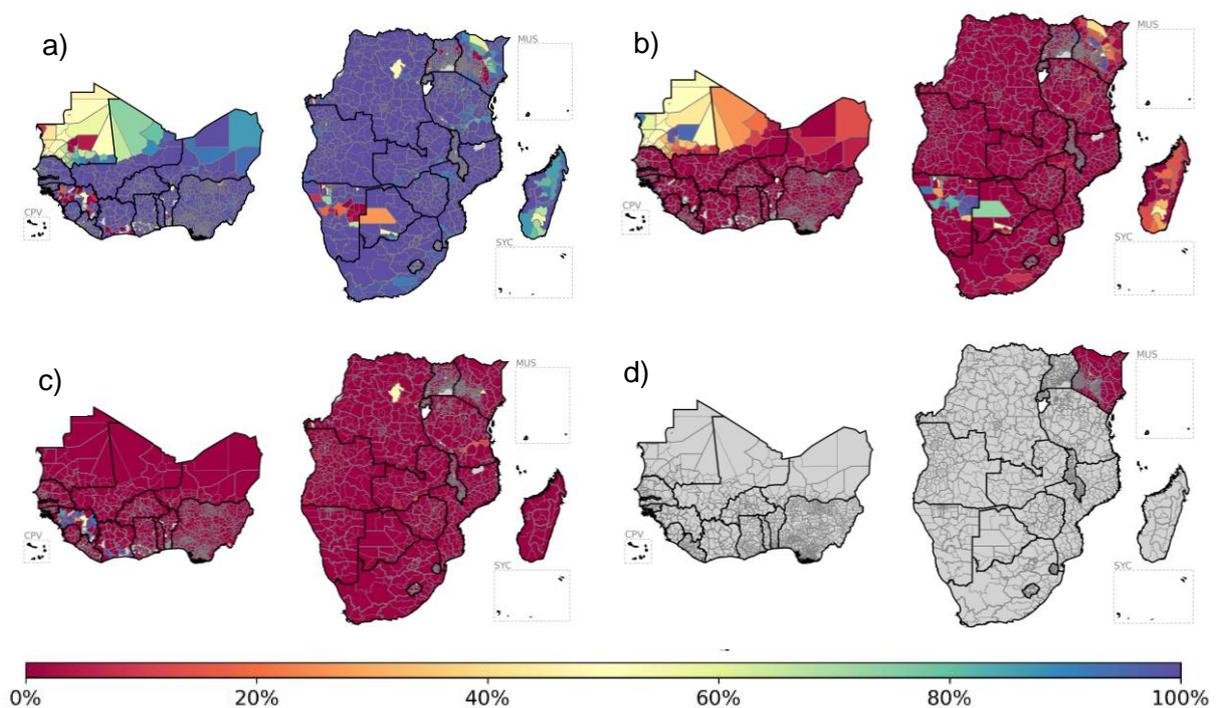

**Figure 24: Share of (a) solar photovoltaics, (b) onshore wind, (c) hydropower and (d) geothermal power in [%] of total renewable generation capacity at 25% potential expansion in the year 2030.**

The most interesting pattern is the tradeoff between solar photovoltaics and wind energy driven by cost reasons, i.e. not due to quantitative potential limits. It can be observed mainly in the Sahara region, in Kenya, Madagascar and to a very limited extent in South Africa. In those regions, attractive wind resources complement solar power especially at night, leading to higher full load hours of the electrolyzer. This reduces required electrolysis investment. Together with higher full load hours of wind farms in these regions, this effect equalizes the higher invest for wind farms and hence reduces levelized cost of hydrogen. The cheapest hydrogen is produced in regions with a wind capacity share of ca. 70-95%. This pattern is most obvious in Mauritania along the northern coastline and on the central Tagant plateau due to very high local wind speeds, but also in Southern Madagascar and in northern and central Kenya. The effect on electrolysis full load hours is significant, as Figure 25 shows on the left. Regions with high wind shares exceed electrolysis full load hours in the best solar-only regions by up to 60%. Similarly high full load hours are achieved in regions that are dominated by reservoir hydropower plants, which can shift or smoothen generation peaks flexibly

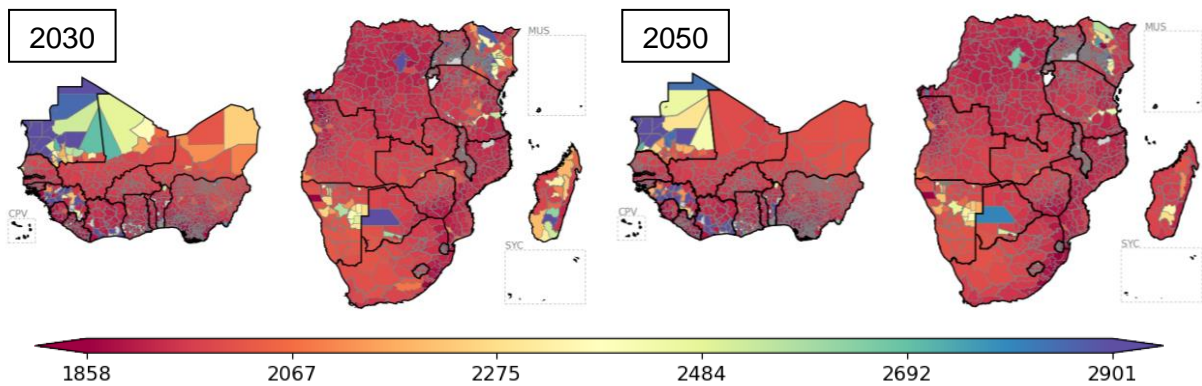

**Figure 25: Electrolysis full load hours [h/a] in the years 2030 (left) and 2050 (right) in comparison.**

However, the right-hand side of Figure 25 reveals a reduction in electrolysis full-load hours towards 2050 in many of the wind regions discussed above. This is rooted in a further shift towards solar photovoltaics as it can be observed in the optimal system designs in 2050 (Figure 26). Besides the wind capacities in the Kalahari that are caused merely by limitations of eligible land for open-field photovoltaics, wind by then plays a significant role only in Mauritania, in Kenya and in two central regions of Madagascar. In Mali and Niger, for example, wind has completely been replaced by open-field photovoltaics in 2050. In Kenya and central Madagascar as well as in several regions in Mauritania, the share of wind has been reduced at least. The reason is the greater cost reduction of photovoltaic technology between 2030 and 2050 compared to wind turbines, which is a more mature technology already. This also affects the electrolysis full load hours, which go down together with the wind share compared to the year 2030. The cost advantage of photovoltaics outweighs the added capacity that is required for electrolysis, given that electrolysis CAPEX is also reduced significantly in 2050 compared to 2030.

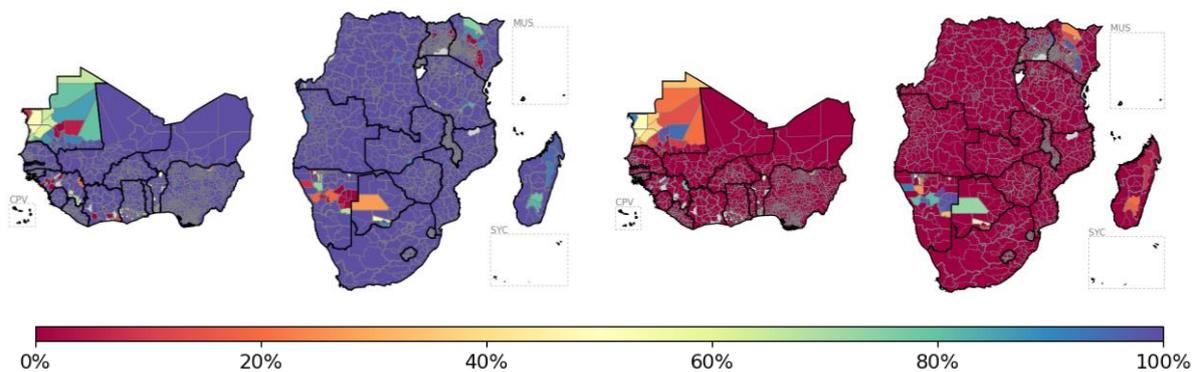

**Figure 26: Share of solar photovoltaics (left) and onshore wind (right) in [%] of total renewable generation capacity at 25% potential expansion in the year 2050.**

It is worth noting that lithium-ion battery capacities were offered as a potential system component, but batteries are not part of the optimal solution. Given the cost assumptions over the coming decades and the local generation profiles, it is more economical to overbuild generation capacities and/or operate electrolysis at reduced utilization rates. This assumes a time-independent lowest-cost hydrogen generation and may change as soon as specific demand profiles of electricity or hydrogen must be met.

## 3.5 Mapping local impact of green hydrogen projects

For this analysis of socio-economic development in Africa, the socio-economic indicator (Figure 27) provides a comprehensive overview of the composite indicator research, with high results indicating potential areas of interest for further green hydrogen and renewable energy projects.

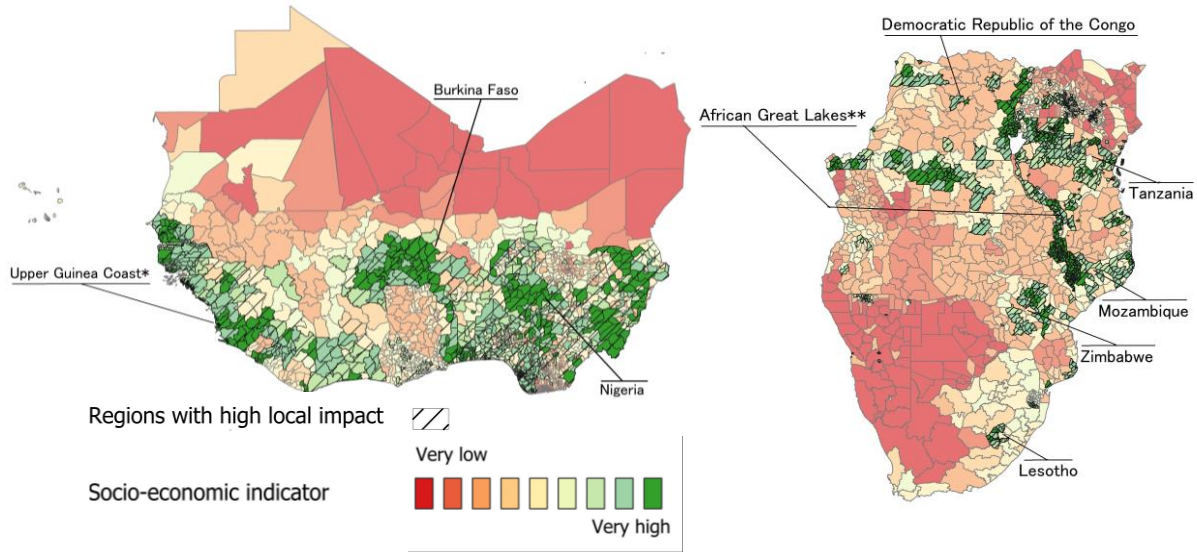

**Figure 27 Socio-economic indicator measuring the impact of green hydrogen project.**

As demonstrated in Figure 27, the most significant local impact (highest values) of green hydrogen and renewable energy projects is evident around the African Great Lakes region of East Southern Africa, including Lake Malawi, Lake Tanganyika, and Lake Victoria. Similar results can be identified in inland regions near the southern and eastern borders of the Democratic Republic of the Congo, northern Tanzania, and parts of Mozambique, Zimbabwe, and Lesotho. For Western Africa, the coastal regions of Upper Guinea and the inland regions encompassing Nigeria and Burkina Faso demonstrate a notable local impact. This high local impact in the described regions is primarily attributable to a combination of factors, including access to energy and macroeconomic potential impacts, as illustrated by the following two sub-indicators:

The local impact in terms of the socio-economic indicator is reflected in the composition of AE, which depicts the energy access indicator in capita/km²; ME, which highlights the macroeconomic effect in Jobs/ (Mwp*km²); and OE, which summarizes the score of the other effect. The underlying indicator results for AE and ME for the two regions are presented in the form of normalized score maps and statistical summaries in the subsections below.

### 3.5.1 Energy Access indicator (AE)

The energy access sub-index considers a combination of regional access to energy, and the access density. A combination of factors, mainly low access to essential services such as electricity and clean fuel, contributes to a high local impact value. The regional distribution is shown in Figure 28, along with the indicator statistics, energy access, and population density average at the national level.

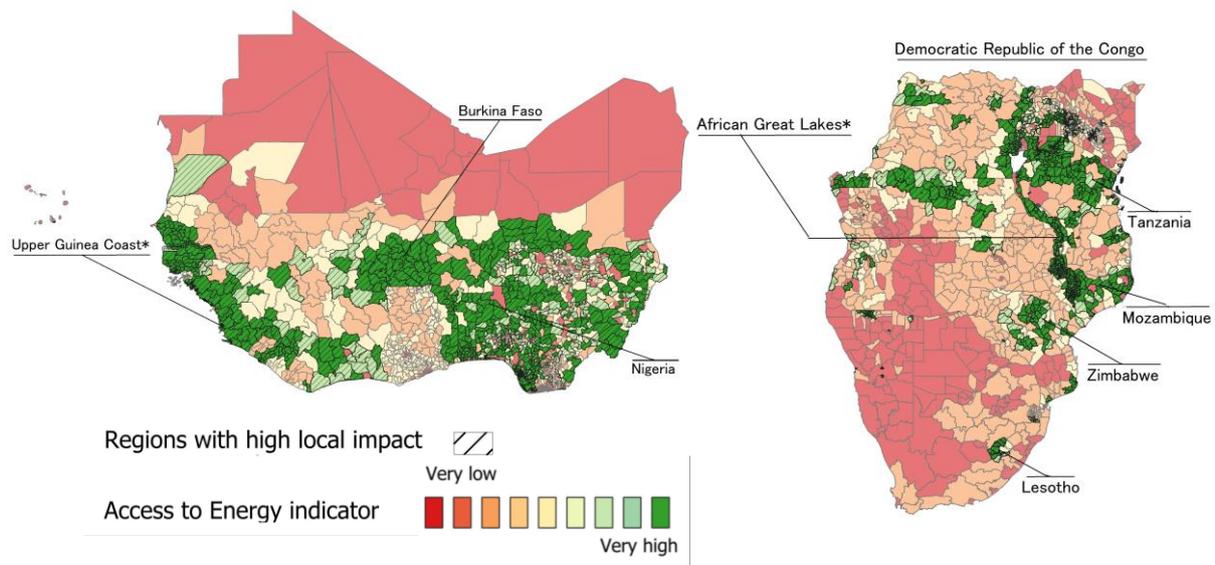

**Figure 28 Mapping the energy access indicator.**

Malawi, the northern regions of Tanzania, the southern areas of Mozambique, and western Zimbabwe, as well as the southern part of the Democratic Republic of Congo, exhibit a significant local impact in this regard. For West Africa, the high impact extends to the entire southern region, except Ghana, the Ivory Coast, and sparsely populated areas in the central region (Figure 28). Notably, countries such as Burkina Faso, Sierra Leone, and Liberia are showing particularly low access rates, with an average of less than 30% access to electricity and less than 20% access to clean fuels. Nevertheless, the significance of considering the population density is exemplified by the case of Niger. In this instance, despite the low access to electricity, which is below 20% on average, the sub-index value remains relatively low for Niger due to the sparsely populated or nomadic areas.

A comparison of the AE indicator statistics at the national level (Table 4) reveals that in the Western region, Nigeria, Gambia, and Sierra Leone, followed by Benin and Burkina Faso, have the greatest potential impact. This impact arises primarily from two factors: the high population density in the first two countries and the low levels of energy access for the remaining. It is also relevant to highlight that the interquartile range (IQR) for Gambia and Nigeria is higher than that of other countries due to regional inequalities in terms of electrification and population distribution, which is particularly evident in the case of Nigeria. Accordingly, it can be argued that these countries would greatly benefit from the implementation of distributed energy projects that aim to enhance renewable energy potential. Conversely, some countries will not benefit from a direct local impact of similar energy projects, such as Cape Verde and Ghana, as they already have high levels of energy access.

In the depicted Southern and East regions (Table 4), due to a combination of low average access to electricity and high population density, Malawi, Rwanda, and Uganda report the highest averages, followed by Tanzania. By contrast, countries such as Botswana and South Africa show lower values due to their high electrification rates.

**Table 4** *Energy access indicator statistics results along national energy access averages for West (W), East and Southern (S-E) regions*

| Country | CN | Region | AE indicator statistics | | | | Energy access | Density |
|---|---|---|---|---|---|---|---|---|
| | | | Median | Q25 | Q75 | IQR | Average in % | Average in c/km2 |
| Benin | BEN | W | 48.3 | 31.8 | 79.4 | 47.7 | 42.0 | 110.0 |
| Burkina Faso | BFA | W | 45.4 | 33.5 | 74.7 | 41.2 | 19.0 | 79.0 |
| Cabo Verde | CPV | W | 5.0 | 3.0 | 9.7 | 6.7 | 95.5 | 139.0 |
| Côte d'Ivoire | CIV | W | 26.0 | 16.4 | 30.7 | 14.3 | 71.1 | 85.0 |
| Gambia | GMB | W | 130.1 | 23.4 | 194.0 | 170.6 | 63.7 | 246.0 |
| Ghana | GHA | W | 19.7 | 7.9 | 24.0 | 16.2 | 86.3 | 139.0 |
| Guinea | GIN | W | 32.2 | 24.0 | 40.4 | 16.4 | 46.8 | 55.0 |
| Guinea-Bissau | GNB | W | 40.8 | 28.0 | 49.8 | 21.8 | 35.8 | 72.0 |
| Liberia | LBR | W | 22.0 | 12.2 | 35.2 | 23.1 | 29.8 | 54.0 |
| Mali | MLI | W | 14.2 | 5.2 | 23.8 | 18.5 | 53.4 | 17.0 |
| Mauritania | MRT | W | 0.7 | 0.3 | 6.9 | 6.6 | 47.7 | 5.0 |
| Niger | NER | W | 38.4 | 15.6 | 63.7 | 48.1 | 18.6 | 20.0 |
| Nigeria | NGA | W | 126.1 | 54.8 | 268.9 | 214.1 | 59.5 | 232.0 |
| Senegal | SEN | W | 39.8 | 12.7 | 54.6 | 41.9 | 68.0 | 89.0 |
| Sierra Leone | SLE | W | 68.5 | 46.8 | 73.3 | 26.5 | 27.5 | 113.0 |
| Togo | TGO | W | 43.0 | 34.3 | 49.7 | 15.4 | 55.7 | 156.0 |
| Angola | AGO | S-E | 6.8 | 1.7 | 17.7 | 16.0 | 48.2 | 27.0 |
| Botswana | BWA | S-E | 0.2 | 0.0 | 1.5 | 1.5 | 73.3 | 4.0 |
| Congo DR | COD | S-E | 14.5 | 0.1 | 30.5 | 30.4 | 20.8 | 41.0 |
| Eswatini | SWZ | S-E | 15.0 | 13.5 | 21.7 | 8.1 | 82.9 | 68.0 |
| Kenya | KEN | S-E | 36.0 | 21.2 | 56.8 | 35.6 | 76.5 | 97.0 |
| Lesotho | LSO | S-E | 37.8 | 27.2 | 44.4 | 17.2 | 50.4 | 71.0 |
| Malawi | MWI | S-E | 149.4 | 101.1 | 254.6 | 153.5 | 14.2 | 208.0 |
| Mozambique | MOZ | S-E | 20.0 | 8.6 | 31.0 | 22.3 | 31.5 | 41.0 |
| Namibia | NAM | S-E | 5.8 | 1.0 | 47.5 | 46.5 | 55.2 | 3.0 |
| Rwanda | RWA | S-E | 397.6 | 256.2 | 465.2 | 209.0 | 48.7 | 538.0 |
| South Africa | ZAF | S-E | 2.5 | 0.0 | 5.2 | 5.1 | 89.3 | 49.0 |
| Tanzania | TZA | S-E | 55.0 | 26.6 | 90.8 | 64.1 | 42.7 | 69.0 |
| Uganda | UGA | S-E | 140.0 | 64.5 | 240.1 | 175.5 | 45.2 | 235.0 |
| Zambia | ZMB | S-E | 11.0 | 7.8 | 15.4 | 7.6 | 46.7 | 25.0 |
| Zimbabwe | ZWE | S-E | 27.4 | 15.0 | 37.1 | 22.2 | 49.0 | 39.0 |

### 3.5.2 Macroeconomic Effects indicator (ME):

The macroeconomic effect sub-index is evaluated in terms of the regional impact on employment, with unemployment data serving as one of the main indicators (Figure 29). This index demonstrates that the most significant possible macroeconomic developments in West Africa are possible in Nigeria, Burkina Faso, and Cape Verde, as well as in regions such as the Upper Guinea Coast. Similarly, Eswatini, Lesotho, Malawi, Uganda, and the north-west of South Africa emerge as key areas with high employment impact in the Southern African context. Nigeria stands out as a state with the highest employment impact, due to its large labor force and dispersed but dense population. Nevertheless, countries such as Angola and the Democratic Republic of Congo exhibit a lower impact on employment levels despite comparable labor forces. This disparity occurs because of a mismatch between the distribution of energy potential and the population. It should be noted that Botswana has the lowest local employment impact, which is largely attributable to a sparse labor force, population distribution disparities, and high labor costs.

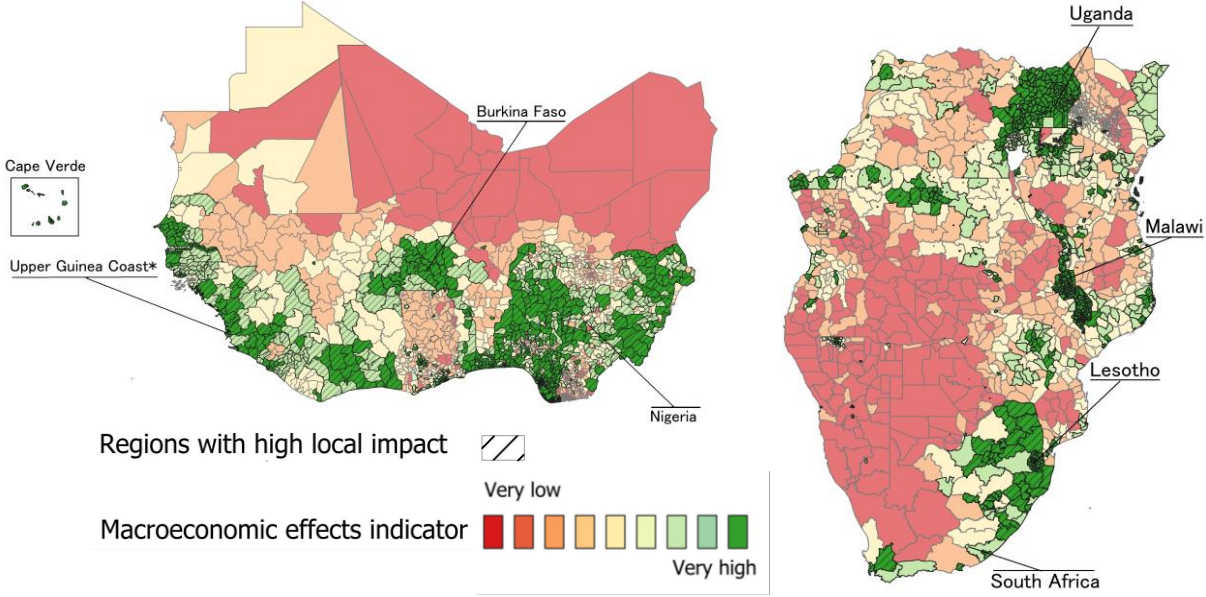

**Figure 29 Mapping of the macroeconomic effects indicator.**

At the national level, the ME indicator statistics (Table 5) demonstrate that in the Western region, the macroeconomic effect values are primarily related to the employment factor of increased electrical power and their potential for job creation. Burkina Faso, Gambia, Ghana, Nigeria, and Togo are notable countries in this region. Cabo Verde, despite having good energy access, exhibits a high macroeconomic effect value due to its high unemployment rate (13.3%). The results indicate that Kenya, Eswatini, Lesotho, Malawi, and Rwanda have a noteworthy potential for job creation from renewable energy projects (high macroeconomic effect). In this regard, for South Africa, it is mainly the unemployment rate that could be tackled with development of green hydrogen projects. In contrast, to other countries, the macroeconomic effect is attributable to the employment factor here in terms of potential job creation per megawatt-peak in these countries higher than 5.8 job/Mwp.

For this analysis, the focus is on the local labor forces in construction or operations and maintenance (OM), with manufacturing jobs being excluded. Nevertheless, the potential for competitive regional solar and wind industries to create an additional 30% and 11% of jobs, respectively, represents a significant opportunity for economic growth. Furthermore, due to data constraints and the fact that the analysis is conducted using local employment factors, indirect and induced jobs are not accounted for. This limitation analysis does not conflict with a local impact indicator that aims to analyze direct local jobs without population displacement.

Whereas Nigeria's potential stems primarily from its large workforce, West Africa's potential stems from its comparatively lower labor costs. Thus, for instance, green hydrogen production from wind, even with a lower employment impact than solar hydrogen production, still generates up to 11 times more jobs per megawatt peak than in the EU due to lower labor cost. In contrast, countries in Southern Africa, such as South Africa, Botswana, and Namibia, have labor costs that are on average only two to three times higher than in the European Union.

Nevertheless, even in these countries, particularly in the northwestern region of South Africa with a high population density, the high labor cost is offset by soaring unemployment rates exceeding 20% over the past two decades. This highlights the necessity for strategic interventions to leverage green hydrogen and renewable energy deployment as a potential employment catalyst.

**Table 5 Macroeconomic indicator statistics results along employment national indicators for West (W), East and Southern (S-E) regions.**

| Country | CN | Region | ME indicator statistics | | | | Unemployment | Employment Factor |
|---|---|---|---|---|---|---|---|---|
| | | | Median | Q25 | Q75 | IQR | Average in % | Average in job/MWp |
| Benin | BEN | W | 6.4 | 2.0 | 14.6 | 12.6 | 1.6 | 5.9 |
| Burkina Faso | BFA | W | 6.7 | 4.8 | 10.8 | 6.0 | 5.0 | 6.0 |
| Cabo Verde | CPV | W | 26.2 | 15.4 | 53.9 | 38.5 | 13.3 | 4.5 |
| Côte d'Ivoire | CIV | W | 5.6 | 3.2 | 8.0 | 4.9 | 2.7 | 5.8 |
| Gambia | GMB | W | 16.3 | 13.2 | 27.4 | 14.2 | 4.6 | 5.9 |
| Ghana | GHA | W | 13.7 | 8.2 | 36.1 | 28.0 | 3.7 | 5.8 |
| Guinea | GIN | W | 6.0 | 4.1 | 9.0 | 4.9 | 5.5 | 5.8 |
| Guinea-Bissau | GNB | W | 5.4 | 3.7 | 6.1 | 2.4 | 3.5 | 6.0 |
| Liberia | LBR | W | 4.1 | 2.0 | 6.1 | 4.0 | 3.4 | 5.7 |
| Mali | MLI | W | 1.0 | 0.4 | 1.9 | 1.5 | 2.4 | 5.9 |
| Mauritania | MRT | W | 1.5 | 0.3 | 3.3 | 2.9 | 10.9 | 5.8 |
| Niger | NER | W | 0.6 | 0.2 | 1.1 | 0.9 | 0.6 | 6.0 |
| Nigeria | NGA | W | 29.1 | 13.3 | 81.9 | 68.6 | 5.6 | 5.4 |
| Senegal | SEN | W | 3.8 | 2.9 | 14.2 | 11.3 | 3.4 | 5.6 |
| Sierra Leone | SLE | W | 11.3 | 7.1 | 12.9 | 5.8 | 3.5 | 5.7 |
| Togo | TGO | W | 7.7 | 5.8 | 14.3 | 8.5 | 3.9 | 5.9 |
| Angola | AGO | S-E | 1.0 | 0.4 | 2.2 | 1.8 | 10.0 | 1.8 |
| Botswana | BWA | S-E | 0.2 | 0.0 | 22.1 | 22.1 | 20.6 | 0.3 |
| Congo DR | COD | S-E | 3.8 | 1.5 | 26.8 | 25.2 | 4.8 | 5.9 |
| Eswatini | SWZ | S-E | 29.4 | 18.7 | 30.7 | 12.0 | 23.9 | 3.1 |
| Kenya | KEN | S-E | 17.0 | 7.8 | 36.0 | 28.1 | 5.2 | 5.8 |
| Lesotho | LSO | S-E | 6.3 | 4.0 | 12.8 | 8.8 | 17.7 | 2.3 |
| Malawi | MWI | S-E | 22.1 | 14.3 | 45.8 | 31.4 | 5.4 | 5.9 |
| Mozambique | MOZ | S-E | 2.4 | 1.1 | 4.2 | 3.1 | 3.7 | 5.9 |
| Namibia | NAM | S-E | 0.7 | 0.1 | 4.7 | 4.6 | 20.6 | 1.4 |
| Rwanda | RWA | S-E | 162.4 | 121.4 | 189.1 | 67.8 | 12.8 | 5.9 |
| South Africa | ZAF | S-E | 7.8 | 4.1 | 16.7 | 12.6 | 26.5 | 1.4 |
| Tanzania | TZA | S-E | 5.1 | 2.2 | 11.5 | 9.3 | 2.5 | 5.9 |
| Uganda | UGA | S-E | 16.4 | 7.8 | 25.9 | 18.1 | 4.0 | 5.8 |
| Zambia | ZMB | S-E | 0.6 | 0.3 | 1.2 | 1.0 | 5.8 | 1.6 |
| Zimbabwe | ZWE | S-E | 3.9 | 1.9 | 6.2 | 4.3 | 7.6 | 4.2 |

## 4 Discussion and conclusions

The obtained results revealed several findings and allowed for various conclusions. Those are presented here first for each analysis step separately and finally for the whole analysis.

### 4.1.1 Land eligibility and renewable electricity potentials

The present study provides valuable insights into the land eligibility for renewable energy technologies in the selected African regions, taking into account technical, sociological, and ecological criteria. The results highlight the distinct inclinations of regional stakeholders, including community members, governmental bodies, and international institutions in every country. It is worth noting that eligibility rates vary per country, ranging from ~0.1% in Seychelles to around 50% in Niger, Mali, and Mauritania. Protected forests and pastures are key constraints to the eligible land areas in both regions, while sand dunes in western Africa also exclude large areas from eligibility.

Regarding the renewable potentials, the results of this study reveal that open-field photovoltaics have the largest installable capacity potential among renewable energy sources in the analyzed regions in Sub-Saharan Africa, especially in the desert regions of the Sahara and the Nama Karoo region in South Africa, Namibia, and Botswana. However, the capacity potential varies greatly across different regions, and even seemingly low-potential regions can have significant absolute potentials resulting in levelized cost of 3.5 Ct$_€$/kWh down to 2 Ct$_€$/kWh by 2030 with a further reduction of roughly 25% until 2050 in the best locations in the Nama Karoo ecoregion in South Africa and Namibia, extending into Botswana. In 2050 with assumed reductions in investment cost a further reduction in LCOE of 25% can be achieved and the spatial difference is narrowed down, resulting in a gradual reduction of the competitive advantage of the desert regions.

Despite the relatively low potential of onshore wind (~13% of PV), it suffices several times the local demand in many countries. In addition, its ability to generate energy in times when PV cannot make it still a viable option to complement and diversify the energy mix. Most of the best locations can be found close and outside intertropical latitudes such as the south of Southern Africa or the North of Eastern Africa. However, scatter potential can nevertheless be found in many countries within the tropics due to terrain characteristics and local wind patterns. In such cases, the land restrictions might be adapted according to the local preference to develop wind energy projects. For hydropower the potential is even more constrained. It is even non-existent in some countries due to their geographical locations, while in others, like the DR Congo it is very abundant and might have the potential to be used as the bases for a robust renewable energy system. Given the very large local impact, its exploitation remains totally dependent on local preferences for this type of energy. Finally, the geothermal energy in Kenya is more expensive compared to local PV but has the advantage of dispatchability. Hence, the full energy system cost is most relevant here.

These findings have significant implications for policymakers and investors. It suggests that the analyzed regions in Sub-Saharan Africa have the potential to become significant contributors to the global renewable energy market in the upcoming years. However, more research is required to explore the economic, social, and environmental impacts of large-scale renewable energy deployment in Africa. It is also important to identify the most effective policy measures and financing mechanisms to support this transition.

### 4.1.2 Sustainable water supply assessment

According to the evaluation of groundwater sustainable yield, we found that water utilization in West and Southern East Africa is sustainable. However, notable variations exist among countries and climate scenarios concerning estimated groundwater sustainable yield. For instance, our analysis indicates that the projected groundwater availability in 2050, particularly under the RCP8.5 climate scenario, is anticipated to be lower than the projected levels observed in 2020 and 2030. Overall, our analysis suggests that the available groundwater resources hold immense potential. They are not only capable of meeting human and environmental water needs in West and Southern East Africa regions but also have the capacity to support green hydrogen production, although under future climate conditions, especially in the pessimistic RCP8.5 scenario there will be some groundwater decline.

Beyond the availability of sustainable groundwater yields seawater desalination exists as a promising option for dry regions best close to shore. However, cost reveals that even long-distance transport is economical and technical possible. Nonetheless, the political feasibility and safety of infrastructure might be the biggest challenges in this regard.

### 4.1.3 Hydrogen cost, potentials and water use

The green hydrogen potential in the project area is huge at 400 PWh/a and more than twice the whole 2021 global primary energy consumption (bp 2022) . They exceed the local projected energy demand of the involved countries by a factor of 20 and, hence, allow export schemes in nearly all involved countries after prioritizing the satisfaction of local demands. The largest and most economical potentials are found in the desert areas of the Sahara, here particularly along the northern Mauritanian coast, and in the border region between South Africa, Namibia and Botswana. Very interesting potentials are also found on Cabo Verde and in Kenya, but also in Madagascar and in several isolated regions where hydropower is available. Whilst the Sahara and Mauritania in particular employ large shares of wind capacity in their optimal systems, hydrogen is produced mainly from solar power in Southern Africa. Hydropower can produce very low-cost hydrogen but must be treated with care as hydropower electricity will be needed for electricity supply and grid stability more urgently. Using only excess energy in turn leads to a reduction of quantitative potentials and increases cost. The use of hydropower for hydrogen should therefore always be assessed in plant-specific assessments.

Entry level cost for solar and wind powered hydrogen are around 2 EUR/kg in 2030, decreasing to 1.6 EUR/kg by 2050. The potential-weighted average for the whole study region is at roughly 2.7 EUR/kg in 2030 and slightly above 1.9 EUR/kg in 2050. The potential of green hydrogen producible under 2.5 EUR/kg in 2030 amounts to roughly 31 TWh/a. By 2050, ~260 TWh/a can be produced under 2 EUR/kg. Unlike wind-based hydrogen, solar-based hydrogen cost is nearly independent of the produced quantity.

In two third of the considered regions, sustainable local groundwater is insufficient to cover the water demand for production of the full technical hydrogen potential, roughly half the regions cannot even provide sufficient water for a moderate expansion degree of 25% of the technical potential. This affects most of the high-potential low-cost regions. The cost of alternative water supply options such as desalination, however, does not exceed ~1% of the levelized cost of hydrogen. It can be concluded that several solutions exist to economically produce sustainable hydrogen beyond the limits of local groundwater potentials. In order to preserve the local water resources for the population, however, attractive levelized cost of hydrogen at large scale often require the construction of additional seawater desalination capacities. Sustainable water supply infrastructure planning should therefore be an integral part of any large-scale sustainable hydrogen project in the region.

### 4.1.4 Mapping local socio-economic impacts of green hydrogen projects

The analysis of the local impacts of green hydrogen projects across selected regions in Africa reveals distinct regional patterns, with significant implications for both energy access and macroeconomic development. The African Great Lakes region, incorporating countries such as Malawi, Tanzania, and Mozambique, as well as areas along the Upper Guinea coast demonstrate the highest local impact. This impact is primarily driven by factors such as low access to essential services like electricity and clean fuel, coupled with high population densities and employment impact.

In West Africa, countries such as Nigeria show potential for job creation due to the substantial labor forces and comparatively lower labor costs. In parallel, in Southern Africa regions, nations Rwanda represent the same pattern, while countries with high unemployment rate among the labor force could benefit from a local hydrogen economy. Nevertheless, intra-country disparities, exemplified by regional discrepancies within Nigeria, present notable challenges. Furthermore, the prospect of competitive regional solar and wind industries mainly is Southern and East Africa could create additional manufacturing jobs opportunities. However, addressing challenges like data limitations and the exclusion of indirect and induced employment metrics remains imperative to ensure a comprehensive understanding of the localized impacts of green hydrogen projects.

### 4.1.5 Green hydrogen potentials in Sub-Saharan Africa derived by a multidisciplinary approach

In the realm of energy production, the competition for land between electricity and hydrogen has garnered attention due to the significant potential for both. In photovoltaic (PV) dominated systems, the costs are less impacted by the expansion rate, opening opportunities for regions with abundant resources. Identifying regions with cheap potential that exceeds their own demand presents the possibility of becoming exporters within and outside of Africa. The hydrogen cost patterns observed in this study, generally align with other studies that evaluated LCOH in African countries. In comparison to Gado et al. (Gado, Nasser, and Hassan 2024), for example, the reported LCOH can be lowered by 2 - 3 EUR/kg in 2030 if a co-optimized PV and wind hybrid system is utilized across the considered countries.

Excessive exploitation of groundwater resources in coastal regions will significantly increase the risks of saltwater intrusion and land subsidence (Galloway and Burbey 2011; Michael et al. 2017). This is in addition to the environmental concerns discussed by Bierkens and Wada (Bierkens and Wada 2019). Therefore, balancing desalination of seawater with groundwater extraction is crucial, particularly in coastal areas. Groundwater exploitation must be planned with caution, even when cost-effective solutions appear viable. Prioritizing desalination over extensive groundwater extraction helps protect these valuable freshwater reserves and ensures a more sustainable and resilient water supply. Furthermore, desalination mitigates risks such as land subsidence and long-term aquifer depletion, which can have significant environmental and economic consequences. Thus, integrating desalination into water resource management strategies is essential for maintaining the integrity of coastal water supplies. Beyond the sole usage for hydrogen, a reasonable oversizing of the desalination and water transport infrastructure might improve local access to affordable and clean water.

Besides the impact of groundwater availability, the overall green hydrogen cost potential has been shown to be driven by the local land eligibility constraints which are tied to the local prevailing policies or regulations. For countries with similar land area, eligibility of such land for green hydrogen production can substantially vary hence leading to substantial variation between countries in levelized cost of hydrogen as indicated.

The coastal regions of Western Africa and Northeastern Africa are promising hubs for hydrogen production, offering not only significant potential for generating green energy but also for delivering substantial local economic benefits. These regions possess abundant natural resources, which, when combined with their strategic geographic locations, make them ideal candidates for large-scale hydrogen production and export. The growing population and the vast labor force presents a unique advantage to support the hydrogen economy, particularly in the early stages of development, where there will be a high demand for construction, installation, operation, and maintenance jobs. However, to fully realize the potential of this emerging technology, it is crucial to invest in capacity building to further develop a local manufacturing industry. In the long term, this development could be driven by the competitive costs of African technologies thanks to The African Continental Free Trade Agreement (AfCFTA) that will enhance cross-border trade and investment in the renewable energy sector.

The current low level of energy access in many African regions could pose challenges such as increasing the risk of energy supply insecurity for hydrogen production and the potential for energy-related conflicts. The developed approach highlighted the potential of green hydrogen projects to enhance energy security and broaden electricity access, thereby addressing long-standing energy deficits. This shift could also help mitigate the risks associated with relying on traditional fossil fuels and imports, contributing to a more sustainable and resilient energy system.

Moreover, the existing presence of fossil fuel and chemical industries in countries like Nigeria, Angola and South Africa offers a solid foundation for hydrogen production and export. These industries already have substantial infrastructure and investment incentives in place, which can be repurposed and upgraded to support hydrogen production. Additionally, the established natural gas pipelines in these regions could be adapted to transport blended green hydrogen, while oil terminal infrastructures might be utilized for liquid organic hydrogen carriers. This approach not only maximizes the use of existing assets but also reduces the initial investment required to develop a hydrogen economy. However, to avoid the potential of technological lock-in and ensure a smooth transition to green hydrogen, it is essential to implement robust decarbonization policies. These policies should aim to gradually phase out reliance on fossil fuels while promoting the adoption of clean, renewable energy sources.

## Acknowledgements

A major part of this work has been carried out within the framework of the $H_2$Atlas-Africa project (03EW0001) funded by the German Federal Ministry of Education and Research (BMBF). Additionally, we acknowledge funding by the European Space Agency (ESA) in the Framework of the Dragon 5 cooperation between ESA and Chinese Ministry of Science and Technology under Projects 59197 and 59316. Parts of this work was supported by the Helmholtz Association under the program "Energy System Design". This work was supported by partners in the focus countries led by the national team leaders who provided comprehensive data and the identification of local preferences: Mr. Chipilica Barbosa (Angola), Prof. Julien Adounkpe (Benin), Dr. Lapologang Magole (Botswana), Prof. Tanga Pierre Zoungrana (Burkina Faso), Prof. Luis Jorge Fernandes (Cape Verde), Mr. Simphiwe Khumalo (Eswaitni), Prof. Wilson A.

Agyare (Ghana), Prof. Konate Souleymane (Ivory Coast), Mr. Joseph Kalowekamo (Malawi), Dr. Yacouba Diallo (Mali), Mr. Mohamed Abdoullah Mohamedou (Mauritania), Dr. Pradeep M. K. Soonarane (Mauritius), Prof. Boaventura Chongo Cuamba (Mozambique), Mr. Panduleni Hamukwaya (Namibia), Prof. Rabani Adamou (Niger), Prof. Apollonia Okhimamhe (Nigeria), Dr Aime Tsinda (Rwanda), Dr. Ibrahima Barry (Senegal), Mr. Crescent Mushwana (South Africa), Mr. Mathew Matimbwi (Tanzania), Prof. Sidat Yaffa (The Gambia), Prof. Agboka Komi (Togo), Mr. Edson Twinomujuni (Uganda), Dr. Martin Mbewe (Zambia), and Dr. Fortunate Farirai (Zimbabwe). Additionally, we would like to acknowledge Mrs. Alberta Aryee, Mrs. Chenai Marangwanda, and Dr. Imasiku Katundu for their valuable support and facilitating interaction with the national teams in the various countries of West and Southern Africa.

## Author contributions

## References

Abiye, Tamiru. 2016. 'Synthesis on Groundwater Recharge in Southern Africa: A Supporting Tool for Groundwater Users'. *Groundwater for Sustainable Development* 2:182–89.

Akinsanola, A. A., K. O. Ogunjobi, I. E. Gbode, and V. O. Ajayi. 2015. 'Assessing the Capabilities of Three Regional Climate Models over CORDEX Africa in Simulating West African Summer Monsoon Precipitation'. *Advances in Meteorology* 2015:1–13. https://doi.org/10.1155/2015/935431.

Alcamo, Joseph, Petra Döll, Thomas Henrichs, Frank Kaspar, Bernhard Lehner, Thomas Rösch, and Stefan Siebert. 2003. 'Development and testing of the WaterGAP 2 global model of water use and availability'. *Hydrological Sciences Journal* 48 (3): 317–37. https://doi.org/10.1623/hysj.48.3.317.45290.

Amjath-Babu, TS, Timothy J Krupnik, Sreejith Aravindakshan, Muhammad Arshad, and Harald Kaechele. 2016. 'Climate Change and Indicators of Probable Shifts in the Consumption Portfolios of Dryland Farmers in Sub-Saharan Africa: Implications for Policy'. *Ecological Indicators* 67:830–38. https://doi.org/10.1016/j.ecolind.2016.03.030.

Ayodele, T. R., and J. L. Munda. 2019. 'Potential and Economic Viability of Green Hydrogen Production by Water Electrolysis Using Wind Energy Resources in South Africa'. *International Journal of Hydrogen Energy* 44 (33): 17669–87. https://doi.org/10.1016/j.ijhydene.2019.05.077.

Ballo, Abdoulaye, Kouakou Valentin Koffi, and Bruno Korgo. 2022. 'Law and Policy Review on Green Hydrogen Potential in ECOWAS Countries'. *Energies* 15 (7). https://doi.org/10.3390/en15072304.

Barron, Manuel, and Maximo Torero. 2017. 'Household Electrification and Indoor Air Pollution'. *Journal of Environmental Economics and Management* 86:81–92.

Bayat, Bagher, Bamidele Oloruntoba, Carsten Montzka, Harry Vereecken, and Harrie-Jan Hendricks Franssen. 2023. 'Implications for Sustainable Water Consumption in Africa by Simulating Five Decades (1965–2014) of Groundwater Recharge'. *Journal of Hydrology* 626 (November):130288. https://doi.org/10.1016/j.jhydrol.2023.130288.

Bierkens, Marc F P, and Yoshihide Wada. 2019. 'Non-Renewable Groundwater Use and Groundwater Depletion: A Review'. *Environmental Research Letters* 14 (6): 063002. https://doi.org/10.1088/1748-9326/ab1a5f.
Blau, M. T., and K.-J. Ha. 2020. 'The Indian Ocean Dipole and Its Impact on East African Short Rains in Two CMIP5 Historical Scenarios With and Without Anthropogenic Influence'. *Journal of Geophysical Research: Atmospheres* 125 (16): e2020JD033121. https://doi.org/10.1029/2020JD033121.
bp. 2022. 'Bp Statistical Review of World Energy 2022'. 71. BP p.l.c. https://www.bp.com/content/dam/bp/business-sites/en/global/corporate/pdfs/energy-economics/statistical-review/bp-stats-review-2022-full-report.pdf.
Brauner, Simon, Amin Lahnaoui, Solomon Agbo, Stefan Böschen, and Wilhelm Kuckshinrichs. 2023. 'Towards Green Hydrogen?–A Comparison of German and African Visions and Expectations in the Context of the H2Atlas-Africa Project'. *Energy Strategy Reviews* 50:101204.
Cardinale, Roberto. 2023. 'From Natural Gas to Green Hydrogen: Developing and Repurposing Transnational Energy Infrastructure Connecting North Africa to Europe'. *Energy Policy* 181:113623. https://doi.org/10.1016/j.enpol.2023.113623.
Chowdhury, Kamal, Ranjit Deshmukh, and Les Armstrong. 2022. 'Enabling a Low-Carbon Electricity System for Southern Africa'. *Joule* 6:1826–44. https://doi.org/10.1016/j.joule.2022.06.030.
Davis, Neil N., Jake Badger, Andrea N. Hahmann, Brian O. Hansen, Niels G. Mortensen, Mark Kelly, Xiaoli G. Larsén, et al. 2023. 'The Global Wind Atlas: A High-Resolution Dataset of Climatologies and Associated Web-Based Application'. *Bulletin of the American Meteorological Society* 104 (8): E1507–25. https://doi.org/10.1175/BAMS-D-21-0075.1.
Döll, Petra, Frank Kaspar, and Bernhard Lehner. 2003. 'A Global Hydrological Model for Deriving Water Availability Indicators: Model Tuning and Validation'. *Journal of Hydrology* 270 (1–2): 105–34.
Döll, Petra, Bernhard Lehner, and Frank Kaspar. 2002. 'Global Modeling of Groundwater Recharge'. In *Proceedings of Third International Conference on Water Resources and the Environment Research, Technical University of Dresden, Germany*, 1:27–31. https://www.academia.edu/download/41391258/Global_Modeling_of_Groundwater_Recharge20160121-1749-1ljhcbk.pdf.
Edmunds, W. M., and C. B. Gaye. 1994. 'Estimating the Spatial Variability of Groundwater Recharge in the Sahel Using Chloride'. *Journal of Hydrology* 156 (1–4): 47–59.
Edmunds, W. M., and E. P. Wright. 1979. 'Groundwater Recharge and Palaeoclimate in the Sirte and Kufra Basins, Libya'. *Journal of Hydrology* 40 (3–4): 215–41.
Egeland-Eriksen, Torbjørn, Amin Hajizadeh, and Sabrina Sartori. 2021. 'Hydrogen-Based Systems for Integration of Renewable Energy in Power Systems: Achievements and Perspectives'. *International Journal of Hydrogen Energy* 46 (63): 31963–83.
Eke, Joyner, Ahmed Yusuf, Adewale Giwa, and Ahmed Sodiq. 2020. 'The Global Status of Desalination: An Assessment of Current Desalination Technologies, Plants and Capacity'. *Desalination* 495 (December):114633. https://doi.org/10.1016/j.desal.2020.114633.
Favreau, G., B. Cappelaere, S. Massuel, M. Leblanc, M. Boucher, N. Boulain, and C. Leduc. 2009. 'Land Clearing, Climate Variability, and Water Resources Increase in Semiarid Southwest Niger: A Review'. *Water Resources Research* 45 (7): 2007WR006785. https://doi.org/10.1029/2007WR006785.
Franzmann, D., H. Heinrichs, F. Lippkau, T. Addanki, C. Winkler, P. Buchenberg, T. Hamacher, M. Blesl, J. Linßen, and D. Stolten. 2023. 'Green Hydrogen Cost-Potentials for Global Trade'. *International Journal of Hydrogen Energy* 48 (85): 33062–76. https://doi.org/10.1016/j.ijhydene.2023.05.012.
GADM. 2023. 'Database of Global Administrative Areas, Version 3.4'. https://gadm.org.
Gado, Mohamed G., Mohamed Nasser, and Hamdy Hassan. 2024. 'Potential of Solar and Wind-Based Green Hydrogen Production Frameworks in African Countries'.

*International Journal of Hydrogen Energy* 68 (May):520–36. https://doi.org/10.1016/j.ijhydene.2024.04.272.

Galloway, Devin L., and Thomas J. Burbey. 2011. 'Review: Regional Land Subsidence Accompanying Groundwater Extraction'. *Hydrogeology Journal* 19 (8): 1459–86. https://doi.org/10.1007/s10040-011-0775-5.

GeoKit. 2024. 'FZJ-IEK3-VSA/Geokit: 1.4.0'. GitHub. https://github.com/FZJ-IEK3-VSA/geokit/tree/v1.4.0.

GLAES. 2024. 'FZJ-IEK3-VSA/Glaes: 1.2.1'. GitHub. https://github.com/FZJ-IEK3-VSA/glaes/tree/v1.2.1.

Groß, Theresa, Kevin Knosall, Maximilian Hoffmann, Noah Pflugtadt, and Detlef Stolten. 2024. 'ETHOS.FINE: A Framework for Integrated Energy System Assessment'. *Submitted to JOSS*. 10.48550/arXiv.2311.05930.

Guendouz, A., A. S. Moulla, W. M. Edmunds, K. Zouari, P. Shand, and A. Mamou. 2003. 'Hydrogeochemical and Isotopic Evolution of Water in the Complexe Terminal Aquifer in the Algerian Sahara'. *Hydrogeology Journal* 11 (4): 483–95. https://doi.org/10.1007/s10040-003-0263-7.

Hafner, Manfred, Simone Tagliapietra, and Lucia de Strasser. 2018. 'Prospects for Renewable Energy in Africa'. In *Energy in Africa*, 47–75. Springer, Cham. ttps://doi.org/10.1007/978-3-319-92219-5_3.

Heinrichs, Heidi, Christoph Winkler, David Franzmann, Jochen Linßen, and Detlef Stolten. 2021. 'Die Rolle von Meerwasserentsalzungsanlagen in Einer Globalen Grünen Wasserstoffwirtschaft'. *14. Aachener Tagung Wassertechnologie*. https://juser.fz-juelich.de/record/905482.

Hersbach, Hans, Bill Bell, Paul Berrisford, Gionata Biavati, András Horányi, Joaquín Muñoz Sabater, Julien Nicolas, et al. 2018. 'ERA5 Hourly Data on Single Levels from 1979 to Present'. *Copernicus Climate Change Service (C3s) Climate Data Store (Cds)* 10 (10.24381).

IRENA, 2024. 2024. 'The Energy Transition in Africa: Opportunities for International Collaboration with a Focus on the G7'. IRENA. https://www.irena.org/-/media/Files/IRENA/Agency/Publication/2024/Apr/IRENA_G7_Energy_transition_Africa_2024.pdf?rev=e7ef89d9b70a4872a84e23aaa7f5e5d4.

Ishmam, Shitab, Heidi Heinrichs, Winkler, C, Bayat, B, Lahnaoui, A, Agbo, S, U. Pena Sanchez, E, et al. 2024. 'Mapping Local Green Hydrogen Cost-Potentials by a Multidisciplinary Approach'. *Submitted to International Journal of Hydrogen Energy*.

Ishmam, Shitab, Heidi Heinrichs, Christoph Winkler, Bagher Bayat, Amin Lahnaoui, Solomon Agbo, Edgar Ubaldo Pena Sanchez, et al. 2024. 'Mapping Local Green Hydrogen Cost-Potentials by a Multidisciplinary Approach'. arXiv. https://doi.org/10.48550/arXiv.2407.07573.

Jülich Systems Analysis, and IBG-3, Forschungszentrum Juelich. 2024. 'H2Atlas-GUI'. 2024. https://africa.h2atlas.de/.

Kamil, Kehinde Ridwan, Bassey Okon Samuel, and Umar Khan. 2024. 'Green Hydrogen Production from Photovoltaic Power Station as a Road Map to Climate Change Mitigation'. *Clean Energy* 8 (2): 156–67. https://doi.org/10.1093/ce/zkae020.

Khandker, Shahidur R, Douglas F Barnes, and Hussain A Samad. 2012. 'The Welfare Impacts of Rural Electrification in Bangladesh'. *The Energy Journal* 33 (1): 187–206.

Khraief, Naceur, Oluwasola E Omoju, and Muhammad Shahbaz. 2016. 'Are Fluctuations in Electricity Consumption per Capita in Sub-Saharan Africa Countries Transitory or Permanent?' *Energy Strategy Reviews* 13–14:86–96. https://doi.org/10.1016/j.esr.2016.08.007.

Kirchherr, Julian, and Katrina J Charles. 2016. 'The Social Impacts of Dams: A New Framework for Scholarly Analysis'. *Environmental Impact Assessment Review* 60:99–114.

Lawal, Adedoyin Isola, Ilhan Ozturk, Ifedolapo O. Olanipekun, and Abiola John Asaleye. 2020. 'Examining the Linkages between Electricity Consumption and Economic Growth in African Economies'. *Energy* 208 (October):118363. https://doi.org/10.1016/j.energy.2020.118363.

Leblanc, Marc J., Guillaume Favreau, Sylvain Massuel, Sarah O. Tweed, Maud Loireau, and Bernard Cappelaere. 2008. 'Land Clearance and Hydrological Change in the Sahel: SW Niger'. *Global and Planetary Change* 61 (3–4): 135–50.

Leduc, Christian, Guillaume Favreau, and P. Schroeter. 2001. 'Long-Term Rise in a Sahelian Water-Table: The Continental Terminal in South-West Niger'. *Journal of Hydrology* 243 (1–2): 43–54.

Litzow, Erin L., Subhrendu K. Pattanayak, and Tshering Thinley. 2019. 'Returns to Rural Electrification: Evidence from Bhutan'. *World Development* 121 (September):75–96. https://doi.org/10.1016/j.worlddev.2019.04.002.

Loehr, Katharina, Custodio Efraim Matavel, Sophia Tadesse, Masoud Yazdanpanah, Stefan Sieber, and Nadejda Komendantova. 2022. 'Just Energy Transition: Learning from the Past for a More Just and Sustainable Hydrogen Transition in West Africa'. *Land* 11 (12): 2193.

L'vovič, Mark I. 1979. *World Water Resources and Their Future*. Raleigh, N.C: American Geophysical Union.

MacDonald, Alan M., R. Murray Lark, Richard G. Taylor, Tamiru Abiye, Helen C. Fallas, Guillaume Favreau, Ibrahim B. Goni, Seifu Kebede, Bridget Scanlon, and James PR Sorensen. 2021. 'Mapping Groundwater Recharge in Africa from Ground Observations and Implications for Water Security'. *Environmental Research Letters* 16 (3): 034012.

Mariita, Nicholas O. 2002. 'The Impact of Large-Scale Renewable Energy Development on the Poor: Environmental and Socio-Economic Impact of a Geothermal Power Plant on a Poor Rural Community in Kenya'. *Africa: Improving Energy Services for the Poor* 30 (11): 1119–28. https://doi.org/10.1016/S0301-4215(02)00063-0.

Michael, Holly A., Vincent E. A. Post, Alicia M. Wilson, and Adrian D. Werner. 2017. 'Science, Society, and the Coastal Groundwater Squeeze'. *Water Resources Research* 53 (4): 2610–17. https://doi.org/10.1002/2017WR020851.

Mukhtar, Mustapha, Humphery Adun, and Dongsheng Cai. 2023. 'Juxtaposing Sub-Sahara Africa's Energy Poverty and Renewable Energy Potential'. *Scientific Reports* 13 (11643). https://doi.org/10.1038/s41598-023-38642-4.

Mutsaka, Farai, and Gerald Imray. 2024. 'Extreme Drought in Southern Africa Leaves Millions Hungry'. *AP News*, 31 March 2024. https://apnews.com/article/southern-africa-drought-hunger-food-climate-2ef702abc386f7182dbc5f8f4192be3c.

Nnachi, Gideon Ude, Coneth Graham Richards, and Yskandar Hamam. 2024. 'A Comprehensive State-of-the-Art Survey on Green Hydrogen Economy in Sub-Saharan Africa'.

Ogola, Pacifica, Brynhildur Davidsdottir, and Ingvar Birgir Fridleifsson. 2012. 'Potential Contribution of Geothermal Energy to Climate Change Adaptation: A Case Study of the Arid and Semi-Arid Eastern Baringo Lowlands, Kenya'. *Renewable and Sustainable Energy Reviews* 16 (6): 4222–46.

Ondraczek, Janosch, Nadejda Komendantova, and Anthony Patt. 2015. 'WACC the Dog: The Effect of Financing Costs on the Levelized Cost of Solar PV Power'. *Renewable Energy* 75:888–98.

Pavlidis, Vasileios, Mahlatse Kganyago, Mxolisi Mukhawana, and Thomas Alexandridis. 2024. 'A Drought Monitoring and Early Warning Service for Food Security in South Africa'. *Climate Services* 34 (April):100463. https://doi.org/10.1016/j.cliser.2024.100463.

Rao, K. R. 2019. 'Global Wind Energy and Power Generation Options: Socioeconomic Factors'. In *Wind Energy for Power Generation: Meeting the Challenge of Practical Implementation*, edited by K. R. Rao, 703–828. Cham: Springer International Publishing. https://doi.org/10.1007/978-3-319-75134-4_3.

Ravillard, Pauline, J Enrique Chueca, Mariana Weiss, and Michelle Carvalho Metanias Hallack. 2021. 'Implications of the Energy Transition on Employment: Today's Results, Tomorrow's Needs'.

Richters, Oliver, Christoph Bertram, Elmar Kriegler, Alaa Al Khourdajie, Ryna Cui, Jae Edmonds, Philip Hackstock, et al. 2022. 'NGFS Climate Scenarios Data Set'. Zenodo. https://doi.org/10.5281/zenodo.7085661.

Richters, Oliver, Christoph Bertram, Elmar Kriegler, Jacob Anz, Thessa Beck, David N Bresch, Molly Charles, et al. 2022. 'NGFS Climate Scenarios Database: Technical Documentation V3. 1'.
Ryberg, David Severin, Dilara Gulcin Caglayan, Sabrina Schmitt, Jochen Linßen, Detlef Stolten, and Martin Robinius. 2019. 'The Future of European Onshore Wind Energy Potential: Detailed Distribution and Simulation of Advanced Turbine Designs'. *Energy* 182 (September):1222–38. https://doi.org/10.1016/j.energy.2019.06.052.
SAPP. 2017. 'SAPP POOL Plan 2017'. Southern Africa Power Pool. https://www.sapp.co.zw/sites/default/files/SAPP%20Pool%20Plan%202017%20Main%20Volume_0.pdf.
Schäffer, Linn Emelie, Magnus Korpås, and Tor Haakon Bakken. 2023. 'Implications of Environmental Constraints in Hydropower Scheduling for a Power System with Limited Grid and Reserve Capacity'. *Energy Systems*, July. https://doi.org/10.1007/s12667-023-00594-z.
Schneider, Tapio, Tobias Bischoff, and Gerald H. Haug. 2014. 'Migrations and Dynamics of the Intertropical Convergence Zone'. *Nature* 513 (7516): 45–53. https://doi.org/10.1038/nature13636.
Serdeczny, Olivia, Sophie Adams, Florent Baarsch, and X X. 2017. 'Climate Change Impacts in Sub-Saharan Africa: From Physical Changes to Their Social Repercussions'. *Reg Environ Change* 17:1585–1600. https://doi.org/10.1007/s10113-015-0910-2.
Shirley, Rebekah, Chih-Jung Lee, Hope Nyambura Njoroge, Sarah Odera, Patrick Kioko Mwanzia, Ifeoma Malo, and Yeside Dipo-Salami. 2019. 'Powering Jobs: The Employment Footprint of Decentralized Renewable Energy Technologies in Sub Saharan Africa'. *Journal of Sustainability Research* 2 (1).
Sovacool, Benjamin K., and Sarah E. Ryan. 2016. 'The Geography of Energy and Education: Leaders, Laggards, and Lessons for Achieving Primary and Secondary School Electrification'. *Renewable and Sustainable Energy Reviews* 58 (May):107–23. https://doi.org/10.1016/j.rser.2015.12.219.
Spalding-Fecher, Randall, Mamahloko Senatla, and Francis Yamba. 2017. 'Electricity Supply and Demand Scenarios for the Southern African Power Pool'. *Energy Policy* 101 (2017): 403–14. http://dx.doi.org/10.1016/j.enpol.2016.10.033.
Sterl, Sebastian, Albertine Devillers, Celray James Chawanda, Ann van Griensven, Wim Thiery, and Daniel Russo. 2021. 'A Spatiotemporal Atlas of Hydropower in Africa for Energy Modelling Purposes'. *Open Research Europe* 1.
Stolten, Detlef, Peter Markewitz, Thomas Schöb, Felix Kullmann, Stanley Risch, and Theresa Groß. 2021. *Neue Ziele Auf Alten Wegen? Strategien Für Eine Treibhausgasneutrale Energieversorgung Bis Zum Jahr 2045*.
Sturchio, N. C., X. Du, R. Purtschert, B. E. Lehmann, M. Sultan, L. J. Patterson, Z.-T. Lu, et al. 2004. 'One Million Year Old Groundwater in the Sahara Revealed by Krypton-81 and Chlorine-36'. *Geophysical Research Letters* 31 (5): 2003GL019234. https://doi.org/10.1029/2003GL019234.
Tester, Jefferson W. 2006. 'The Future of Geothermal Energy: Impact of Enhanced Geothermal Systems (EGS) on the United States in the 21st Century'. Edited by Idaho National Laboratory.
Tramberend, Sylvia, Robert Burtscher, Peter Burek, Taher Kahil, Günther Fischer, Junko Mochizuki, Peter Greve, et al. 2021. 'Co-Development of East African Regional Water Scenarios for 2050'. *One Earth* 4 (3): 434–47. https://doi.org/10.1016/j.oneear.2021.02.012.
Vernet, Antoine, Jane N.O. Khayesi, Vivian George, Gerard George, and Abubakar S. Bahaj. 2019. 'How Does Energy Matter? Rural Electrification, Entrepreneurship, and Community Development in Kenya'. *Energy Policy* 126 (March):88–98. https://doi.org/10.1016/j.enpol.2018.11.012.
Welder, Lara, D.Severin Ryberg, Leander Kotzur, Thomas Grube, Martin Robinius, and Detlef Stolten. 2018. 'Spatio-Temporal Optimization of a Future Energy System for Power-to-Hydrogen Applications in Germany'. *Energy* 158 (September):1130–49. https://doi.org/10.1016/j.energy.2018.05.059.

West African Power Pool. 2024. 'WAPP Ongoing Projects - Under Implementation'. http://pipes.ecowapp.org/en/projects/on-going-projects/under-Implementation.

World Bank Group, and Solargis s.r.o. 2023. 'Global Solar Atlas 2.0'. globalsolaratlas.info.

World Resources Institute (WRI). 2023. 'Aqueduct 4.0 Current and Future Global Maps Data'. https://www.wri.org/data/aqueduct-global-maps-40-data.

Xu, Yongxin, and Hans E. Beekman. 2003. *Groundwater Recharge Estimation in Southern Africa*. Vol. 64. Unesco. https://unesdoc.unesco.org/ark:/48223/pf0000132404.

# 5 Appendix

## 5.1 Renewable Energy Potential Values

Table 6: Technical renewable electricity capacity potentials by country and technology

| Country | Technical capacity potential [GW] by technology | | | |
|---|---|---|---|---|
| | Open-field PV | Onshore wind | Hydropower in 2030 | Hydropower in 2050 |
| **Angola** | 21169 | 1328 | 8 | 8 |
| **Benin** | 39 | 7.5 | 0.32 | 0.58 |
| **Burkina Faso** | 3489 | 249 | 0.06 | 0.11 |
| **Botswana** | 3677 | 2071 | - | - |
| **Côte d'Ivoire** | 126 | 3.3 | 1.6 | 2.4 |
| **Democratic Republic of the Congo** | 24430 | 911 | 22 | 22 |
| **Comoros** | 2.7 | 0.04 | - | - |
| **Cape Verde** | 43 | 11 | - | - |
| **Ghana** | 774 | 117 | 1.9 | 2.3 |
| **Guinea** | 15 | 2.3 | 2.3 | 4.2 |
| **Gambia** | 114 | 8.7 | - | - |
| **Guinea-Bissau** | 224 | 2.1 | 0.02 | 0.02 |
| **Kenya** | 7581 | 895 | 1.6 | 1.6 |
| **Liberia** | 104 | 0.09 | 0.4 | 0.77 |
| **Lesotho** | 120 | 15 | 0.27 | 0.27 |
| **Madagascar** | 2706 | 953 | 0.45 | 0.45 |
| **Mali** | 30480 | 4950 | 0.44 | 1.2 |
| **Mozambique** | 13992 | 1153 | 6.1 | 6.1 |
| **Mauritania** | 29274 | 4058 | - | - |
| **Mauritius** | 6 | 0.13 | - | - |
| **Malawi** | 850 | 0.38 | 1.3 | 1.3 |
| **Namibia** | 3748 | 2760 | 0.65 | 0.65 |
| **Niger** | 29504 | 4618 | 0.35 | 0.35 |
| **Nigeria** | 10112 | 957 | 5.9 | 10 |
| **Rwanda** | 76 | 38 | 0.33 | 0.33 |
| **Senegal** | 2139 | 339 | - | 0.13 |
| **Sierra Leone** | 292 | 6.4 | 0.47 | 0.94 |
| **Swaziland** | 117 | 4.4 | 0.19 | 0.19 |
| **Seychelles** | 0.03 | - | - | - |
| **Togo** | 955 | 22 | 0.18 | 0.19 |
| **Tanzania** | 8135 | 190 | 4.6 | 5.3 |
| **Uganda** | 1957 | 115 | 3.3 | 3.9 |
| **South Africa** | 21298 | 3059 | 0.74 | 0.74 |
| **Zambia** | 9227 | 454 | 6.4 | 6.4 |
| **Zimbabwe** | 4520 | 684 | 3.5 | 3.5 |

Notes: values are rounded up to the nearest decimal. Zero values represent values under 0.05; "-" denotes countries for which the use of current input data did not result in any potential. For hydropower it denotes no hydropower plants with the status 'Existing', 'Committed', 'Candidate', 'Planned' with a capacity larger than 10MW were found for the corresponding year according to Sterl et al. (Sterl et al. 2021). There is no capacity potential change between 2030 and 2050 for onshore wind and open-field PV because the local preferences in the land eligibility analysis that determine the installable capacity remain fixed across the evaluated years. Consequently, the same amount of land, installable capacity, and ultimately electricity generation also remains the same. Onshore wind potentials lean more towards a 2050 scenario since a synthetic turbine with improved efficiency was used.

**Table 7: Technical annual renewable electricity generation potentials by country and technology**

| Country | Technical annual generation potential [TWh/a] by technology | | | |
| --- | --- | --- | --- | --- |
| | Open-field PV | Onshore wind | Hydropower in 2030 (Sterl et al. 2021) | Hydropower in 2050 (Sterl et al. 2021) |
| Angola | 45869 | 1562 | 27 | 27 |
| Benin | 80 | 10 | 0.94 | 2 |
| Burkina Faso | 7346 | 454 | 0.11 | 0.22 |
| Botswana | 8891 | 3726 | - | - |
| Côte d'Ivoire | 239 | 3 | 6.5 | 10 |
| Democratic Republic of the Congo | 48209 | 716 | 152 | 152 |
| Comoros | 5.1 | 0.04 | - | - |
| Cape Verde | 97 | 30 | - | - |
| Ghana | 1493 | 119 | 6.3 | 7.4 |
| Guinea | 31 | 2.3 | 9 | 15 |
| Gambia | 242 | 15 | - | - |
| Guinea-Bissau | 466 | 2.8 | 0.05 | 0.05 |
| Kenya | 14517 | 2277 | 5.5 | 5.5 |
| Liberia | 187 | 0.09 | 1.3 | 2.4 |
| Lesotho | 290 | 34 | 1.2 | 1.2 |
| Madagascar | 5688 | 1841 | 2.3 | 2.3 |
| Mali | 67306 | 13117 | 1.8 | 4.3 |
| Mozambique | 28606 | 1655 | 20 | 20 |
| Mauritania | 66552 | 15394 | - | - |
| Mauritius | 11 | 0.34 | - | - |
| Malawi | 1800 | 0.48 | 5.1 | 5.1 |
| Namibia | 9520 | 5050 | 2.6 | 2.6 |
| Niger | 69641 | 12714 | 1.1 | 1.1 |
| Nigeria | 20564 | 1595 | 20 | 34 |
| Rwanda | 140 | 25 | 1.3 | 1.3 |
| Senegal | 4562 | 688 | - | 0.53 |
| Sierra Leone | 551 | 6 | 1.9 | 3.6 |
| Swaziland | 234 | 7.8 | 0.61 | 0.61 |
| Seychelles | 0.05 | - | - | - |
| Togo | 1827 | 24 | 0.43 | 0.48 |
| Tanzania | 16442 | 304 | 23 | 25 |
| Uganda | 3840 | 140 | 20 | 22 |
| South Africa | 52100 | 6357 | 0.81 | 0.81 |
| Zambia | 20529 | 640 | 16 | 16 |
| Zimbabwe | 10210 | 1004 | 13 | 13 |

Notes: values are rounded up to the nearest decimal. Zero values represent values under 0.05; "-" denotes countries for which the use of current input data did not result in any potential. For hydropower it denotes no hydropower plants with the status 'Existing', 'Committed', 'Candidate', 'Planned' with a capacity larger than 10MW were found for the corresponding year according to Sterl et al. (Sterl et al. 2021). There is no electricity generation potential change between 2030 and 2050 for onshore wind and open-field PV because the local preferences in the land eligibility analysis that determine the installable capacity remain fixed across the evaluated years. Consequently, the same amount of land, installable capacity, and ultimately electricity generation also remains the same. Onshore wind potentials lean more towards a 2050 scenario since a synthetic turbine with improved efficiency was used.

## 5.2 Local Green Hydrogen Potential Values

**Table 8: National hydrogen potentials and energy demands**

| Country | Potential | Energy demand | | |
|---|---|---|---|---|
| Full name | $H_2$ potential [TWh/a] | Electricity demand 2050 (H2 equivalent) [TWh/a] | $H_2$ demand 2050 [TWh/a] | Total demand/potential share [%] |
| Republic of Angola | 32892 | 59.0 | 6.7 | 0.2% |
| Republic of Benin | 61 | 19.1 | 1.8 | 34.3% |
| People's Republic of Burkina Faso | 5302 | 19.7 | 1.9 | 0.4% |
| Republic of Botswana | 8980 | 8.5 | 1.0 | 0.1% |
| Republic of Côte d'Ivoire | 168 | 64.1 | 6.2 | 41.8% |
| Democratic Republic of the Congo | 34058 | 46.5 | 4.5 | 0.2% |
| Union of the Comoros | 4 | 1.8 | 0.1 | 48.4% |
| Republic of Cabo Verde | 92 | 1.1 | 0.1 | 1.3% |
| Republic of Ghana | 1113 | 42.3 | 4.1 | 4.2% |
| Republic of Guinea | 25 | 33.0 | 3.2 | 144.5% |
| Republic of the Gambia | 180 | 3.8 | 0.4 | 2.3% |
| Republic of Guinea-Bissau | 324 | 1.8 | 0.2 | 0.6% |
| Republic of Kenya | 12611 | 138.1 | 8.7 | 1.2% |
| Republic of Liberia | 128 | 3.5 | 0.3 | 3.0% |
| Kingdom of Lesotho | 233 | 4.1 | 0.5 | 2.0% |
| Republic of Madagascar | 5330 | 49.1 | 3.1 | 1.0% |
| Republic of Mali | 55099 | 11.0 | 1.1 | 0.02% |
| Republic of Mozambique | 21318 | 33.2 | 3.8 | 0.2% |
| Islamic Republic of Mauritania | 57272 | 8.2 | 0.8 | 0.02% |
| Republic of Mauritius | 8 | 5.7 | 0.4 | 76.3% |
| Republic of Malawi | 1266 | 27.8 | 3.2 | 2.4% |
| Republic of Namibia | 10135 | 10.9 | 1.2 | 0.1% |
| Republic of the Niger | 56220 | 7.8 | 0.8 | 0.02% |
| Federal Republic of Nigeria | 15417 | 548.8 | 52.7 | 3.9% |
| Rwandese Republic | 112 | 39.4 | 2.5 | 37.4% |
| Republic of Senegal | 3683 | 31.1 | 3.0 | 1.0% |
| Republic of Sierra Leone | 385 | 4.4 | 0.4 | 1.3% |
| Kingdom of Eswatini | 170 | 3.3 | 0.4 | 2.2% |
| Republic of Seychelles | 0.036 | 0.6 | 0.03 | 1632.7% |
| Togolese Republic | 1273 | 5.0 | 0.5 | 0.4% |
| United Republic of Tanzania | 11265 | 104.9 | 12.0 | 1.0% |
| Republic of Uganda | 2742 | 110.4 | 7.0 | 4.3% |
| Republic of South Africa | 41273 | 397.3 | 76.0 | 1.2% |
| Republic of Zambia | 14599 | 63.8 | 7.3 | 0.5% |
| Republic of Zimbabwe | 7855 | 14.6 | 1.7 | 0.2% |
| Total | 401593 | 1923.5 | 216.95 | 0.05% |

**Table 9: Potential-weighted region-average LCOH per country for different expansion steps and years**

| Country | Levelized cost of hydrogen (LCOH) (region averages) | | | |
|---|---|---|---|---|
| Full name | 1% expansion (2030) | 25% expansion (2030) | 1% expansion (2050) | 25% expansion (2050) |
| Republic of Angola | 2.7 € | 2.8 € | 2.0 € | 2.0 € |
| Republic of Benin | 2.9 € | 2.9 € | 2.0 € | 2.1 € |
| People's Republic of Burkina Faso | 2.8 € | 2.8 € | 2.0 € | 2.0 € |
| Republic of Botswana | 2.6 € | 2.7 € | 1.8 € | 1.9 € |
| Republic of Côte d'Ivoire | 3.0 € | 3.0 € | 2.1 € | 2.1 € |
| Democratic Republic of the Congo | 2.9 € | 3.0 € | 2.1 € | 2.1 € |
| Union of the Comoros | 2.9 € | 2.9 € | 2.0 € | 2.1 € |
| Republic of Cabo Verde | 2.4 € | 2.6 € | 1.9 € | 1.9 € |
| Republic of Ghana | 2.8 € | 3.0 € | 2.0 € | 2.1 € |
| Republic of Guinea | 2.6 € | 2.6 € | 1.9 € | 1.9 € |
| Republic of the Gambia | 2.8 € | 2.8 € | 2.0 € | 2.0 € |
| Republic of Guinea-Bissau | 2.9 € | 2.9 € | 2.1 € | 2.0 € |
| Republic of Kenya | 2.8 € | 2.9 € | 2.1 € | 2.1 € |
| Republic of Liberia | 3.2 € | 3.2 € | 2.2 € | 2.3 € |
| Kingdom of Lesotho | 2.6 € | 2.6 € | 1.8 € | 1.9 € |
| Republic of Madagascar | 2.6 € | 2.7 € | 1.9 € | 2.0 € |
| Republic of Mali | 2.7 € | 2.7 € | 1.9 € | 1.9 € |
| Republic of Mozambique | 2.8 € | 2.9 € | 2.0 € | 2.0 € |
| Islamic Republic of Mauritania | 2.2 € | 2.4 € | 1.7 € | 1.8 € |
| Republic of Mauritius | 2.9 € | 3.0 € | 2.1 € | 2.1 € |
| Republic of Malawi | 2.8 € | 2.9 € | 2.0 € | 2.0 € |
| Republic of Namibia | 2.6 € | 2.7 € | 1.8 € | 2.0 € |
| Republic of the Niger | 2.6 € | 2.6 € | 1.8 € | 1.9 € |
| Federal Republic of Nigeria | 2.8 € | 2.9 € | 2.0 € | 2.0 € |
| Rwandese Republic | 3.0 € | 3.1 € | 2.1 € | 2.2 € |
| Republic of Senegal | 2.8 € | 2.8 € | 2.0 € | 2.0 € |
| Republic of Sierra Leone | 2.7 € | 3.1 € | 2.1 € | 2.2 € |
| Kingdom of Eswatini | 3.0 € | 3.1 € | 2.1 € | 2.2 € |
| Republic of Seychelles | 3.2 € | 3.2 € | 2.2 € | 2.3 € |
| Togolese Republic | 3.0 € | 3.0 € | 2.1 € | 2.1 € |
| United Republic of Tanzania | 2.9 € | 2.9 € | 2.1 € | 2.1 € |
| Republic of Uganda | 2.9 € | 3.0 € | 2.1 € | 2.1 € |
| Republic of South Africa | 2.6 € | 2.6 € | 1.8 € | 1.8 € |
| Republic of Zambia | 2.7 € | 2.8 € | 2.0 € | 2.0 € |
| Republic of Zimbabwe | 2.8 € | 2.8 € | 2.0 € | 2.0 € |
| Total | 2.63 € | 2.70 € | 1.90 € | 1.94 € |

## 5.3 Groundwater sustainable yield national results

The national-level summary of groundwater sustainable yield estimates is provided for the years 2020 (2015–2035), 2030 (2015–2045), and 2050 (2036–2065) under the RCP2.6 and RCP8.5 climate scenarios for conservative (Table 10), medium (Table 11), and extreme (Table 12) environmental scenarios. In planning for water supply, three environmental scenarios are considered based on the percentage of annual recharge reserved for environmental flow (preserving nature itself). The extreme scenario allocates 30% of the recharge to environmental flow, leaving 70% for supplementary uses like green hydrogen production. The medium scenario sets aside 60% for environmental flow, resulting in 40% for other uses. The conservative scenario dedicates 90% to environmental flow, leaving just 10% for supplementary uses. These scenarios are considered for better comparison and to promote sustainability.

Table 10. The conservative groundwater sustainable yield estimation for 2020 (2015 – 2035), 2030 (2015 – 2045), and 2050 (2036 – 2065) for West and South-eastern African countries under RCP2.6 and RCP8.5 climate scenarios.

| Country | Groundwater sustainable yield (conservative scenario) [mm yr$^{-1}$] | | | | | |
| --- | --- | --- | --- | --- | --- | --- |
| | 2020 | | 2030 | | 2050 | |
| | RCP2.6 | RCP8.5 | RCP2.6 | RCP8.5 | RCP2.6 | RCP8.5 |
| Angola | 29.8 | 28.7 | 29.6 | 28.6 | 27.3 | 26.4 |
| Botswana | 13.0 | 11.3 | 13.0 | 11.2 | 11.3 | 10.7 |
| Democratic Republic of the Congo | 28.7 | 27.4 | 27.7 | 27.0 | 25.2 | 24.9 |
| Comoros | 91.6 | 79.4 | 86.4 | 80.1 | 85.4 | 83.8 |
| Lesotho | 52.1 | 51.3 | 51.7 | 50.0 | 48.0 | 49.1 |
| Madagascar | 30.2 | 31.7 | 28.5 | 31.0 | 27.2 | 28.3 |
| Mozambique | 11.3 | 9.9 | 10.2 | 9.8 | 9.0 | 8.8 |
| Mauritius | 0.0 | 0.0 | 0.0 | 0.0 | 0.0 | 0.0 |
| Malawi | 7.2 | 5.7 | 5.5 | 4.9 | 3.5 | 3.1 |
| Namibia | 8.1 | 7.8 | 8.3 | 7.6 | 7.5 | 7.2 |
| Swaziland | 22.5 | 23.8 | 22.4 | 22.3 | 19.6 | 19.7 |
| Seychelles | - | - | - | - | - | - |
| Tanzania | 8.6 | 8.7 | 8.1 | 8.4 | 7.5 | 7.3 |
| South Africa | 15.7 | 14.8 | 15.7 | 14.5 | 13.9 | 13.9 |
| Zambia | 14.5 | 13.3 | 13.2 | 13.0 | 11.7 | 11.1 |
| Zimbabwe | 6.8 | 6.0 | 6.1 | 5.8 | 5.3 | 5.2 |
| Kenya | 3.9 | 3.7 | 3.3 | 3.3 | 2.4 | 3.1 |
| Rwanda | 9.3 | 7.5 | 6.8 | 6.0 | 3.5 | 3.5 |
| Uganda | 5.0 | 4.5 | 4.1 | 3.7 | 2.6 | 2.5 |
| Benin | 5.2 | 5.0 | 4.1 | 4.2 | 2.2 | 2.0 |
| Burkina Faso | 3.1 | 3.7 | 2.7 | 3.0 | 1.6 | 1.3 |
| Côte d'Ivoire | 11.6 | 10.3 | 10.6 | 10.1 | 8.9 | 8.6 |
| Cape Verde | - | - | - | - | - | - |
| Ghana | 11.8 | 10.6 | 10.5 | 9.8 | 8.2 | 6.6 |
| Guinea | 29.5 | 27.3 | 28.5 | 26.5 | 25.3 | 24.4 |
| Gambia | 2.5 | 1.3 | 1.4 | 0.9 | 0.3 | 0.3 |
| Guinea-Bissau | 13.1 | 11.2 | 12.4 | 10.0 | 9.8 | 7.6 |
| Liberia | 56.8 | 50.8 | 55.8 | 51.5 | 51.4 | 50.0 |
| Mali | 2.5 | 2.4 | 2.3 | 2.3 | 1.9 | 1.7 |
| Niger | 4.0 | 4.1 | 3.8 | 3.8 | 3.2 | 2.8 |
| Nigeria | 10.9 | 10.3 | 9.2 | 8.7 | 6.1 | 5.2 |
| Senegal | 2.6 | 2.0 | 1.9 | 1.6 | 1.1 | 1.0 |
| Sierra Leone | 69.8 | 63.5 | 69.9 | 63.6 | 65.2 | 62.4 |
| Togo | 7.4 | 7.1 | 6.2 | 6.4 | 4.3 | 3.9 |
| Mauritania | 1.3 | 1.1 | 1.2 | 1.1 | 1.2 | 0.9 |

Note: "-" indicates a lack of reliable input data in the simulations, not an absence of potential groundwater sustainable yield.

**Table 11. The medium groundwater sustainable yield estimation for 2020 (2015 – 2035), 2030 (2015 – 2045), and 2050 (2036 – 2065) for West and South-eastern African countries under RCP2.6 and RCP8.5 climate scenarios.**

| Country | Groundwater sustainable yield (medium scenario) [mm yr$^{-1}$] | | | | | |
|---|---|---|---|---|---|---|
| | 2020 | | 2030 | | 2050 | |
| | RCP2.6 | RCP8.5 | RCP2.6 | RCP8.5 | RCP2.6 | RCP8.5 |
| Angola | 120.8 | 116.2 | 120.4 | 116.5 | 112.6 | 108.7 |
| Botswana | 52.4 | 45.4 | 52.1 | 45.2 | 45.6 | 43.1 |
| Democratic Republic of the Congo | 116.2 | 111.2 | 113.1 | 110.3 | 105.4 | 104.3 |
| Comoros | 366.7 | 317.7 | 345.7 | 320.4 | 341.8 | 335.4 |
| Lesotho | 216.5 | 213.3 | 216.6 | 209.4 | 206.0 | 209.6 |
| Madagascar | 137.7 | 144.8 | 132.4 | 143.6 | 132.7 | 137.7 |
| Mozambique | 46.9 | 41.2 | 42.8 | 41.0 | 39.0 | 38.5 |
| Mauritius | - | - | - | - | - | - |
| Malawi | 41.9 | 34.6 | 35.6 | 32.9 | 26.6 | 24.4 |
| Namibia | 32.9 | 31.7 | 33.5 | 30.7 | 30.4 | 29.2 |
| Swaziland | 107.6 | 113.0 | 108.1 | 107.8 | 99.7 | 99.4 |
| Seychelles | - | - | - | - | - | - |
| Tanzania | 38.7 | 38.9 | 37.4 | 38.6 | 37.3 | 36.7 |
| South Africa | 70.4 | 66.7 | 70.4 | 65.7 | 63.4 | 63.4 |
| Zambia | 62.0 | 57.2 | 57.0 | 56.5 | 52.1 | 49.7 |

| Country | | | | | | |
|---|---|---|---|---|---|---|
| Zimbabwe | 31.9 | 28.3 | 29.1 | 28.1 | 26.3 | 25.8 |
| Kenya | 21.2 | 20.1 | 19.0 | 19.1 | 16.0 | 19.4 |
| Rwanda | 59.7 | 53.2 | 52.7 | 50.1 | 40.5 | 41.2 |
| Uganda | 27.4 | 25.5 | 24.7 | 22.6 | 18.7 | 17.9 |
| Benin | 28.3 | 27.5 | 25.7 | 26.0 | 20.8 | 18.8 |
| Burkina Faso | 17.8 | 20.6 | 16.8 | 18.6 | 13.7 | 12.1 |
| Côte d'Ivoire | 53.2 | 48.0 | 50.5 | 48.4 | 46.5 | 45.0 |
| Cape Verde | 0.0 | 0.0 | 0.0 | 0.0 | 0.0 | 0.0 |
| Ghana | 56.9 | 52.0 | 53.0 | 49.9 | 46.3 | 38.9 |
| Guinea | 129.5 | 120.6 | 127.0 | 118.7 | 118.1 | 114.3 |
| Gambia | 18.7 | 12.6 | 14.5 | 11.3 | 6.9 | 7.8 |
| Guinea-Bissau | 61.2 | 53.6 | 59.6 | 49.9 | 53.8 | 44.4 |
| Liberia | 232.0 | 208.1 | 229.8 | 212.6 | 219.4 | 214.5 |
| Mali | 11.1 | 11.0 | 10.6 | 10.6 | 9.5 | 8.4 |
| Niger | 17.2 | 17.5 | 16.5 | 16.7 | 14.9 | 13.0 |
| Nigeria | 63.0 | 60.6 | 58.0 | 55.9 | 47.0 | 41.8 |
| Senegal | 13.3 | 10.7 | 11.0 | 9.3 | 7.9 | 7.5 |
| Sierra Leone | 300.3 | 274.8 | 302.6 | 277.3 | 290.9 | 280.4 |
| Togo | 43.4 | 41.8 | 39.6 | 40.2 | 33.2 | 30.2 |
| Mauritania | 5.2 | 4.5 | 4.9 | 4.4 | 4.8 | 3.9 |

Note: "-" indicates a lack of reliable input data in the simulations, not an absence of potential groundwater sustainable yield.

**Table 12. The extreme groundwater sustainable yield estimation for 2020 (2015 – 2035), 2030 (2015 – 2045), and 2050 (2036 – 2065) for West and South-eastern African countries under RCP2.6 and RCP8.5 climate scenarios.**

| Country | Groundwater sustainable yield (extreme scenario) [mm yr$^{-1}$] | | | | | |
|---|---|---|---|---|---|---|
| | 2020 | | 2030 | | 2050 | |
| | RCP2.6 | RCP8.5 | RCP2.6 | RCP8.5 | RCP2.6 | RCP8.5 |
| Angola | 211.9 | 203.8 | 211.3 | 204.4 | 198.0 | 191.2 |
| Botswana | 91.7 | 79.5 | 91.3 | 79.2 | 79.8 | 75.5 |
| Democratic Republic of the Congo | 203.8 | 195.0 | 198.6 | 193.7 | 185.8 | 183.9 |
| Comoros | 641.7 | 556.0 | 605.0 | 560.8 | 598.1 | 587.1 |
| Lesotho | 381.1 | 375.6 | 381.7 | 369.0 | 364.2 | 370.3 |
| Madagascar | 248.2 | 260.9 | 239.4 | 259.3 | 241.6 | 250.6 |
| Mozambique | 82.7 | 72.6 | 75.6 | 72.4 | 69.2 | 68.4 |
| Mauritius | - | - | - | - | - | - |
| Malawi | 78.5 | 65.6 | 68.6 | 64.0 | 55.5 | 51.5 |

| Country | | | | | | |
|---|---|---|---|---|---|---|
| Namibia | 57.6 | 55.5 | 58.7 | 53.9 | 53.4 | 51.3 |
| Swaziland | 193.2 | 202.6 | 194.5 | 194.0 | 181.1 | 180.6 |
| Seychelles | - | - | - | - | - | - |
| Tanzania | 69.2 | 69.5 | 67.1 | 69.2 | 68.1 | 67.0 |
| South Africa | 126.1 | 119.6 | 126.1 | 117.9 | 114.1 | 114.1 |
| Zambia | 110.0 | 101.5 | 101.2 | 100.2 | 92.9 | 88.7 |
| Zimbabwe | 57.4 | 51.1 | 52.5 | 50.8 | 47.8 | 46.9 |
| Kenya | 39.6 | 37.7 | 36.1 | 36.3 | 31.7 | 37.7 |
| Rwanda | 113.3 | 102.3 | 103.9 | 99.8 | 90.8 | 92.5 |
| Uganda | 50.7 | 47.5 | 46.9 | 43.4 | 38.7 | 37.2 |
| Benin | 52.1 | 50.5 | 48.1 | 48.8 | 41.9 | 38.2 |
| Burkina Faso | 33.0 | 37.9 | 31.7 | 35.1 | 27.9 | 25.2 |
| Côte d'Ivoire | 95.2 | 85.9 | 90.8 | 87.1 | 85.1 | 82.5 |
| Cape Verde | 0.0 | 0.0 | 0.0 | 0.0 | 0.0 | 0.0 |
| Ghana | 103.1 | 94.4 | 96.9 | 91.6 | 87.1 | 73.8 |
| Guinea | 230.3 | 214.9 | 226.2 | 211.7 | 211.7 | 205.1 |
| Gambia | 35.5 | 24.8 | 28.9 | 23.1 | 16.9 | 18.7 |
| Guinea-Bissau | 109.6 | 96.3 | 107.2 | 90.2 | 98.2 | 81.6 |
| Liberia | 407.1 | 365.4 | 403.8 | 373.8 | 387.6 | 379.1 |
| Mali | 19.9 | 19.7 | 19.0 | 19.1 | 17.3 | 15.4 |
| Niger | 30.4 | 31.1 | 29.4 | 29.7 | 26.8 | 23.7 |
| Nigeria | 117.5 | 113.3 | 110.3 | 106.8 | 94.4 | 84.9 |
| Senegal | 24.6 | 19.8 | 20.8 | 17.6 | 15.7 | 15.0 |
| Sierra Leone | 531.1 | 486.5 | 535.5 | 491.3 | 516.8 | 498.9 |
| Togo | 80.5 | 77.6 | 74.6 | 75.6 | 65.1 | 59.8 |
| Mauritania | 9.1 | 8.0 | 8.7 | 7.7 | 8.4 | 6.9 |

Note: "-" indicates a lack of reliable input data in the simulations, not an absence of potential groundwater sustainable yield.